\documentclass[onecolumn]{aa} 

\usepackage{graphicx}
\usepackage{txfonts}
\usepackage{xcolor}
\begin{document}

   \title{Near-IR and optical radial velocities of the active M dwarf star Gl~388 (AD~Leo)  with SPIRou at CFHT and SOPHIE at 
   	  OHP\thanks{Based on observations obtained with the spectropolarimeter SPIRou at the Canada-France-Hawaii Telescope (CFHT)
	  and the spectrograph SOPHIE 
          at the Observatoire de Haute-Provence (OHP) in France.}:
          }

   \subtitle{A 2.23 day rotation period and no evidence for a corotating planet.
          }

   \author{A. Carmona
          \inst{1}\fnmsep
          \and
          X. Delfosse \inst{1}
          \and
          S. Bellotti \inst{2,13}
          \and
          P. Cort\'es-Zuleta \inst{3}
           \and
          M. Ould-Elhkim \inst{2}
          \and
          N. Heidari \inst{3}
          \and 
          L. Mignon \inst{1,9}
          \and
          J.F. Donati \inst{2}
          \and
          C. Moutou \inst{2}
          \and
          N. Cook \inst{4}
          \and
          E. Artigau \inst{4,5}
          \and
          P. Fouqu\'e \inst{2}
          \and 
          E. Martioli \inst{6,7}
          \and
          C. Cadieux \inst{4}
          \and
          J. Morin \inst{8}
          \and
          T. Forveille \inst{1}
          \and
          I. Boisse \inst{3}
          \and
          G. H\'ebrard \inst{7}
          \and
          R. F. D\'iaz \inst{10}  
          \and
          D. Lafreni\`ere \inst{4}
          \and
          F. Kiefer \inst{7}
           \and
          P. Petit \inst{2}
           \and
          R. Doyon \inst{4}
          \and
          L. Acu\~{n}a \inst{3}
          \and
          L. Arnold \inst{11}
          \and 
          X. Bonfils \inst{1}
          \and
          F. Bouchy \inst{9}
          \and 
          V. Bourrier \inst{9}
          \and 
          S. Dalal \inst{7}
          \and
          M. Deleuil \inst{3}
          \and
          O. Demangeon \inst{12}
          \and 
          X. Dumusque \inst{9}
          \and
          N. Hara \inst{9}
          \and
          S. Hoyer \inst{3}
          \and
          O. Mousis \inst{3}
          \and
          A. Santerne \inst{3}
          \and
          D. S\'egrasan \inst{8}
          \and
          M. Stalport \inst{9}
          \and
          S. Udry \inst{9}
          }
 
   \institute{Univ. Grenoble Alpes, CNRS, IPAG, 38000 Grenoble, France\\
    \email{andres.carmona@univ-grenoble-alpes.fr}     
    \and
     Universit\'e de Toulouse, UPS-OMP, IRAP, 14 avenue E. Belin, Toulouse, F-31400, France
     \and
     Aix Marseille Univ, CNRS, CNES, Institut Origines, LAM, Marseille, France
     \and
     Institute for Research on Exoplanets, Universit\'e de Montr\'eal, D\'epartement de Physique, C.P. 6128 Succ. Centre-ville, Montr\'eal, QC H3C 3J7, Canada
     \and 
     Observatoire du Mont-M\'egantic, Universit\'e de Montr\'eal, D\'epartement de Physique, C.P. 6128 Succ. Centre-ville, Montr\'eal, QC H3C 3J7, Canada
     \and 
     Laborat\'{o}rio Nacional de Astrof\'{i}sica, Rua Estados Unidos 154, 37504-364, Itajub\'{a} - MG, Brazil
     \and
      Sorbonne Universit\'e , CNRS, UMR 7095, Institut d'Astrophysique de Paris, 98 bis bd Arago, F-75014 Paris, France
      \and
      Universit\'e de Montpellier, CNRS, LUPM, F-34095 Montpellier, France
      \and
      Observatoire Astronomique de l'Universit\'e de Gen\`eve, Chemin Pegasi 51b, 1290 Versoix, Switzerland
      \and
      International Center for Advanced Studies and ICIFI (CONICET),
    ECyT-UNSAM, Campus Miguelete, 25 de Mayo y Francia, (1650), Buenos Aires, Argentina
          \and
     Canada France Hawaii Telescope (CFHT) Corporation, UAR2208 CNRS-INSU, 65-1238 Mamalahoa Hwy, Kamuela 96743 HI, USA
     \and
     Instituto de Astrof\'isica e Ci\^encias do Espa\c co, Universidade do Porto, CAUP, Rua das Estrelas, 4150-762 Porto, Portugal
     \and
     Science Division, ESA/ESTEC, Keplerlaan 1, 2201 AZ, Noordwijk, The Netherlands \\
 }

   \date{}

\titlerunning{SPIRou and SOPHIE radial velocities of Gl 388 (AD Leo)}
 
  \abstract
  {The search for extrasolar planets around the nearest M dwarfs is a crucial step toward identifying the nearest Earth-like planets.
  One of the main challenges in this search is that M dwarfs can be magnetically active and stellar activity can produce radial velocity (RV) signals 
  that could mimic those of a planet. 
  } 
   {We aim to investigate whether the 2.2 day period observed in optical RVs of the nearby active M dwarf star Gl~388 (AD~Leo) 
   is due to stellar activity or to a planet that corotates with the star as suggested in the past.}
   {We obtained quasi-simultaneous RVs of Gl~388 from 2019 to 2021 with SOPHIE, 
   the optical \'echelle spectrograph  ($R\sim$75k) at the Observatoire de Haute-Provence in France,
   and RV and Stokes V measurements with SPIRou, 
   the near-infrared spectropolarimeter at the Canada France Hawaii Telescope ($R\sim$70k).
   }
   {
   The SOPHIE RV time series (precision of  3 to 5 m s$^{-1}$ per visit) displays a periodic signal with a 2.23$\pm$0.01 day period and 
   23.6$\pm$0.5 m s$^{-1}$ amplitude, which is consistent with previous HARPS observations obtained in 2005$-$2006.
   The SPIRou RV time series (precision of 2 m s$^{-1}$ per visit) is flat at 5 m s$^{-1}$ rms  and displays no periodic signals. 
   RV signals of amplitude higher than 5.3~m~s$^{-1}$ at a period of 2.23 days can be excluded with 
   a confidence level higher than 99\%.    
   Using the modulation of the longitudinal magnetic field (B$_\ell$) measured with SPIRou as a proxy of stellar rotation, 
   we measure a rotation period of 2.2305$\pm$0.0016 days.}
   { 
   SPIRou RV measurements provide solid evidence that the periodic variability of the optical RVs of Gl~388 
   is due to stellar activity rather than to a corotating planet. 
   The magnetic activity nature of the optical RV signal is further confirmed by the modulation of 
   B$_\ell$ with the same period. 
   The SPIRou campaign on Gl~388 demonstrates the power of near-infrared RV to confirm or infirm planet candidates discovered in the optical 
   around active stars. 
   Our SPIRou observations additionally reiterate how effective spectropolarimetry is at determining the stellar rotation period from the variations of B$_\ell$. 
   }


   \maketitle

\twocolumn

\section{Introduction}
A crucial milestone in the long-term goal of identifying life outside the Solar System
is to identify and characterize rocky planets orbiting the nearest stars which 
could have liquid water on the surface.
As the great majority of main sequence stars within 25 pc are M dwarfs \citep[$\sim$ 75\%,][]{Henry2006},
there has been a growing interest in making a census of their planets. 
In addition to their number, M dwarfs are particularly exciting targets since
transit studies in field stars  \cite[e.g.,][]{DressingCharbonneau2015, Ballard2016, Gaidos2016, Hsu2020}
and radial velocity (RV) surveys \cite[e.g.,][]{Bonfils2013,Sabotta2021}
show that rocky planets occur more often around M dwarfs than around Sun-like stars.
Furthermore, as M dwarfs have low masses, Earth-like planets in their habitable zone can induce a RV signal of a few m s$^{-1}$, 
which can be detected with current ground-based velocimeters.

Detecting low-mass planets of M dwarfs, however, faces several challenges.
Late-type M dwarfs, even when nearby, are faint in the optical domain, 
which makes obtaining the high signal-to-noise (S/N) 
high-resolution spectra required for precise RV challenging with moderate aperture telescopes. 
M dwarfs are also frequently  magnetically active. 
Stellar activity and the spots and plages it creates on the stellar surface
can generate signatures in the RV time series that can be confusingly 
similar to those produced by planets \citep[e.g.,][]{Queloz2001,Klein2021}.

Activity indicators, such as the integrated line flux or the equivalent width of activity sensitive lines, 
or the change in the symmetry of the cross-correlation function (e.g.,~bisector) and its 
(anti) correlation with the RV jitter can shed light on whether 
a RV signal is due to planets or to stellar activity \citep[e.g.,][]{Desort2007}.
A more powerful test, however, is whether
the RV signal has the same amplitude and phase in optical and near-infrared (near-IR) observations.
The RV signature of a planet is achromatic, 
while activity induces a different RV signature in the near-IR.
Furthermore,
the activity RV signature is expected to be weaker in the near-IR if it is dominated by temperature features because 
the brightness contrast between the stellar photosphere and spots or plages decreases at longer wavelengths \citep[e.g.,][]{Huelamo2008, Mahmud2011, Bailey2012}.  

In this paper, we investigate the magnetically active M dwarf star Gl~388 (AD~Leo), 
as a test of how much SPIRou can improve the RV  performance in active stars compared to optical velocimeters. 
Optical RV measurements of Gl~388 show a stable periodic signal 
\cite[][]{Morin2008,Bonfils2013,Reiners2013},
which has mostly been interpreted as due to activity but which \cite{Tuomi2018} claim is caused by a 0.24 M$_J$ (76 M$_\oplus$) 
planet that corotates with the star.
Recent RV work in the optical and near-IR by \citet[][]{Carleo2020} and  \citet[][]{Kossakowski2022} 
have provided further additional evidence against the planet suggested by  \cite{Tuomi2018}.
However, these works have not been able to firmly reject the hypothesis of the corotating planet.
\citet[][]{Kossakowski2022} find, for instance, that a mixed model of a stable 
($K=$16.6 m~s$^{-1}$, 3$\sigma$ upper limit on the putative planet of 27 M$_\oplus$) 
plus a quasi-periodic red-noise  model best explains their RV data.
Near-IR RV measurements with a higher accuracy can settle
the issue of the corotating planet around Gl 388 for good, 
which is what we set to do in this paper.

We obtained quasi-simultaneous RV measurements from 2019 to 2021
in the optical, with SOPHIE\footnote{Spectrographe pour l'Observation des Ph\'enom\`enes des Int\'erieurs stellaires et des Exoplan\`etes, meaning, spectrograph for the observation 
of the phenomena of the stellar interiors and of the exoplanets.} 
at the OHP, and in the near-IR, 
with SPIRou\footnote{Spectro-Polarim\`etre~InfraRouge, meaning, infrared spectropolarimeter.} at the CFHT.
SPIRou observations were obtained in the framework of the SPIRou Legacy Survey, 
a large observing program of 310 nights with SPIRou dedicated to study planetary 
systems around nearby M dwarfs as well as magnetized stars and planet formation.
We use the SOPHIE and SPIRou RVs and activity indicators, 
and SPIRou measurements of
the longitudinal magnetic field (B$_\ell$), to probe whether the RV
signal detected in the optical is indeed due to a planet, 
or instead whether it reflects stellar activity. 

The paper is organized as follows.
Section 2 presents Gl~388 and reviews previous works on the star.
Section 3 describes the new observations with SOPHIE and SPIRou, 
the data reduction methods, 
the details of the measurements of RV, 
and the technique used to measure B$_\ell$ with SPIRou.
In Section 4, we present the RV analysis.  
In Section 5, we perform the activity analysis.
In Section 6, we discuss our results and present our final conclusions.
 
\begin{table}
\begin{center}

\caption{Stellar properties of Gl~388}
\begin{tabular}{l|cc}
\hline\hline
Parameter & Value & Ref. \\[1mm]
\hline
Sp. Type             & M3.4              & 1 \\[1mm]
T${\rm eff}$ [K]     & 3370$\pm$60       & 1 \\[1mm]
d~[pc]               & 4.9651$\pm$0.0007 & 2 \\[1mm]
Mass~[M$_{\odot}$]   & 0.41$\pm$0.04     & 1 \\[1mm]
Radius~[R$_{\odot}$] & 0.43$\pm$0.02     & 1 \\[1mm]
L~[L$_{\odot}$]  & 0.0227 +- 0.0003 & 1,2 \\[1mm]
[Fe/H]               & +0.15$\pm$0.08    & 1\\[1mm]
Age~[Myr]            & 25$-$300          & 3\\[1mm]
Rot. Period~[days]   & 2.2305$\pm$0.0016 & 4\\[1mm]
i~[$^\circ$]         & $\simeq$20        & 5\\[1mm]
\hline

\end{tabular}
\tablefoot{ 
(1) \citet[][]{Mann2015};
(2) \citet[][]{Gaia2022,Gaia2021};
(3) \citet[][]{Shkolnik2009};
(4) this work;
(5) \citet[][]{Morin2008}.}
\label{stellar-properties}
\end{center}
\end{table}

\section{Gl 388}
\label{Gl388}
Gl~388 is a well-known active flare M dwarf star, at a distance of 4.96 pc~\citep[][]{Gaia2022,Gaia2021}.
Table~\ref{stellar-properties} summarizes its fundamental stellar properties.
Its optical spectra display strong chromospheric emission lines, including a prominent H$\alpha$ emission with central absorption, 
Na D lines with emission cores,
a \ion{Ca}{ii}  triplet heavily filled with emission,
and  He I $\lambda$ 5876  in emission \citep[e.g.,][]{Pettersen1981}.
\citet[][]{Morin2008} studied Gl 388 with the ESPaDOnS\footnote{\'Echelle SpectroPolarimetric Device for the Observation of Stars.}  spectropolarimeter, 
and report an effective Rossby number of  4.7$\times10^{-2}$, a $R~{\rm sin}~i$ of 0.13 R$_\odot$, and an inclination of 20$^\circ$. 
They find a strong negative polarimetric signature (i.e., a longitudinal field directed toward the object)
and that the RV smoothly varies by about 40~m~s$^{-1}$ (30 m s$^{-1}$ average accuracy per measurement) over the rotational cycle.
Assuming solid body rotation, Zeeman Doppler imaging (ZDI) analysis indicated a rotational period of 2.2399 $\pm$ 0.0006 ($3\sigma$) days.
Using a projected rotational velocity of $\upsilon$~sin~$i$ of 3.0~km~s$^{-1}$,
ZDI  revealed an average magnetic flux $B\simeq$ 0.2 kG and a strong polar spot with a radial field with a maximum magnetic flux $B\simeq$ 1.3 kG. 
The prominent component observed is the radial component of a dipole almost 
aligned with the rotational axis which indicated a strongly axisymmetric magnetic
topology with 90\% of the large-scale energy in the $m=$~0 modes.

The \citet[][]{Bonfils2013} M dwarf survey 
with the HARPS\footnote{High-Accuracy Radial velocity Planet Searcher \citep[][]{Mayor2003}.} instrument report high-precision RVs of Gl~388. 
Their RV time series periodogram contains a significant power excess at 1.8 and a 2.2 day periods, 
which are 1-day aliases of each other, and the latter which is consistent with the stellar rotational period previously identified through ZDI. 
The authors performed a one-planet fit to the data obtaining a period of $2.2267\pm0.0001$ days, 
a semiamplitude $K=30.5\pm1.3$ m s$^{-1}$, and an eccentricity of $0.1$.
\citet[][]{Bonfils2013} show that the  bisector span has a periodogram with broad power at shorter periods 
and that it is anticorrelated with the RV, which strongly suggests that stellar activity is responsible 
for the periodic RV variation.
\citet[][]{Reiners2013} analyzed the same HARPS data set with an alternative template-matching
technique \citep[HARPS-TERRA\footnote{Template-Enhanced Radial velocity Re-analysis Application.} software,][]{Anglada-Escude2012}, 
and independently confirm  the conclusions of \citet[][]{Bonfils2013}. 
Analyzing the RV behavior in seven sections of the spectra, centered at 419, 450, 486, 528, 583, 631, and 665~nm, 
they found the same 2.2~day period for each section (as would also be expected for a Doppler signal) 
but that the semiamplitude $K$ decreases with increasing wavelength, 
providing independent evidence that the periodic RV signal is due to stellar activity.
When analyzing the full spectrum, \citet[][]{Reiners2013} also found that the cross-correlation function (CCF) and the template-matching technique 
yield very different semi-amplitudes, $K$=30.5$\pm$1.0 and 23.4$\pm$0.9 m s$^{-1}$, respectively.
This difference suggests, according to the authors, ``that the changes in the line shapes affect each method in a very different way,
further indicating that the measured RV offsets are due to changes in the line profile shapes rather than real Keplerian signals.''

\citet[][]{Tuomi2018} revived interest in Gl~388 in the context of RV planet searches 
by claiming the detection of a 0.24 $M_{\rm J}$ planet in 1/1 spin-orbit resonance (i.e., in a 2.2 day period orbit). 
They analyzed HARPS and HIRES\footnote{HIgh Resolution Echelle Spectrometer \citep[][]{Vogt1994}.} 
optical RV time series together with optical photometry time series from ASAS\footnote{All-Sky Automated Survey \citep[][]{ASAS1997}.} 
and MOST\footnote{Microvariability and Oscillations of STars \citep[][]{MOST2012}.}.
\citet[][]{Tuomi2018} recovered a period of 2.22579 days in the RV time series,
which is consistent with previous measurements, 
and similar to \citet[][]{Bonfils2013} report a correlation between the RV signal and the bisector span (BIS). 
Contrary to \citet[][]{Reiners2013}, \citet[][]{Tuomi2018} report the RV signal to be independent of the wavelength range, 
besides the bluest 24 HARPS orders, which the authors claim to be heavily contaminated by activity.
Their photometric time series displays the 2.23 day period too, but with considerable variability: 
the photometric signal was recovered in the ASAS-N and MOST data but not in the ASAS-S data,
and its period and phase changed from the first to the second half of the MOST time series.
The photometric signal therefore appears to vary on short timescales, 
while the RV signal is stable over thousands of days.
\citet[][]{Tuomi2018} tested multiple star-spot scenarios and found simultaneously explaining a rapidly 
varying photometric signal and a stable RV signal very challenging. 
They interpret the RV signal as a 0.24 M$_J$ planet because they find the signal independent of the wavelength range 
and because it is stable and invariant over time. 
They consider it unlikely that this is caused by stellar activity 
and stellar spots corotating on the stellar surface.

A Doppler RV signal is wavelength independent,
and therefore should be recovered in observations carried in the near-IR. 
To test this,
\citet[][]{Carleo2020} obtained simultaneous observations of Gl~388 with spectrographs 
HARPS-North\footnote{High-Accuracy Radial velocity Planet Searcher - North \citep[][]{Consentino2012}.}
in the optical (0.39 $-$ 0.68 $\mu$m, R=115000) 
and GIANO\footnote{\citet[][]{Oliva2006}.} in the near-IR (0.97 $-$ 2.45 $\mu$m, $R\sim$50,000) 
at the TNG\footnote{Telescope Nazionale Galileo.} from April to June 2018. 
Consistent with previous optical RV measurements, 
they find their HARPS-N optical RVs to vary with a period of 2.2231 days
and a semiamplitude $K$ of 33 m s$^{-1}$ and that the RV signal and the BIS are strongly anticorrelated.
Similar to \citet[][]{Reiners2013}, they report different RV amplitudes for optical RVs measured by a cross-correlation method  
(DRS\footnote{Data Reduction System.} pipeline) and by a template-matching method (TERRA pipeline), 
with a smaller amplitude for the template-matching method.
The periodogram of their near-IR RV  time series shows no peak around 2.2 days, 
and just a low significance periodicity at 11.5 days. 
With a Gaussian prior of 2.225 $\pm$ 0.02 days on the period and 0.3 for the eccentricity,
\citet[][]{Carleo2020} set a 3$\sigma$ upper limit of 23~m~s$^{-1}$ on the near-IR semiamplitude, 
which is in strong tension with the 33~m~s$^{-1}$ derived from the simultaneous optical data. 
Furthermore, their phase-folded optical and near-IR RVs do not overlap. 

In a recent work, \citet[][]{Kossakowski2022} investigated Gl~388 at optical and near-IR wavelengths 
with CARMENES\footnote{Calar Alto high-Resolution search for M dwarfs with Exoearths 
with Near-infrared and optical \'Echelle Spectrographs.} \citep[][]{Quirrenbach2014}.
The 2.23 day periodicity was recovered in the CARMENES optical RVs and a
strong anticorrelation between the chromatic index (CRX), the bisector (BIS), and 
the RV was observed, providing additional evidence of the stellar activity origin of the signal.
\citet[][]{Kossakowski2022} modeled a large ensemble of RVs in the optical 
and near-IR (HARPS, HIRES, CARMENES, HARPS-N, GIANO-B, HPF\footnote{Habitable Zone Planet Finder.}).
They find that the RV signal exhibits amplitude fluctuations and phase shifts in time,
a behavior that is not compatible with a Keplerian origin of the signal.
However, the authors report that a mixed model of a stable signal 
plus a quasi-periodic red-noise model is the one that best describes the data. 
The stable component in the model of \citet[][]{Kossakowski2022} 
has a semiamplitude $K$=16.6$\pm$2.2~m~s$^{-1}$,
which translates into  a 3$\sigma$ upper limit to the candidate planet of 27 M$_\oplus$ (=0.084 M$_J$) in a 2.23 day period.
The GIANO and CARMENES near-IR measurements provide strong, but not overwhelming, 
evidence that the RV periodic signal observed in Gl~388 is due to stellar activity.
Measurements in the near-IR with higher precision are required to 
test whether the stable component suggested by the Gaussian process analysis of
\citet[][]{Kossakowski2022} is real.

We obtained new RV observations of Gl~388
with the SOPHIE (optical) and SPIRou (near-IR) instruments between 2019 and 2021.
To investigate variations on longer timescales,
we also included archival optical data obtained  from 2005 to 2016 with HARPS.
In the following subsections, 
we describe each data set.

\begin{figure}
\centering
\includegraphics[width=0.5\textwidth]{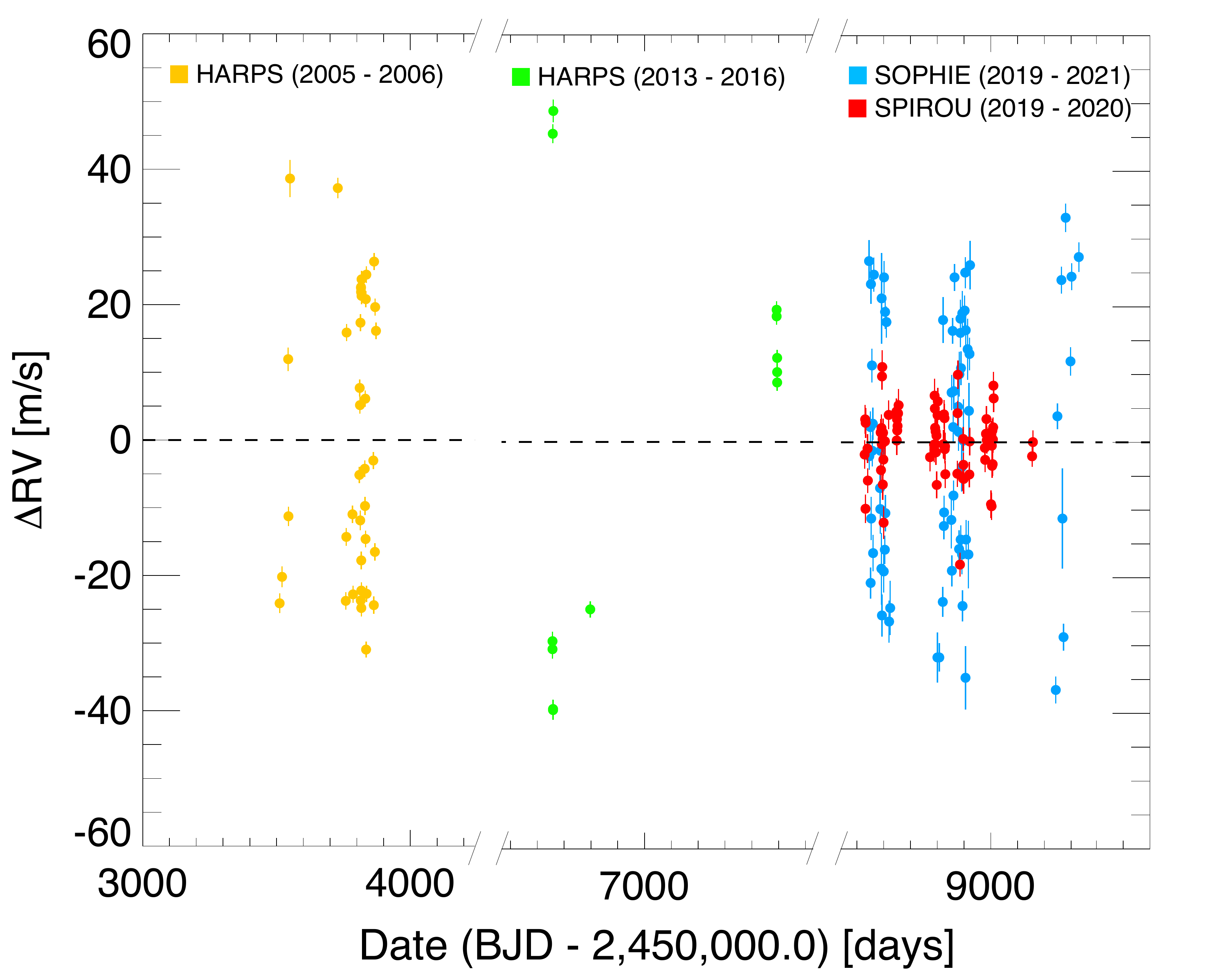}
\caption{New SPIRou and SOPHIE RVs, together with the archival HARPS data used in this work.
}
         \label{HARPS_SOPHIE_SPIROU_time_coverage}
\end{figure}

\section{Observations and data reduction}

\subsection{HARPS}
Gl~388 was observed from 2005 to 2016 with the HARPS
high-precision velocimeter \citep[][]{Mayor2003,Pepe2004} installed 
on the 3.6 m telescope at La Silla Observatory in Chile.
The HARPS data set consists of 51 measurements, of which
39 were acquired between 2005 and 2006 (HARPS1 data set) 
and previously reported in  \citet[][]{Bonfils2013}
and 12 are new measurements obtained between 2013 and 2016 (HARPS2 data set). 
Each spectrum covers the 380-690 nm spectral range with 71 orders at a 
median resolving power of 115000. 
Spectra were wavelength calibrated to a high accuracy using a simultaneous ThAr hollow cathode 
or Fabry-P\'erot calibration lamp.
Science-ready optical spectra were obtained from the ESO\footnote{European Southern Observatory.}-archive facility and
we measured RVs using the template-matching algorithm described in \citet[][]{Astudillo2015}.
The summary of the HARPS1 and HARPS2 RV measurements is given in the Appendix.
The median per visit precision of these RVs is 1.3 m s$^{-1}$, 
with a minimum of 1.2 and a maximum of 2.7~m~s$^{-1}$. 

\subsection{SOPHIE}
We obtained 71 RV measurements of Gl 388 
between 
March  2,  2019 and April 23, 2021, with the SOPHIE 
high-resolution optical spectrograph \citep[][]{Perruchot2008,Bouchy2013}
installed on the 1.93m Telescope of the Observatoire de Haute Provence (OHP)
in southern France.
The observations used the high-resolution mode ($R\sim$75000) and a simultaneous Fabry-P\`erot calibration lamp. 
SOPHIE covers a wavelength domain from 3872 to 6943 \AA~ across 38 spectral orders.
Data were reduced with the SOPHIE data-reduction system \citep[][]{Bouchy2009a,Heidari2022thesis}.
RVs and their error bars were computed using a CCF technique with an empirical M5 mask, with
a RV zero-point nightly correction based on measurements of RV standards \citep[][]{Boisse2010,Heidari2022}.
The summary of the SOPHIE RVs is given in the Appendix.
The per visit precision of these RVs ranges from  2 to 7 m s$^{-1}$,
with a median of 2.6 m s$^{-1}$.

\begin{figure}
\centering
\includegraphics[width=0.47\textwidth]{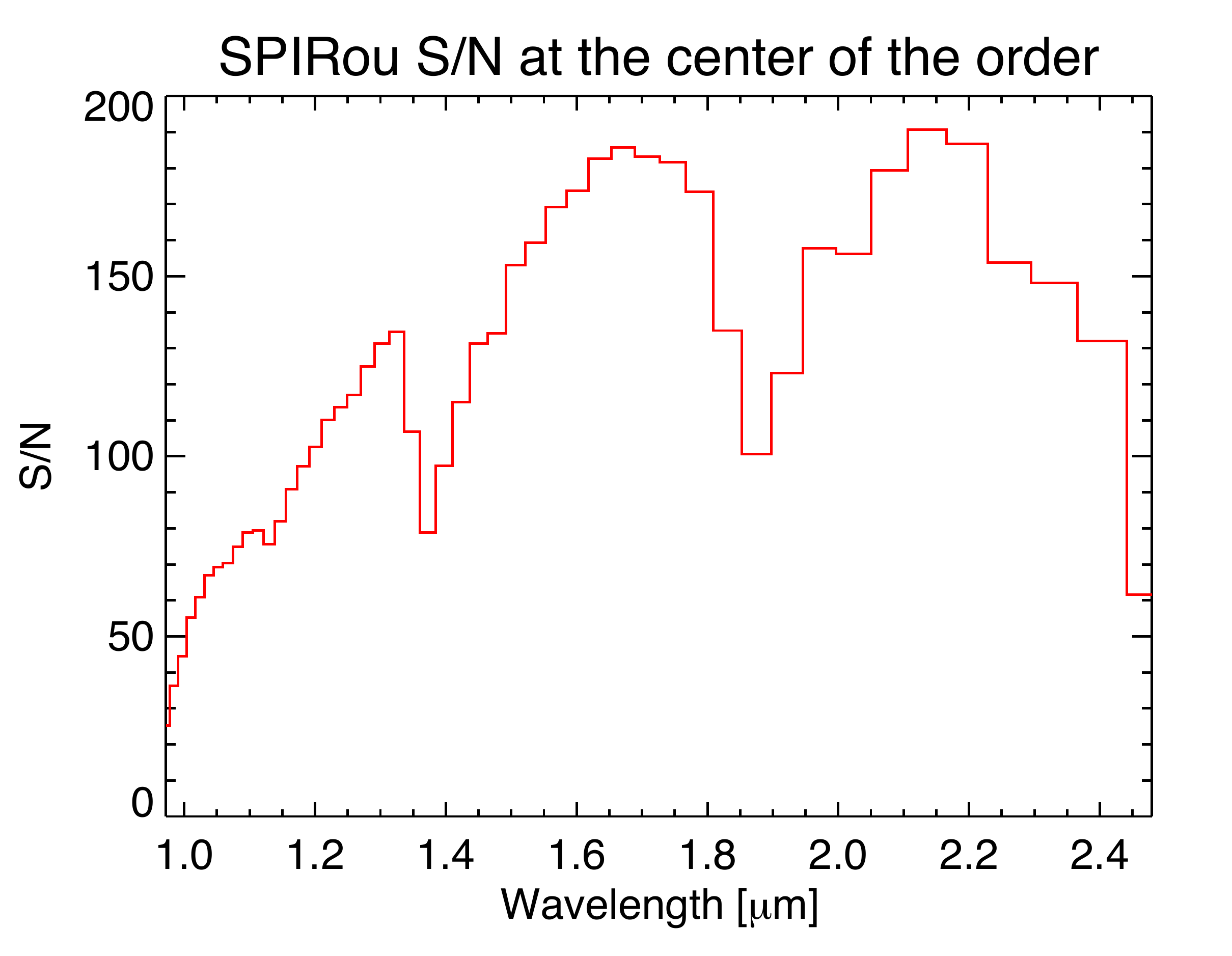}
      \caption{Example of the S/N per pixel at the center of the order achieved in the extracted 1D SPIRou spectrum of Gl~388.
                 }
         \label{SPIROU_SN_plot}
\end{figure}

\begin{figure*}
   \centering
\begin{tabular}{cc}
\includegraphics[width=0.37\textwidth]{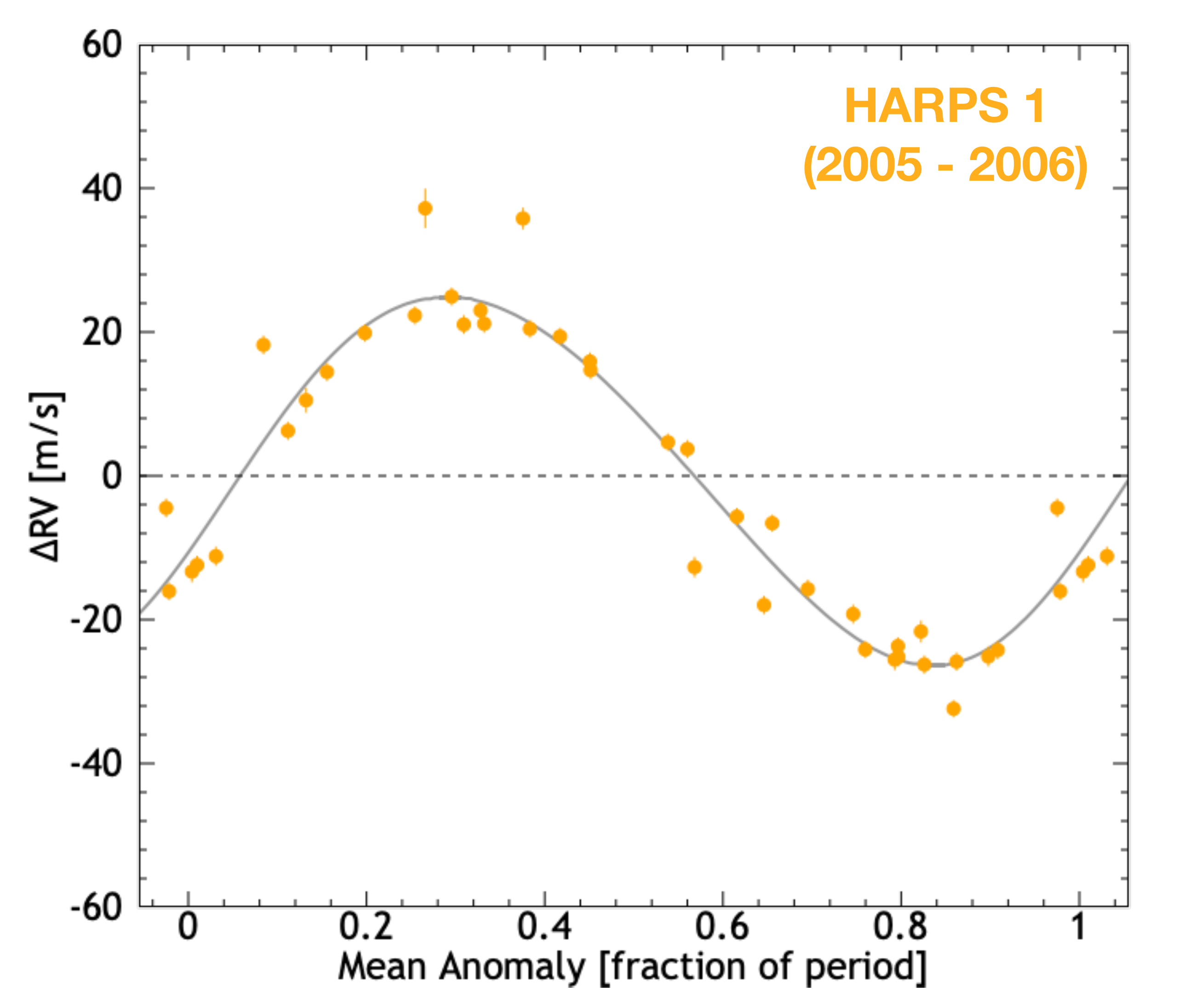} & \includegraphics[width=0.37\textwidth]{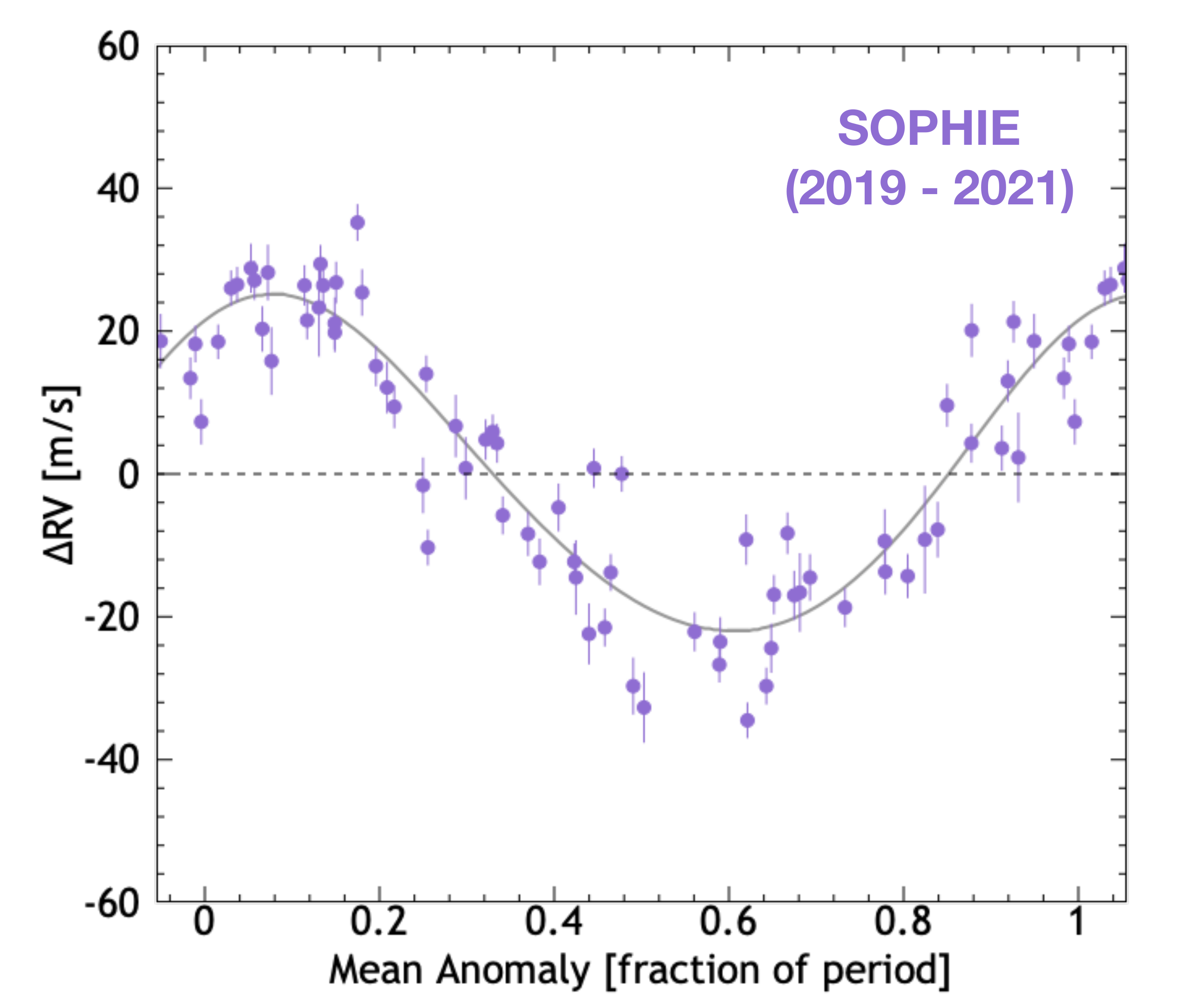} \\
\includegraphics[width=0.37\textwidth]{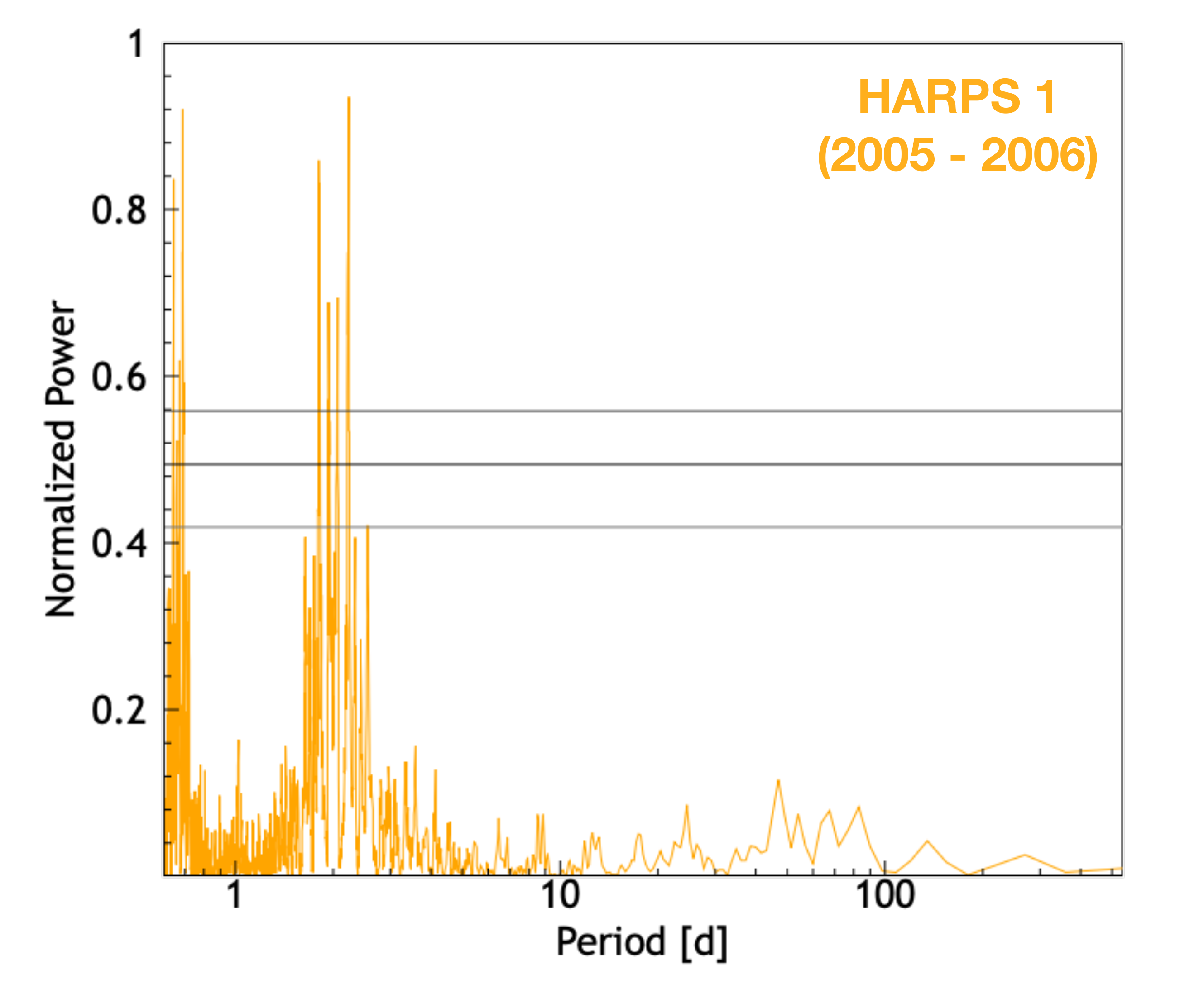} & \includegraphics[width=0.37\textwidth]{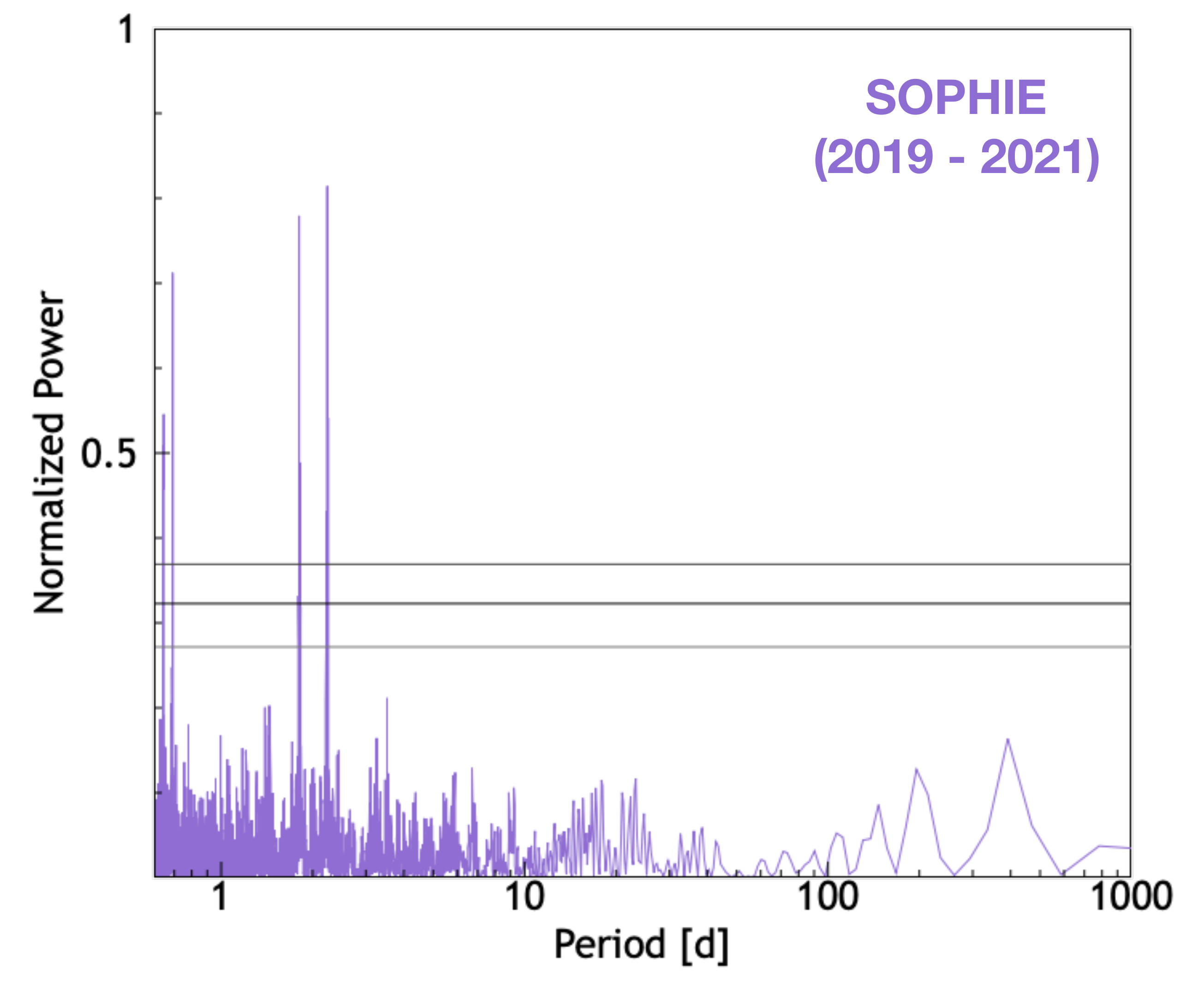} \\
\includegraphics[width=0.37\textwidth]{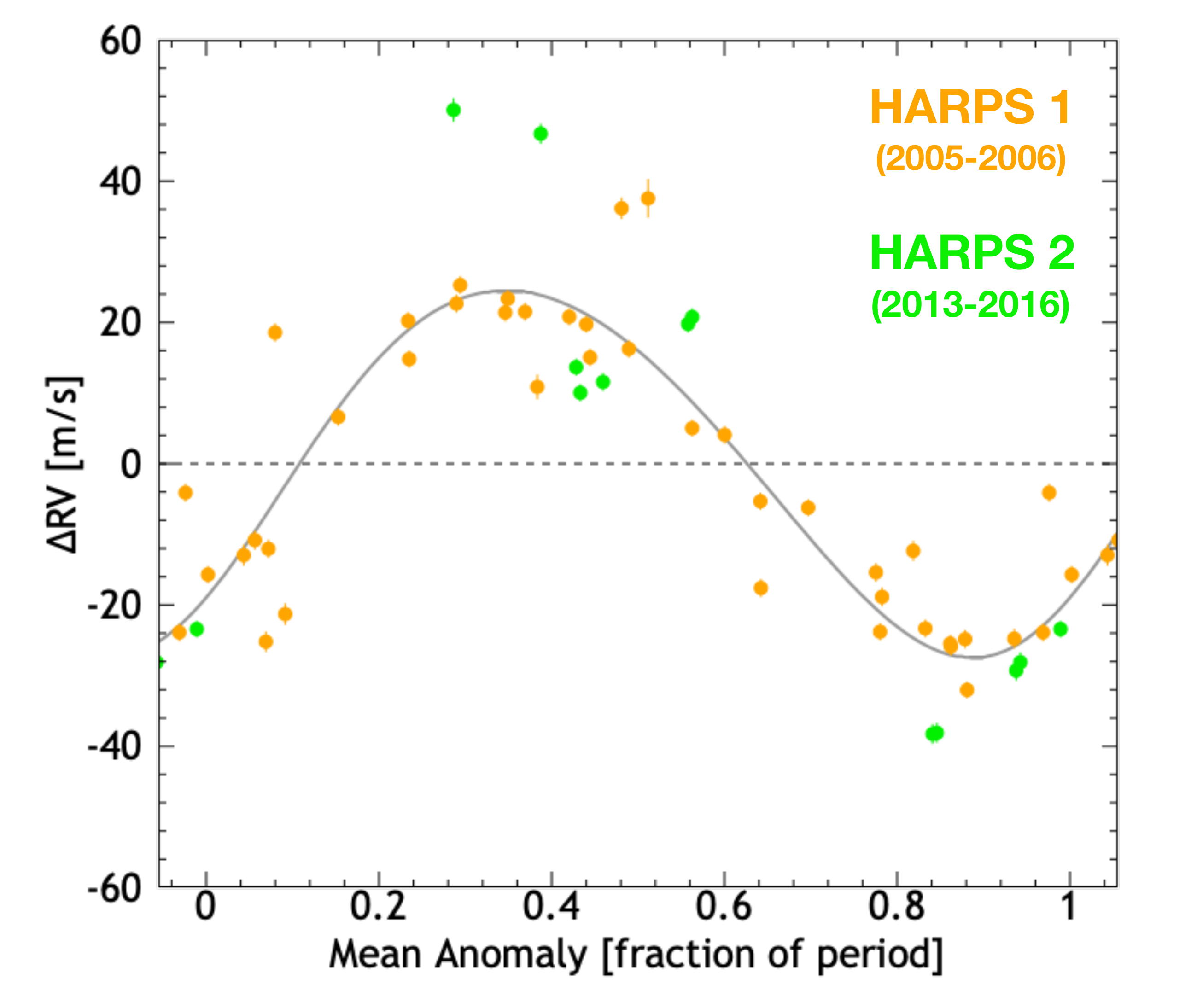} & \includegraphics[width=0.37\textwidth]{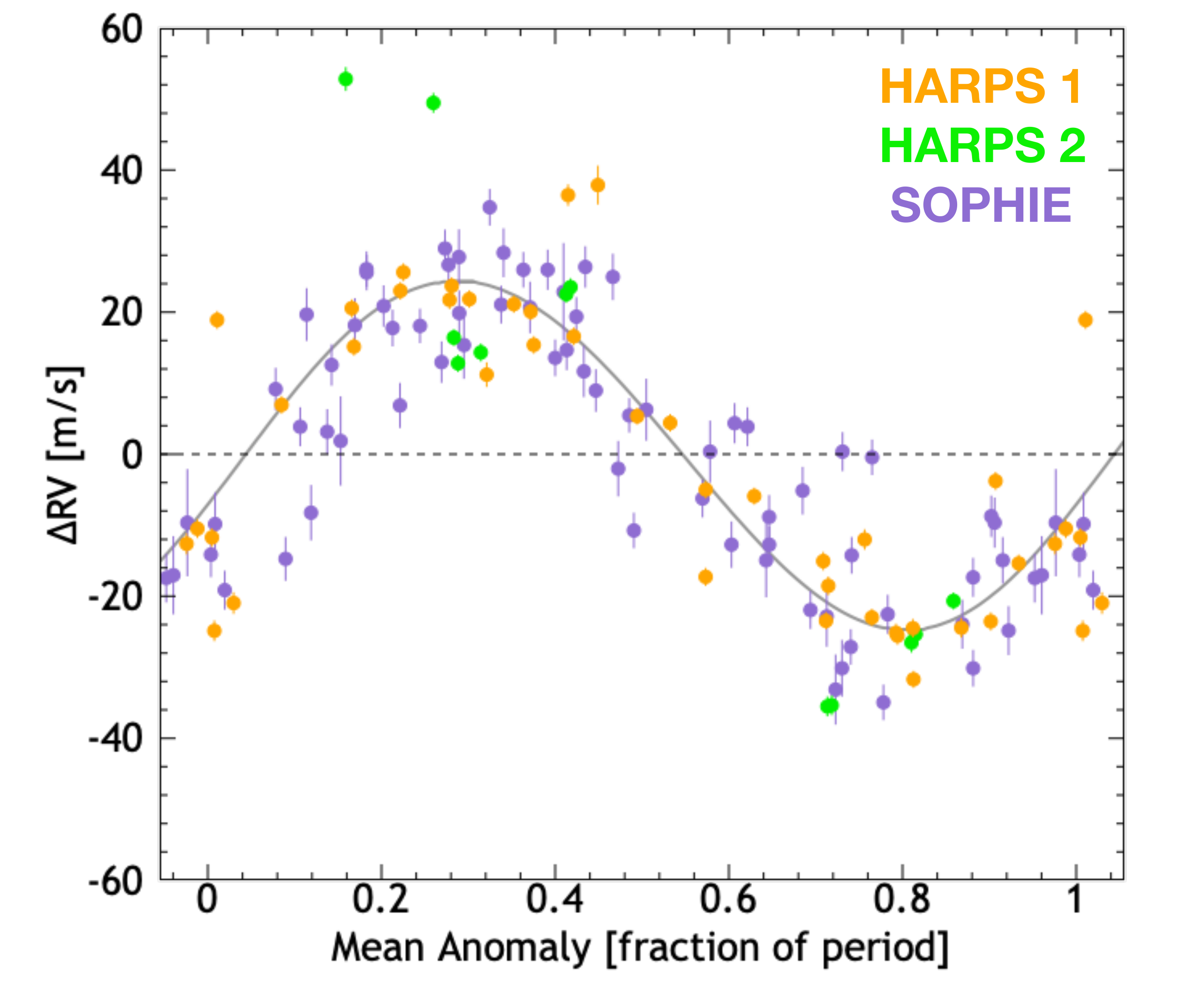} \\
\includegraphics[width=0.37\textwidth]{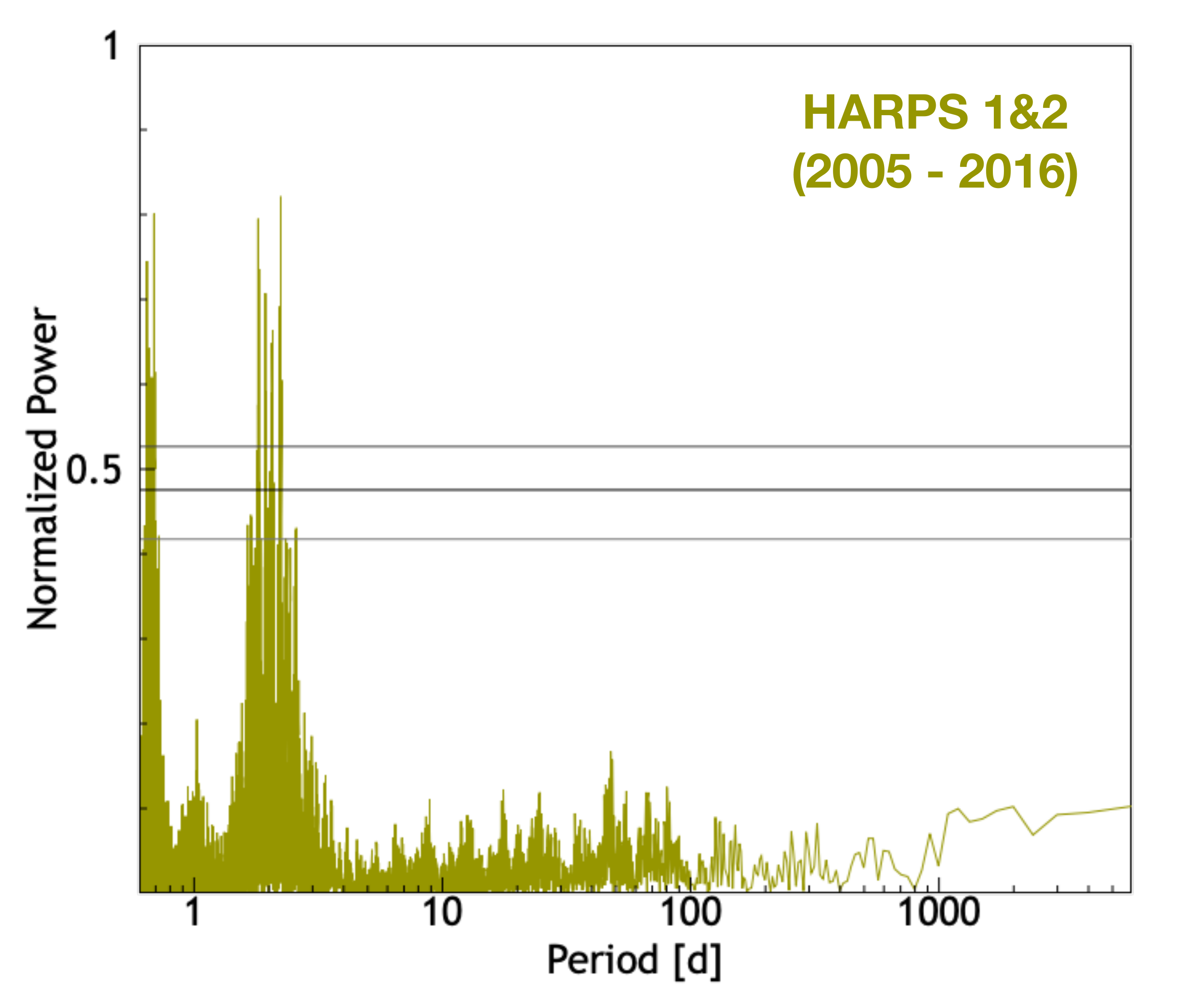} & \includegraphics[width=0.37\textwidth]{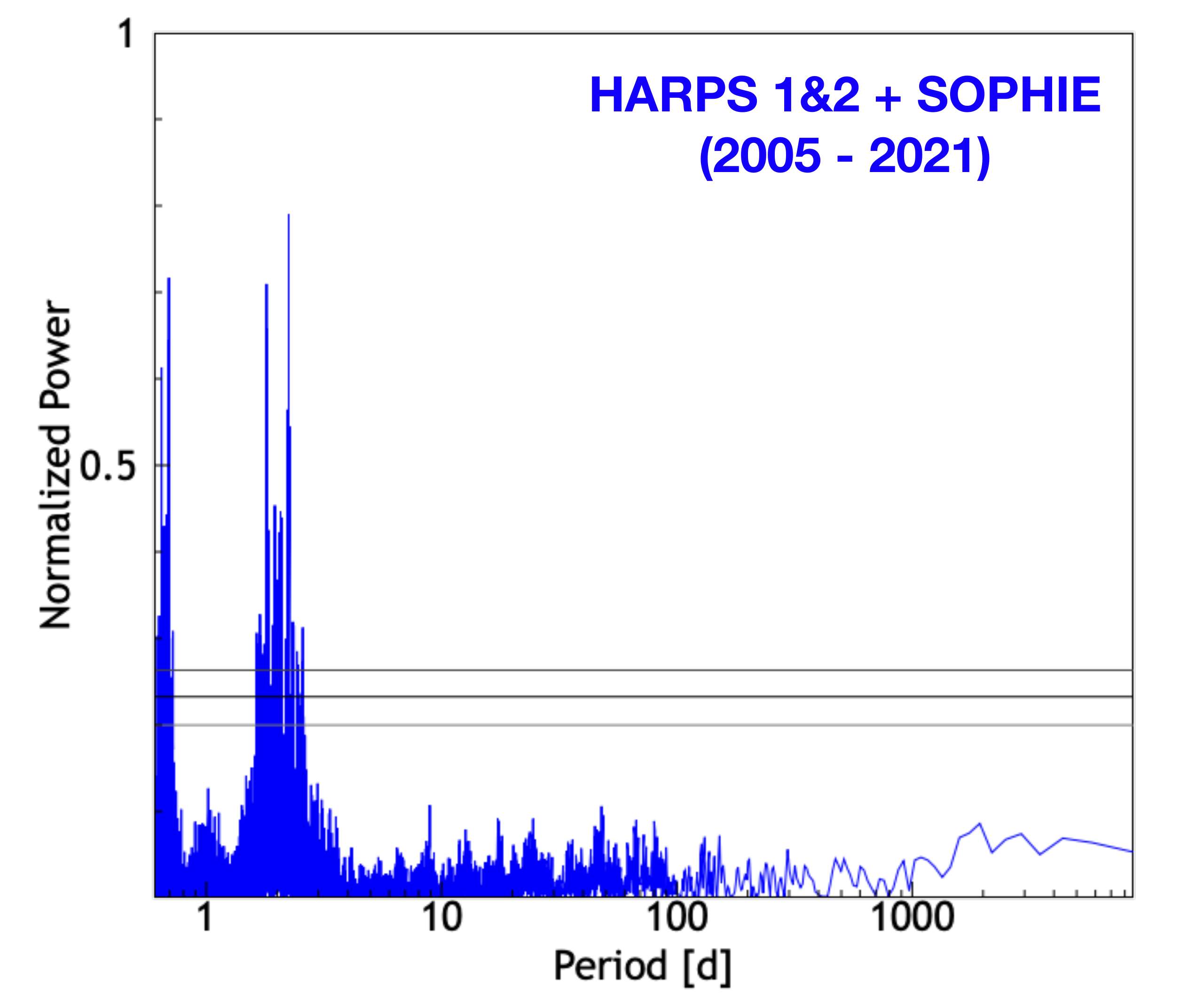} \\
\end{tabular}
\caption{Periodograms and phase-folded RVs of the HARPS1, HARPS2, and SOPHIE data sets. 
The horizontal lines mark FAP levels of 0.1, 1, and 10\%.
A prominent peak is observed in all the periodograms at 2.23 days,
the rotational period of the star, with
aliases at 1.81 and 0.69 days. 
The DACE Keplerian model with period $P=$2.23 days used to phase-fold each data set is 
described in Table~\ref{table:HARPS_SOPHIE_periodogram}.
}
\label{Periodograms_HARPS_SOPHIE}
\end{figure*}

\begin{table*}
\caption{HARPS and SOPHIE RV time series analysis.}
\begin{center}
\begin{tabular}{l c c c c c}     
\hline\hline      \\ 
                       & HARPS1   & HARPS1+2 & SOPHIE & HARPS1+2 \& SOPHIE \\ 
                       & $(2005-2006)$ & $(2005-2016)$ & $(2019 - 2021)$ & $(2005 - 2021)$ \\
\hline   
norm. power$^\dagger$ 			& 0.92              & 0.87                  & 0.84                & 0.79 \\
log$_{\rm 10}$(FAP)$_{\rm Analytical}^\dagger$  	 
                        & $<-15.95$           & $<-12.4$            & $<-15.95$           & $<-15.95$  \\
$P$ [days]              & 2.2270$\pm$0.0001  & 2.23099$\pm$0.00001  & 2.23001$\pm$0.0001 & 2.231023$\pm$0.000004\\
$K$ [m s$^{-1}$ ]       & 25.6$\pm$0.3       & 25.9$\pm$0.3         & 23.6$\pm$0.5        & 24.6$\pm$0.2\\
ecc                     & 0.08$\pm$0.01      & 0.08$\pm$0.01        & 0.08$\pm$0.02       & 0.03$\pm$0.01\\
$\omega$ [deg]          & 247$\pm$8          & 227$\pm$7            & 327$\pm$16          & 254$\pm$20\\
$M_{0}$  [deg]          & 242$\pm$17         & 138$\pm$7            & 151$\pm$28          & 101$\pm$20\\
O$-$C rms [ m s$^{-1}$ ] & 4.8               & 9.2                  & 7.5                 & 9.5 \\
\hline
\end{tabular}
\tablefoot{ 
$^\dagger$  The normalized power and the FAP are given at the 2.23 days periodogram peak.} 

\end{center}
\label{table:HARPS_SOPHIE_periodogram} 
\end{table*}

\subsection{SPIRou}   
We observed Gl~388 on 74 epochs between  February 14, 2019 and November 3, 2020,
with SPIRou, the near-IR spectropolarimeter  \citep[][]{Donati2020} 
installed on the 3.6 m Canada-France-Hawaii Telescope (CFHT) 
atop Mauna Kea (Hawaii). The SPIRou spectra cover from 0.96 to 2.50 $\mu$m at a resolving power of $R\sim70000$.
Figure~\ref{HARPS_SOPHIE_SPIROU_time_coverage} shows the time coverage of the HARPS, SOPHIE, and SPIRou observations.
Each SPIRou visit obtains a Stokes V polarimetric sequence consisting in four 61s exposures with distinct positions of the Fresnel rhombs.  
The raw data were reduced using version v06132 of the SPIRou data reduction pipeline APERO\footnote{A PipelinE to Reduce Observations.} \citep[][]{Cook2022},
installed at the SPIRou Data Centre hosted at the Laboratoire d'Astrophysique de Marseille. 
Each 2D SPIRou spectrum consists of two science spectra,  A and B, one per polarimetric channel, 
and one spectrum of a Fabry-P\'erot  \'{e}talon in the calibration channel C.
Our RV analysis does not use the polarimetric information and uses combined extraction of the  A and B science channels.
The spectra were extracted, dark- and flat-field corrected, and wavelength calibrated using the nightly calibrations.
Details of the wavelength calibration algorithm, which uses both a Uranium-Neon lamp and Fabry-P\'erot exposures,  
is described in \citet[][]{Hobson2021} and \citet[][]{Cook2022}.
The 1D spectrum was then corrected for telluric absorption using a principal component analysis (PCA) method 
and a large library of spectra of telluric standard stars from nightly observations
since 2018. The principle of the method is described in \citet[][]{artigau2014}.
Figure~\ref{SPIROU_SN_plot} displays the typical S/N per pixel achieved at the center of the order in one exposure. 
A S/N higher than 150 was achieved typically in the center of the H and K bands.

Measurements of RVs were extracted from the spectra with a line-by-line (LBL) algorithm 
described in \citet[][]{Artigau2022}.
They were referred to the barycenter of the Solar System using the  astropy routine barycorrpy, 
and corrected for the drift of the Fabry-P\'erot and for changes in the RV zero point.
The RV zero point correction was derived from observations of a set of RV standards
that have been observed by SPIRou nightly since 2018.
The final RV for a visit is the average of the RVs measured in each of the four spectra of the Stokes V sequence, 
and has a typical precision of 2 m s$^{-1}$. The summary of the SPIRou RVs is given in the Appendix.

Stokes $I$ and $V$ spectra were obtained from the same raw data employing the polarimetric reduction module of
APERO.  We then applied least-squares deconvolution (LSD; \citealt{Donati1997,Kochukhov2010}) 
to those spectra to extract average Stokes $I$ (polarized) and $V$ (circularly polarized) profiles. 
The line mask for LSD was generated via the Vienna Atomic Line Database \citep[VALD,][]{Gustafsson2008} from a synthetic spectrum 
with $T_{\mathrm{eff}}=3500$ K, $\log g=$ 5.0 cm~s$^{-2}$, and $v_{\mathrm{micro}}=$ 1 km~s$^{-1}$, 
and it contains 1400 atomic lines between 950 and 2600~nm deeper than 3\% of the continuum level and with a known 
Land\'e factor (g$_\mathrm{eff}$, i.e., the Zeeman effect sensitivity). 
The longitudinal magnetic field  was computed from the average Stokes profiles using  
\begin{equation}
B_\ell\;[G] = \frac{-2.14\cdot10^{11}}{\lambda_0 \mathrm{g}_{\mathrm{eff}}c}\frac{\int vV(v)dv}{\int(I_c-I)dv} \,,
\label{eq:Bl}
\end{equation}
where $\lambda_0$ and $\mathrm{g}_\mathrm{eff}$ are the normalization wavelength and the average Land\'e factor of the Stokes profiles, 
$c$ is the speed of light in vacuum, $v$ is the point-wise RV of the profile in the star's rest frame, and $I_c$ is the continuum level \citep{Donati1997}.
The integral in Eq.~\ref{eq:Bl} was computed within $\pm$ 50 km s$^{-1}$ of the profile center
to encompass the full line width in both Stokes $I$ and $V$.
We exclude six February 2019 observations from the magnetic analysis presented here
because an optical component of the instrument was malfunctioning 
at very low temperatures and affected those six circular polarization spectra.
The summary of the SPIRou $B_\ell$ measurements is given in the Appendix.

\section{Radial velocity analysis}

\subsection{HARPS and SOPHIE optical radial velocities}
\label{optical_RV}

Using the  DACE\footnote{Data Analysis Center for Exoplanets.} 
online interface hosted by the University of Geneva\footnote{https://dace.unige.ch/radialVelocities/},
~~~we~~calculated~~least-square~~periodograms~~for~~~the \\
HARPS1, HARPS2, and SOPHIE time series, both individually and combined.
Figure~\ref{Periodograms_HARPS_SOPHIE} summarizes the periodograms obtained.
All optical data sets produced a prominent peak at 2.23 days 
with analytical false alarm probabilities\footnote{Analytical FAP values are estimated following \citet[][]{Baluev2008}.} (FAP) well below 0.01\% (see Table~\ref{table:HARPS_SOPHIE_periodogram}).
We used DACE to fit a Keplerian model to each data set, using the 2.23 day peak to set
the initial conditions\footnote{The initial conditions of the Keplerian model were computed using the formalism of \citet[][]{Delisle2016}.}.
Table~\ref{table:HARPS_SOPHIE_periodogram} summarizes the properties of these Keplerian fits and 
Figure~\ref{Periodograms_HARPS_SOPHIE} displays the corresponding phase-folded RVs.
Even though the 2013$-$2016 HARPS2 data set has a larger dispersion than 
the 2005$-$2006 HARPS1 data set (see Figure~\ref{HARPS_SOPHIE_SPIROU_time_coverage}),
the combined HARPS1 + HARPS2 data set can be described by a similar period 
and semiamplitude to that of HARPS1, with the main difference being, 
$\omega$, the argument of periastron, and $M_0$, the mean anomaly at the reference time $t=0$. 
The 2019$-$2021 SOPHIE RV signal is very similar to the earlier HARPS1+2 signal, 
with a  2.23 day period retrieved in both data sets, but
with a  2.3$\pm$0.6~m~s$^{-1}$ difference 
between the semiamplitude of the Keplerian fit.
The different orbital solutions in Table~\ref{table:HARPS_SOPHIE_periodogram} 
converge to periods that differ very slightly but beyond their error bars. 
This could be due to a subtle change in the effective rotation 
period because of the combination of differential stellar rotation 
with a change in the latitude of the main spot. 
However, the 2.227, 2.230, and 2.231~day periods are aliases of each other 
for samplings of $\sim$3300 and $\sim$5000 days, which correspond to the average interval 
between the HARPS1 and HARPS2 acquisition periods and between the HARPS1 and SOPHIE ones 
(Figure~\ref{HARPS_SOPHIE_SPIROU_time_coverage}), respectively. 
We therefore tentatively interpret these period differences as aliasing effects.

As illustrated by Figure~\ref{Periodograms_HARPS_SOPHIE},
a single global Keplerian solution can describe all HARPS and SOPHIE RVs. 
It does, however, have larger residuals than separate fits to the HARPS or SOPHIE data alone (Table~\ref{table:HARPS_SOPHIE_periodogram}), 
indicating that although the optical RV signal is globally stable from 2005 to 2021, 
small variations are nevertheless present.
The amplitude and phase differences between the HARPS and SOPHIE data and the relative constant period suggest that the optical
RV signal is dominated by rotational modulation of a slightly variable stellar activity.
Because the SOPHIE and SPIRou data were obtained quasi-simultaneously, 
we use only the SOPHIE data in our comparison with the near-IR RVs.

\subsection{SPIRou radial velocities}
The summary of the SPIRou RV measurements is provided in the Appendix.
Figure~\ref{SPIRou_SOPHIE_RV_plot} shows the phase-folded SPIRou and SOPHIE RVs at $P$=2.23 days. 
We can clearly see that the near-IR RVs do not match the periodic signal observed in the optical data, 
and that they are much flatter.
Figure~\ref{Periodogram_SPIRou} displays the periodogram of the SPIRou RV data.
There are no peaks with significance above the FAP 1\% level, and most notably, 
there is no signal whatsoever at $P$=2.23 days.
The dispersion of the SPIRou time series is just 5~m~s$^{-1}$  rms;
it clearly excludes  the periodic signal of 24~m~s$^{-1}$ semiamplitude observed in the optical RVs
{and the 16.6~m~s$^{-1}$ stable RV component suggested in \citet[][]{Kossakowski2022}.
}
\begin{figure}
   \includegraphics[width=0.5\textwidth]{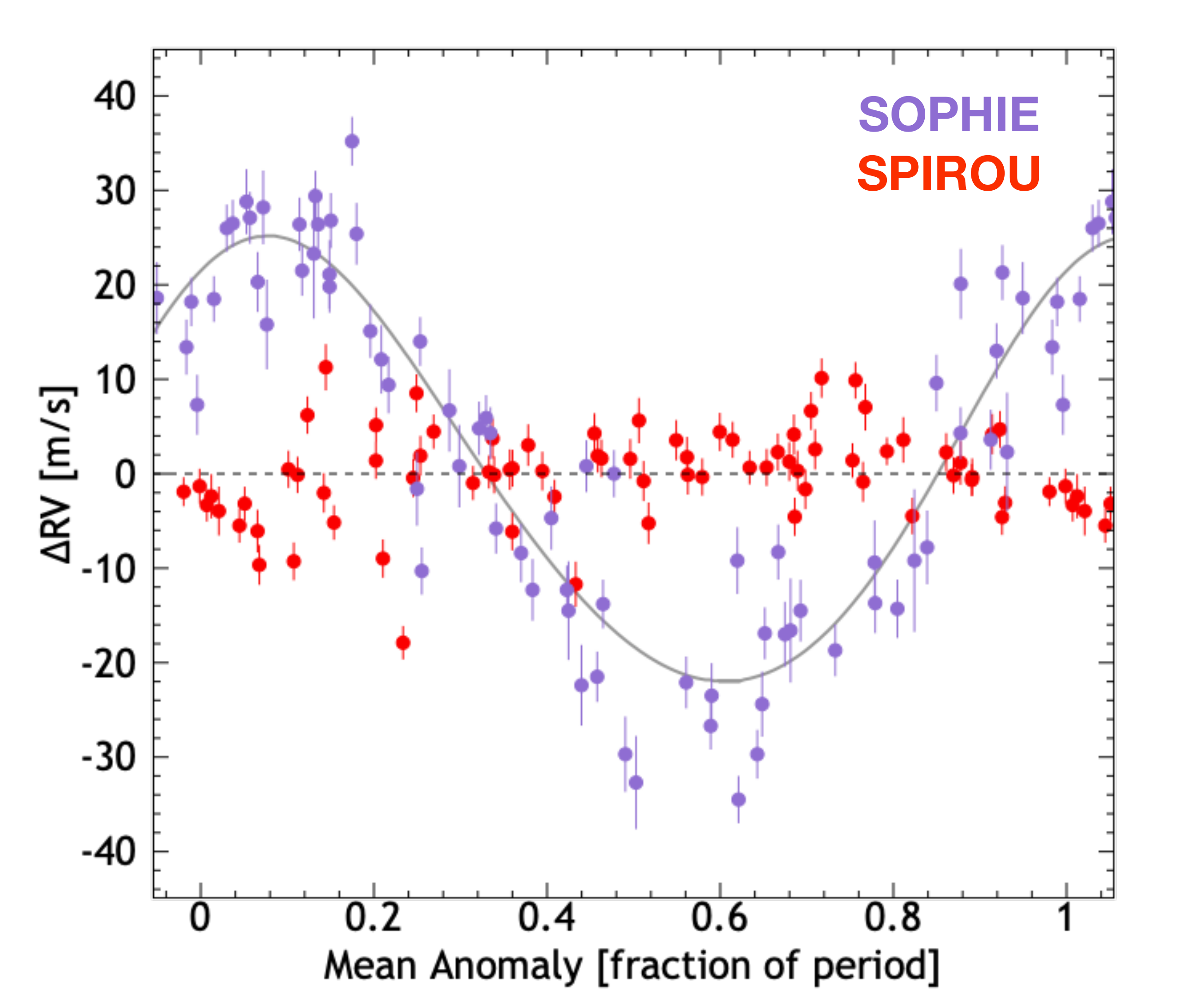}
      \caption{Phase-folded RV measurements of Gl~388 with SPIRou in the near-IR (red dots) and with SOPHIE (blue dots) in the optical.  
      The gray line represents the Keplerian model with a period of 2.23 days and $K=$ 23.6 m s$^{-1}$ derived from the SOPHIE data (Table~\ref{table:HARPS_SOPHIE_periodogram}).
      }
  \label{SPIRou_SOPHIE_RV_plot}
   \centering        
   \includegraphics[width=0.5\textwidth]{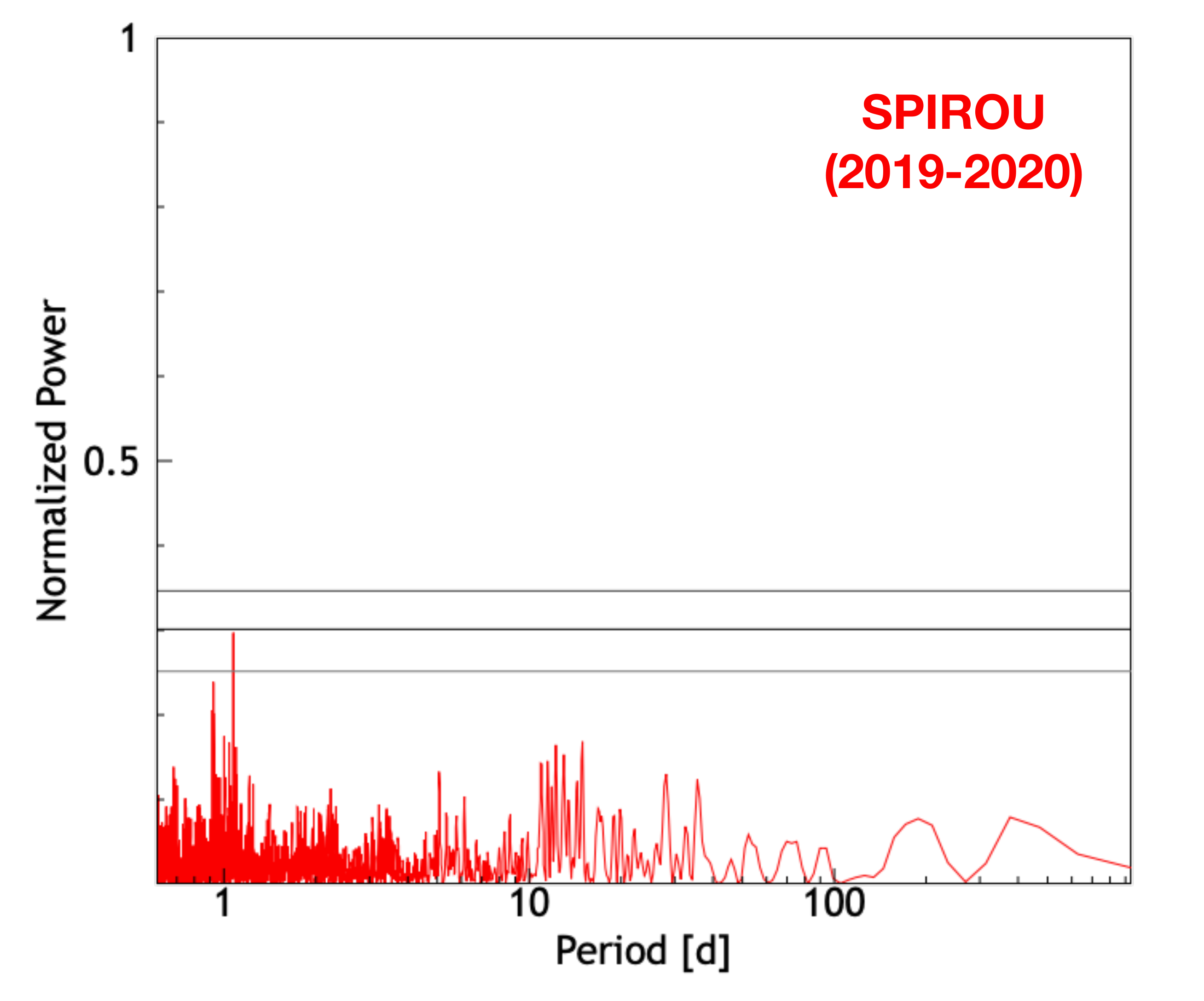}
    \caption{Periodogram of the SPIRou RV series. The horizontal lines mark FAP levels of 0.1, 1, and 10\%. No peak is present at 2.23 days.}
    \label{Periodogram_SPIRou}
\end{figure}

Previous RV observations with SPIRou show that a Keplerian signal of 20 m s$^{-1}$ is easily detected by the instrument. 
For example, in \citet[][]{Moutou2020}, SPIRou detected  the $\sim$25~m~s$^{-1}$ RV signal of the Rossiter-McLaughlin 
effect in the transiting extrasolar planet HD~189733b previously measured in the optical by HARPS \citep[][]{Triaud2009}. 
SPIRou observations, using the LBL technique applied in this work,
have detected planetary signals with amplitudes below 5 m~s$^{-1}$, for example, 
in the star TOI-1759 \citep[2.4~m s$^{-1}$,][]{Martioli2022}, 
and have shown flat RVs at a precision of 2.6~m~s$^{-1}$ rms in Barnard's star 
\citep[][]{Artigau2022}, 
a star with the same spectral type as Gl 388. 

The stability of the SPIRou measurements~ in Gl 388 
provides solid evidence rejecting the hypothesis that 
the RV modulation seen in the optical has a Doppler origin.
We conclude that the periodic signal observed with 
HARPS and SOPHIE is mainly due to stellar activity,
with a negligible contribution from any orbiting planet in a spin-orbit resonance.

\subsection{Planet detection limits from the SPIRou radial velocities}
\label{upperlimits}
We computed planet detection limits from the SPIRou RVs  using \citet{Zechmeister2009} 
generalized Lomb-Scargle (GLS) periodograms as in \cite{Bonfils2013}. 
The method obtains the upper limits by injecting signals in the time series at each period and
determining the maximum amplitude that we can still miss with a given probability.
We assume that the SPIRou time series contains no genuine planetary signal and for each period
we inject a sinusoidal Keplerian signal at 12 equi-spaced phases.
The amplitude (i.e., the planet signal strength) is then increased until the corresponding peak of the 
GLS periodogram reaches the 1\% FAP level.
We tested periods from 0.8 to 10000~days and derived the upper limit on the signal 
amplitude (and the corresponding projected mass) as a function of the period. 
The 1\% FAP confidence level was computed from the levels of the highest peaks of 1000 permuted data sets, 
and the 1\% upper limits' function was obtained by sorting the 
power of peaks of all periodograms at each period.

\begin{figure*}
   \centering  
    \includegraphics[width=0.97\textwidth]{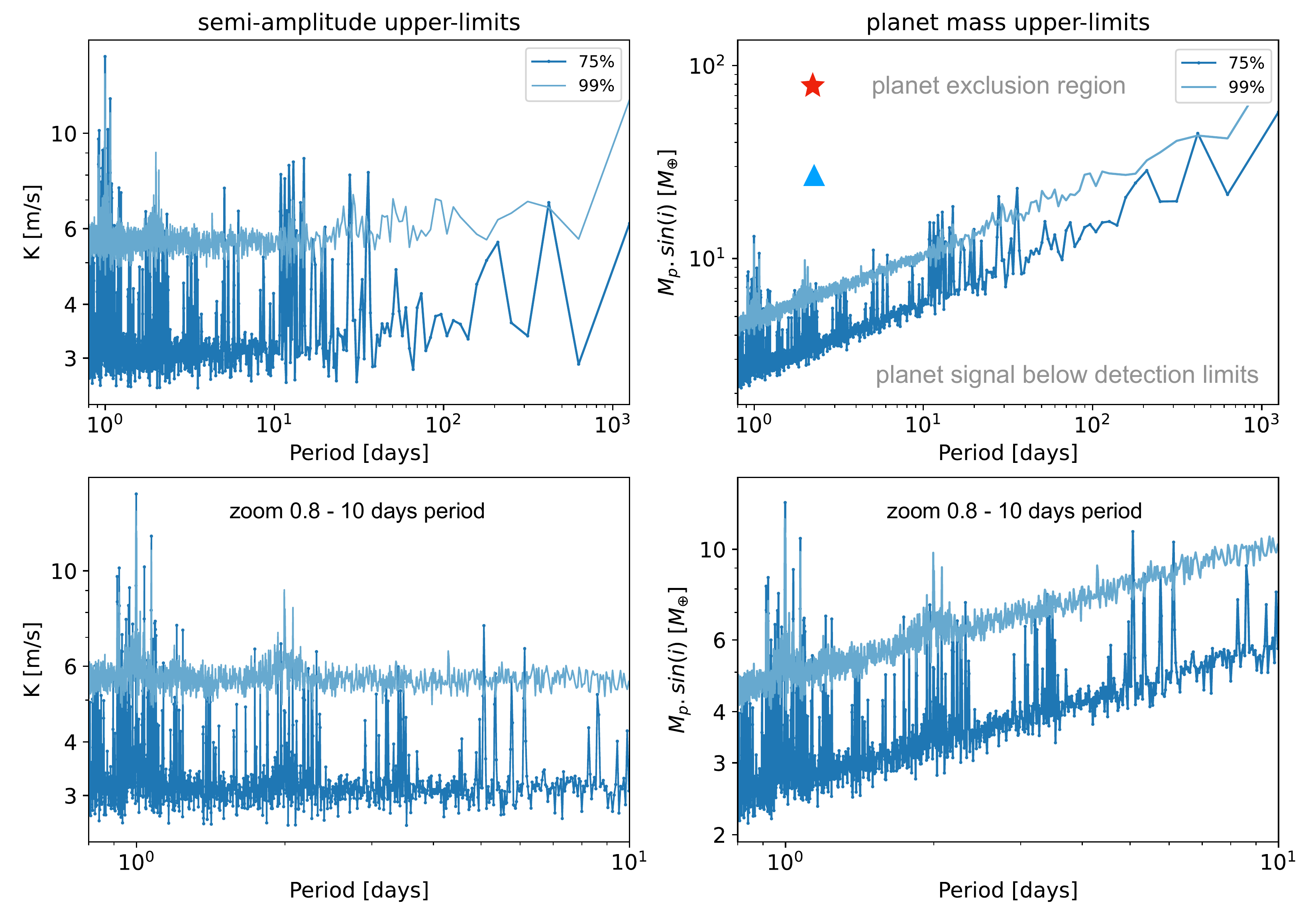}
    \caption{Detection limits in the SPIRou RV time series.  
    {\it Left panels: }  Upper limits in m s$^{-1}$ of the semiamplitude $K$ of the sinusoidal signal.
     {\it Right panels:} Upper limits in terms of the projected mass m~{\it sin(i)}. 
     Objects with a projected mass above the limit are ruled out by the observed GLS periodogram 
     with 99\% (light blue) and 75\% (dark blue) confidence levels  (see text for details). 
     {\it Upper panels:} Upper limits for periods from 0.8 to 1000 days.
     {\it Lower panels:} Zoom on the 0.8 to 10~day range. 
     The red star indicates the location of the 0.24 M$_J$ (76 M$_\oplus$) planet in corotation claimed by \citet[][]{Tuomi2018}.
     The blue triangle shows the planet-mass equivalent (27 M$_\oplus$) of the stable RV component in \citet[][]{Kossakowski2022}.  
     Both are ruled out by the SPIRou data with a confidence higher than 99\%.
     }
    \label{detection_limit}
\end{figure*}

For each period, the upper limit is the highest signal amplitude (at the worst phase) that 
could be present in the time series without creating a peak above the 1\% FAP level. 
Figure~\ref{detection_limit} shows the projected mass as a function of the period 
at which 99\%  and 75\% of the planets injected would be detected in the GLS periodogram of the SPIRou data.
As described in \cite{Bonfils2013}, the degraded upper limits for periods around one and two days are 
due to the data sampling and an unavoidable consequence of observing at night from a single location. 
Our analysis assumes circular orbits and uses the stellar parameters of Section~\ref{Gl388}. 
The SPIRou observations would have detected a 2.23~day period planet of 8 and 4 Earth-masses
with 99\% and 75\% confidence levels, respectively.

\subsection{Optical and near-IR radial velocity chromaticity effects.}

Spots on the stellar surface can induce RV signals modulated by the 
stellar rotation.
We have demonstrated that the Gl~388 RV signal is strongly 
chromatic between optical and near-IR wavelengths. 
In this section, we explore how the RV signals vary
when the RV is measured selecting specific regions in the spectrum.
Our objective is to explore whether there could be a 
chromatic dependence of the RV signal inside the optical or the near-IR domain.

In a first test, we computed the RV selecting the regions in the spectra
based on the expected flux contrast between photospheric and spot model spectra.
For this test, we used
the ratio between PHOENIX \citep[][]{Husser2013}\footnote{PHOENIX~ACES~AGSS~COND~2011~R~10000, 
available at  http://phoenix.astro.physik.uni-goettingen.de}
medium resolution synthetic stellar spectra  
with temperatures of 3400~K and 3000~K, representing the stellar photosphere and the spot 
(see Figures  \ref{SOPHIE_high_low_regions} and  \ref{SPIROU_spot_3400_3000} in the Appendix). 
We selected a 400~K temperature difference between the photosphere and the spot as it is typical 
for a mid-M star \citep[][]{berdyugina2005, barnes2015}.  
To analyze the SOPHIE data, 
we constructed two cross-correlation masks starting from the M-star mask used to compute the RVs: 
a "high-contrast" mask, which contains the lines that fall in the regions where the flux ratio is higher than its median value,
and a  "low-contrast" mask with the lines that fall in the complementary regions where the flux ratio is  below its median value.
The "high" and "low" contrast  regions of SOPHIE are illustrated in  Figure~\ref{SOPHIE_high_low_regions} in the Appendix.
To analyze the SPIRou data,
we defined the low and high contrast regions in an analogous manner as illustrated in 
Figure~\ref{SPIROU_spot_3400_3000} in the Appendix.
The LBL method calculates a RV for each "line" in the spectrum 
and then computes a final RV as a weighted average of the lines'~RVs. 
We filtered the LBL file of each visit selecting the lines that fell to the low and high contrast regions,
and then computed the final RV for the low and for the high case. 
The SPIRou RVs were corrected for the RV zero point as previously described.

The properties of SOPHIE's high, low, and "full" RV time series and periodograms are given in Table~\ref{table:SOPHIE}.
The three masks  produce similar individual RV measurements uncertainties of about 3 m~s$^{-1}$, 
with the high-contrast mask giving a slightly better precision of 2.8 m~s$^{-1}$.
The RV zero points of the low-contrast and high-contrast masks differ slightly from that of the full mask, 
$-12$ m~s$^{-1}$ for the low-contrast mask and by +10 m~s$^{-1}$ for the high-contrast mask. 
After correcting the three time series from their RV zero point, 
we found largely equivalent results for the three masks. 
All of them produce periodograms with a peak at the same 2.23 day period and have a similar phase for the signal.
The amplitude, however, does differ by a small but significant 2.5 m s$^{-1}$ 
between the RVs calculated with the low and the high mask.
Figure \ref{SOPHIE_high_low_RV} in the Appendix displays the phase-folded 
SOPHIE RV time series obtained with the three masks, they differ a little.

The properties of SPIRou's high, low, and full RV time series are summarized in Table~\ref{table:SPIROU_chromatic}. 
The  phase-folded RVs are given in Figure~\ref{SPIROU_high_low_RV} in the Appendix.
We find no significant differences between the SPIRou
low, high and full RV time series:
all three gave the same median RV point,
and in terms of RV performances, 
the difference between the high and low time series is below 1 m~s$^{-1}$. 
The best performance is obtained when the full wavelength range is taken into account ($5$ m~s$^{-1}$ rms). 
\begin{table*}
\caption{SOPHIE RV time series properties obtained with the full, low-contrast, and high-contrast masks.}             
\label{table:SOPHIE}      
\centering          
\begin{tabular}{l c c c }     
\hline\hline       
   & full & high & low \\
    \hline                    
median RV [km s$^{-1}$]			& 12.505 	                         & 12.515                             & 12.493\\
mean $\sigma_{\rm RV}$ [m s$^{-1}$] & 3.0	                                  & 2.8		                  & 3.2 \\
norm. power $^\dagger$			        & 0.84                                & 0.81                                 & 0.75\\
log$_{\rm 10}$(FAP)$_{\rm Analytical}$$^\dagger$  & $<-15.95$        &  $<-15.95$                       &  $<-15.95$  \\
$P$ [days]                                 & 2.2300$\pm$0.0001       &  2.2301$\pm$0.0001       & 2.2300$\pm$0.0001\\
$K$ [m s$^{-1}$]                       & 23.6$\pm$0.5        		& 24.7$\pm$0.5 		  & 22.4$\pm$0.5 \\
ecc                                            & 0.08$\pm$0.02       		& 0.04$\pm$0.02                 & 0.16$\pm$0.02\\
$\omega$ [deg]       		       & 327$\pm$16          		& 317$\pm$28                     & 340$\pm$8\\
$L_{0}$                   		       & 120$\pm$21          		& 133$\pm$21                     & 105$\pm$20\\
O$-$C rms [m s$^{-1}$] 	       & 7.5                			& 8.4	                                    & 8.0\\
\hline                  
\end{tabular}
\tablefoot{ 
$^\dagger$  The normalized power and the FAP are given at the 2.23 days periodogram peak.} 
\end{table*}

\begin{table}     
\centering  
\caption{SPIRou RV time series properties for the high, low, and full regions, and the Y, H, J, and K bands.}
\begin{tabular}{c c c c c c c}     
\hline\hline   
Band   & median RV &  rms   \\
    & [km s$^{-1}$]    & [m s$^{-1}$] \\
\hline
high & 12.696   & 6.7 \\[1mm]
low  & 12.696   & 5.9 \\[1mm]
full & 12.696   & 5.0 \\[1mm]
Y    & 12.695   & 37.4  \\[1mm]
J    & 12.695   & 22.2	\\[1mm]
H    & 12.696   & 6.2   \\[1mm]
K    & 12.696   & 9.2  \\[1mm]
\hline
\end{tabular}
\label{table:SPIROU_chromatic} 
\end{table}   
\begin{table}
\centering  
\caption{Confidence interval (1$\sigma$) on the semiamplitude $K$ 
of a sinusoidal fit, with a fixed period (2.23 days) and phase, 
to the RV time series per photometry band.}
\begin{tabular}{l c c c c c c}     
\hline\hline   
Band   & $K$ sin. fit \\
       & [m s$^{-1}$] \\
\hline
B (SOPHIE)  & $24.9_{-3.1}^{+3.1}$ \\[1mm]
V (SOPHIE)  & $23.0_{-3.4}^{+3.4}$ \\[1mm]
R (SOPHIE)  & $23.0_{-3.7}^{+3.7}$ \\[1mm]
full BVR    & $22.7_{-3.4}^{+3.4}$ \\[1mm]
Y (SPIRou)  & $17.1_{-20}^{+20}$ \\[1mm]
J (SPIRou)  & $6.3_{-14.4}^{+14.4}$ \\[1mm]
H (SPIRou)  & $-1.2_{-3.9}^{+3.9}$ \\[1mm]
K (SPIRou)  & $-4.9_{-5.5}^{+5.5}$ \\[1mm]
full YJHK   & $-2.1_{-2.8}^{+2.8}$ \\[1mm]
\hline
\end{tabular}
\label{table:SOPHIE_SPIROU_K_sinfit} 
\end{table}

{In a second test for RV chromaticity,
we examined the RV time series obtained per photometry band.
In the case of the SOPHIE observations,
we calculated the RVs using the same M dwarf mask,
but this time we selected the orders used to build the CCF.
For the B band, 
we selected the orders 3 to 18 ($4008.5 - 4962.7$ \AA), 
for the V band the orders 19 to 31 ($4934.9 - 6092$ \AA),
and for the R band the orders 32 to 38 ($6077 - 6943$ \AA).
For SPIRou,
we analyzed the RV time series obtained by applying the LBL algorithm to individual near-IR bands within the SPIRou spectral domain:
Y ($900.0 - 1113.40$ nm),
J ($1153.59 - 1354.42$ nm),
H ($1462.89 - 1808.54$ nm), and
K ($1957.79 - 2343.10$ nm) bands.
Table~\ref{table:SPIROU_chromatic} displays the median and rms of the RV time series for each SPIRou band.
Figure~\ref{RV_chromaticity_sinfit} in the Appendix
shows the SOPHIE and SPIRou RV time series per band phase-folded using a period of 2.23 days.
In all data sets, 
we used the same T$_0$ such that the SOPHIE RVs have phase~=~0 at $v=0$~m~s$^{-1}$.
In each RV time series, 
we fit a sinusoidal signal with a fixed period of 2.23 days and a phase~=~0.
Table~\ref{table:SOPHIE_SPIROU_K_sinfit} shows the retrieved best-fit semiamplitude $K$ in m s$^{-1}$ 
with a $1\sigma$ error bar.
   
In the case of the SOPHIE data, 
the B, V, and R RV time series display, within 1$\sigma$, the same sinusoidal signal with a very similar amplitude.
A slight decrease of $-2$~m~s$^{-1}$ in the semiamplitude with respect to the B band is seen 
in the V and R bands; however, the difference is smaller than the $3 - 4$~m~s$^{-1}$ $1\sigma$ uncertainty in the semiamplitude.
In the SPIRou data, the RV uncertainty per visit and the RV dispersion differs from band to band, 
as expected from their respective RV information content  \citep{artigau2018}.
The Y and J bands display the highest RV dispersion (37 and 22 m s$^{-1}$ rms, respectively), 
followed by the  K band with 9.2 m s$^{-1}$ rms, and the H band performs best with 6.2 m s$^{-1}$ rms.
The best overall performance is obtained using the full spectral domain (5.0 m s$^{-1}$ rms).
The sinusoidal fits in the Y and J band hint at a periodic signal similar to that observed in 
the optical but with a weaker amplitude.
However, 
when the periodogram analysis is performed (see Figure$~\ref{SPIROU_periodogram_bands}$ in the Appendix), 
none of the near-IR RV time series per band display a significant peak at $P$=2.23 days.
The semiamplitudes' error shows that $K$ is compatible with 0 m s$^{-1}$ within the 1$\sigma$ uncertainty.
The sinusoidal model is in fact fitting the noise dispersion.
The sinusoidal fit enables, nevertheless, the upper limit of the semiamplitude to be lowered to 17, 6, 1, 5, and 2 m s$^{-1}$ 
for the Y, J, H, K, and the full near-IR SPIRou domain, respectively (see Table~\ref{table:SOPHIE_SPIROU_K_sinfit}).
The difference in semiamplitude between the H and K bands could suggest a difference in RV dispersion due to the onset 
of magnetic effects (i.e., the Zeeman broadening), but the difference can also be due to a difference in the intrinsic performances between 
the K and the H band due to the telluric absorption and the S/N of the spectra. 
The increase in RV semiamplitude at $\lambda>1500$ nm of tens of m s$^{-1}$ predicted by some 
theoretical models (see for example Fig. 5 in \citet[][]{Kossakowski2022}) is not observed in our SPIRou's Gl~388 data.

In summary, we detect a strong chromatic effect in the RV between the optical and near-IR wavelengths taken as a whole;
however,  we do not observe strong chromatic effects within the individual photometry bands.
A small but significant difference is found in the RV analysis between 
high and low CCF masks of SOPHIE. No difference has been found in other tests.

\begin{figure*}	
    \centering
    \begin{tabular}{c c}
    \includegraphics[width=0.35\textwidth]{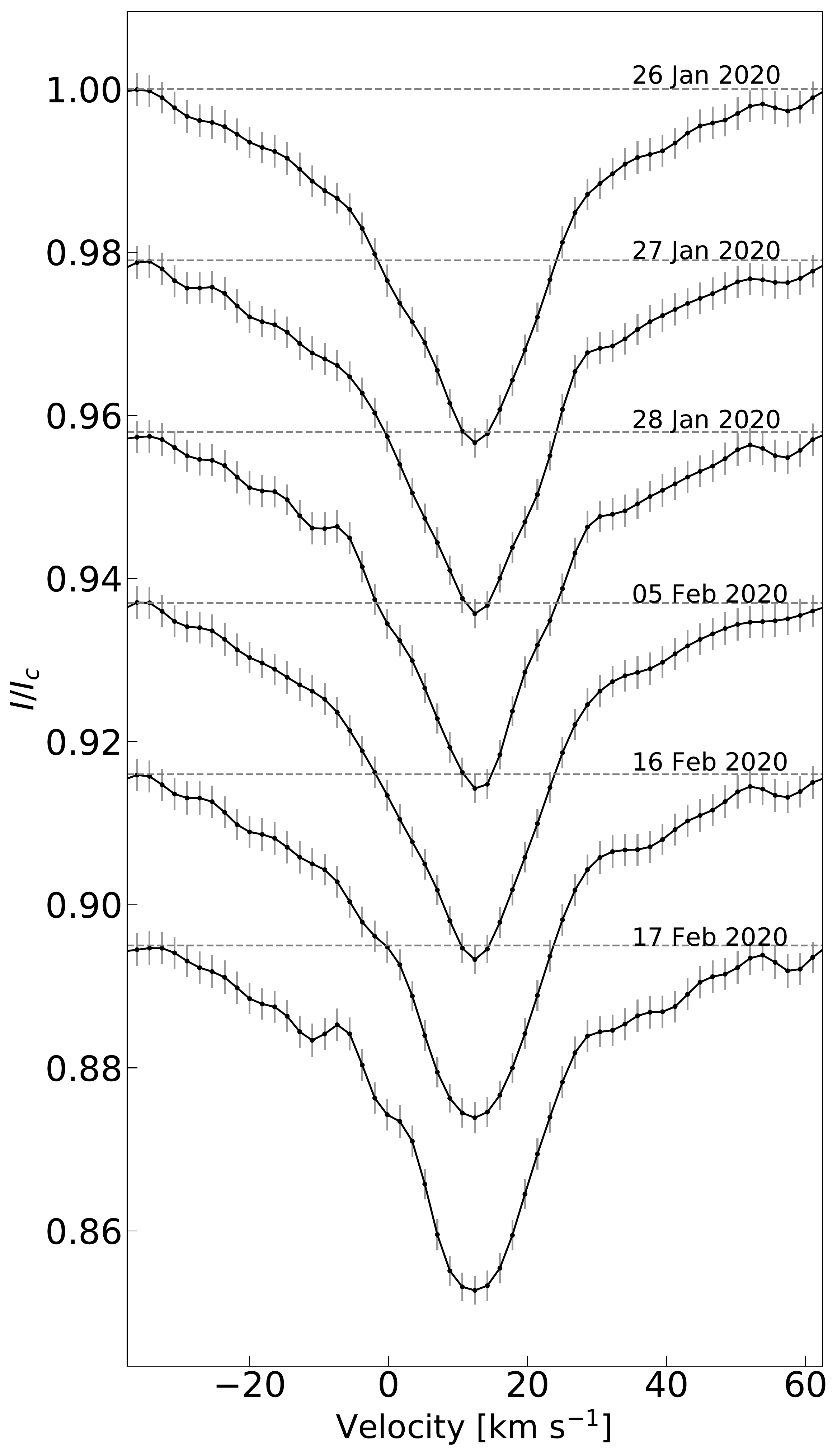} &
    \includegraphics[width=0.35\textwidth]{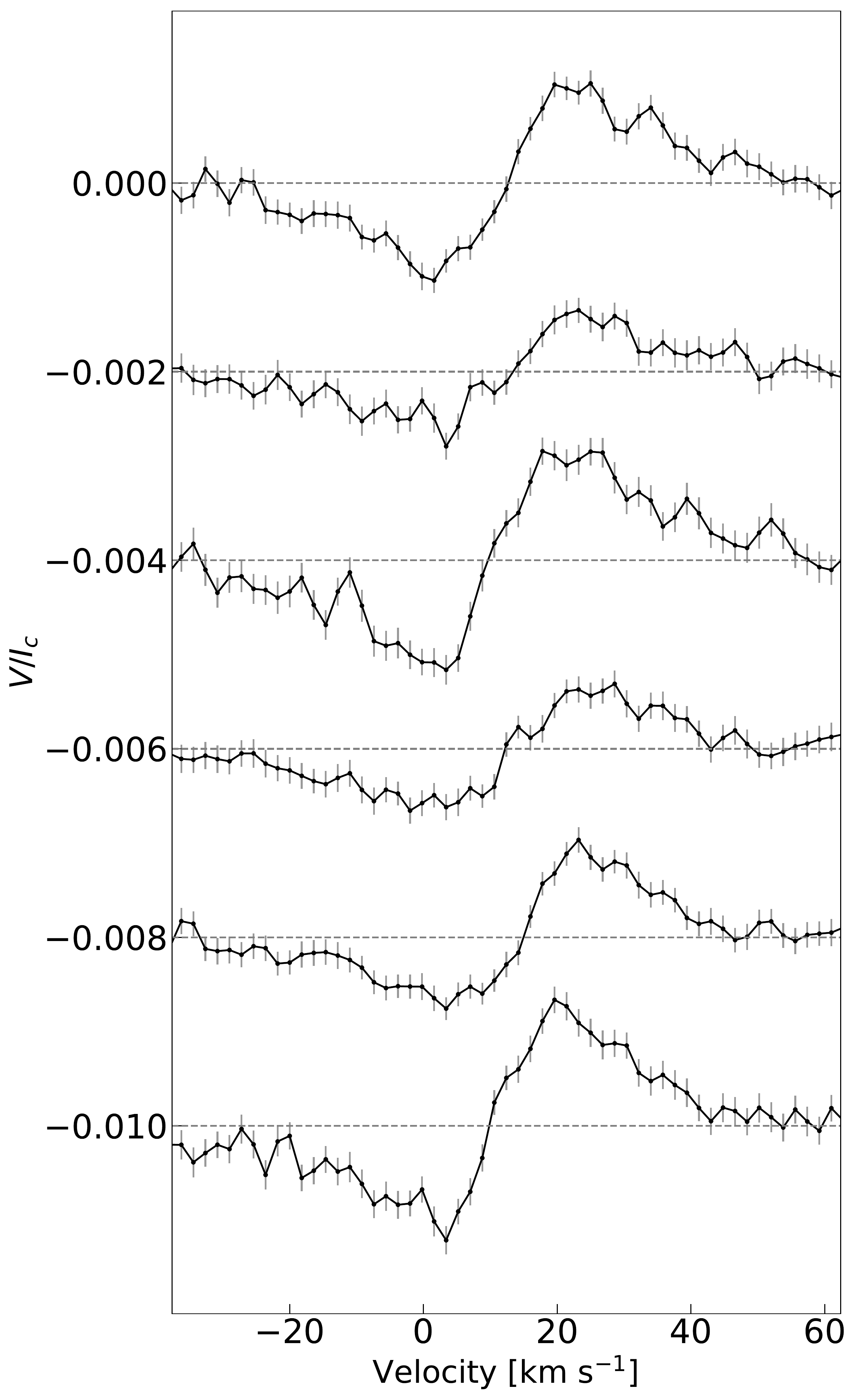} \\
    \end{tabular}
    \caption{Six representative continuum normalized Stokes $I$ (left) and $V$ (right) profiles from SPIRou observations collected in early 2020. 
    The profiles are vertically shifted for visualization purposes.}
    \label{Bl_examples}%
\end{figure*}

  \begin{figure}
   \includegraphics[width=0.5\textwidth]{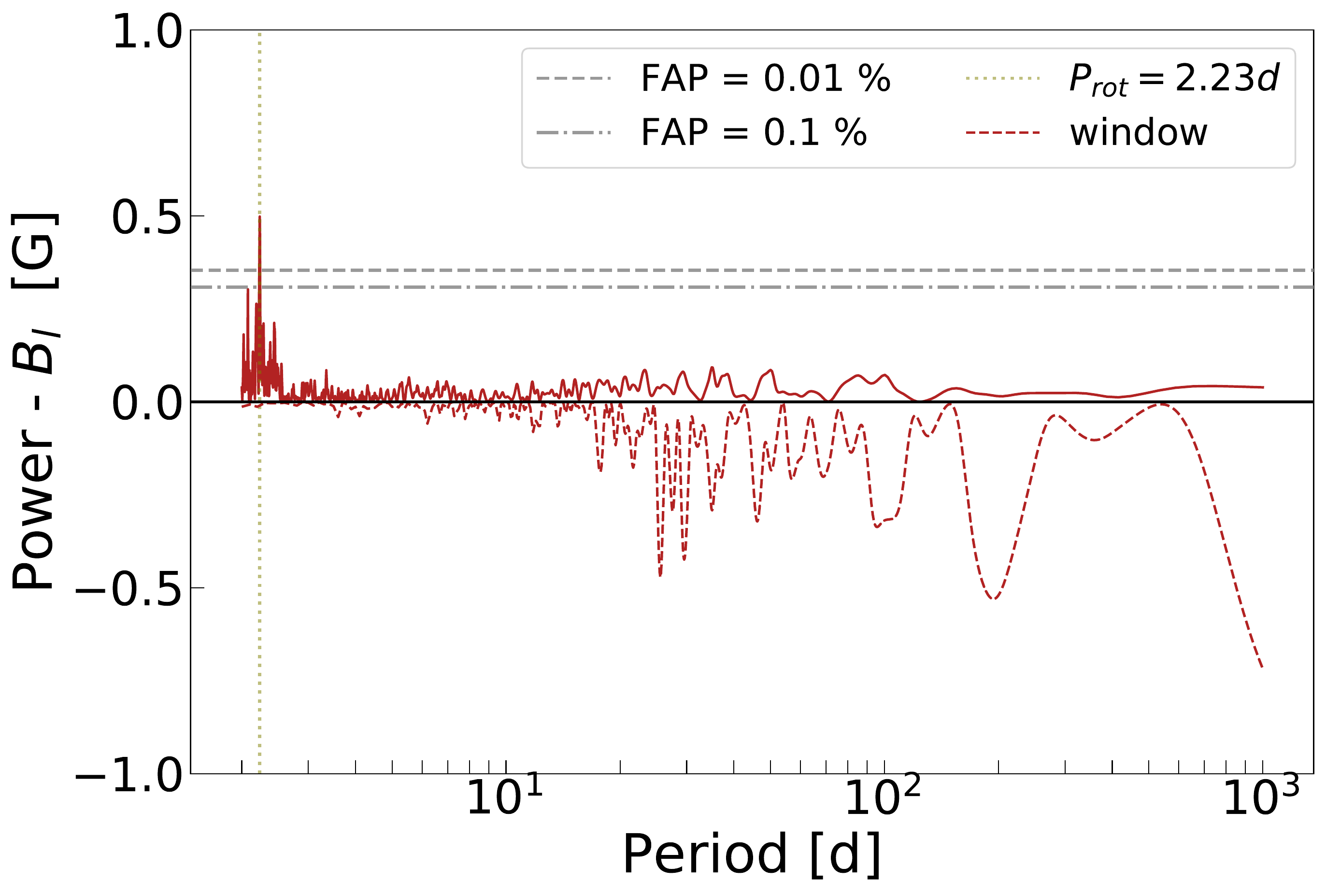}
    \caption{Generalized Lomb-Scargle periodogram and window function of the  SPIRou $B_\ell$ time series.  
    A long-term variation of the $B_\ell$ data set was filtered out beforehand to prevent it from perturbing the period analysis. 
    The highly significant peak  (FAP$<$ 0.01\%) at 2.23$\pm$0.01 days matches the rotation period reported in the literature 
    and the period observed in the SOPHIE RV data.
    }
    \label{fig:mask}%
   \end{figure}

\section{Activity analysis}
  
\subsection{SPIRou longitudinal magnetic field measurements}\label{sec:magfield}
Figure~\ref{Bl_examples} presents examples of the measured SPIRou Stokes $I$ and Stokes $V$ profiles, 
which a future publication (Bellotti et al. in prep) will use for a full ZDI analysis.
Table~\ref{SPIRouRVBlTable} summarizes the $B_\ell$ measurements  obtained from those profiles.
We note that $B_\ell$ is detected at every epoch and varies from $-55$~G to $-277$~G,
with a median of $-194$~G, and a median noise of 15~G.
Figure~\ref{fig:mask} displays the GLS periodogram of the $B_\ell$ time series. 
Although stellar activity signals sometimes lose  coherency over long time spans, 
the periodogram here shows a single, clear, narrow, and highly significant (FAP$<$ 0.01\%) peak at 2.23$\pm$0.01 days.
This period is consistent with those derived from both the optical spectropolarimetric observations 
of \citet{Morin2008} and the SOPHIE, HARPS, HARPS-N, and HIRES RV measurements.
The periodogram structure at longer periods is likely due to aliases of the observation window.
Previous studies have shown that $B_\ell$ is a reliable magnetic activity tracer \citep{Folsom2016, Hebrard2016} 
and that it varies on the stellar rotation period. 
The peak in the $B_\ell$ periodogram is thus further evidence that the rotational period of Gl~388 is 2.23 days and that the RV signal in the optical is due to stellar activity.
Thus, while the stellar rotation signal of Gl~388 is not detected in the SPIRou RV data,
it is clearly detected in the  $B_\ell$ time series.

\begin{table*}[]
\centering
  \caption[]{Prior and posterior distributions of the quasi-periodic GP model of the SPIRou $B_\ell$ measurements 
  and the HARPS and SOPHIE RV time series.}
  \label{GP_table}
 \begin{tabular}{llllll}
  \hline
  \hline
  \multicolumn{1}{c}{\rm Parameter} & \multicolumn{2}{l}{\rm Prior distribution} & \multicolumn{2}{c}{\rm Posterior distribution} \\
   & & & ln units & linear units \\
   \hline 
   \noalign{\smallskip}
    GP SPIRou $B_\ell$\\
    \hline
    Amplitude  [G]                         & $\ln{A}$               & $\mathcal{U}\left(2,6\right)$                                             & $4.38^{+0.45}_{-0.34}$          & $80^{+45}_{-23}$              \\[1mm]
    Decay time [days]           	 & $\ln{l}$                 & $\mathcal{U}\left(2,8\right)$                                             & $5.37^{+0.32}_{-0.29}$          & $216^{+82}_{-54}$            \\[1mm]
    Smoothing parameter             & $\ln{\Gamma}$    & $\mathcal{U}\left(-5,3\right)$                                            & $-1.47^{+0.71}_{-0.85}$         & $0.23^{+0.24}_{-0.13}$          \\[1mm]
    Cycle length (period) ~[days]  & $\ln{P}$               & $\mathcal{U}\left(\ln{\left(2.20\right)},\ln{\left(2.25\right)}\right)$   & $0.80221^{+0.00070}_{-0.00058}$ & $2.2305^{+0.0016}_{-0.0013}$ \\[1mm]
    Uncorr. Noise [G]                     & $\ln{\sigma}$      & $\mathcal{U}\left(-30,15\right)$                                         & $-3.7^{+4.1}_{-4.3}$         & $0.03^{+1.52}_{-0.02}$        \\[1mm] 
    \hline
    \noalign{\smallskip}
     GP HARPS \& SOPHIE optical RVs \\
    \hline
    Offset [km s$^{-1}$]          	& $V_0$            	& median(RV) + $\mathcal{U}\left(-10,10\right)$            &                                                         & $0.00^{+0.01}_{-0.01}$    \\[1mm] 
    Amplitude  [km s$^{-1}$]   	& $\ln{A}$         		& $\mathcal{U}\left(-5,5\right)$                                        & $-3.30^{+0.24}_{-0.22}$                  & $0.037^{+0.010}_{-0.007}$    \\[1mm] 
    Decay time [days]            	& $\ln{l}$         		& $B_\ell$ Posterior                                                        & $5.32^{+0.16}_{-0.17}$                    & $205^{+36}_{-32}$            \\[1mm]
    Smoothing parameter            & $\ln{\Gamma}$	& $B_\ell$ Posterior                                                        & $-0.69^{+0.40}_{-0.40}$           	     & $0.50^{+0.26}_{-0.17}$           \\[1mm]
    Cycle length (period) [days]   & $\ln{P}$         	& $B_\ell$ Posterior                                                        & $0.80146^{+0.00037}_{-0.00041}$  & $2.2288^{+0.0008}_{-0.0009}$  \\[1mm]
    Uncorr. Noise [km s$^{-1}$]   & $\ln{\sigma}$  	& $\mathcal{U}\left(-10,5\right)$                                     & $-5.61^{+0.13}_{-0.13}$                    & $0.0037^{+0.0005}_{-0.0004}$   \\[1mm] 
    \hline
 \end{tabular}
\end{table*}

\subsection{Gaussian process analysis of the optical HARPS and SOPHIE data using SPIRou's $B_\ell$ as an activity proxy.}

\subsubsection{Gaussian process modeling with a variable decay time.}
Gaussian process (GP) modeling of RV time series is nowadays a standard procedure to correct for 
the stellar activity RV jitter and to search for Keplerian signals \citep[e.g.,][]{Rajpaul2015, SuarezMascareno2020, Klein2021, Kossakowski2022}.
The measurements of the $B_\ell$ obtained with SPIRou are quasi-simultaneous  
with the RV measurements obtained with SOPHIE;
$B_\ell$  therefore provides an excellent proxy of stellar activity for modeling the optical RVs within the GP framework.
The  use of $B_\ell$ as a stellar activity indicator is motivated by the close physical link between stellar magnetic activity and
the presence of spots and plages, and the results of spectropolarimetry 
ZDI studies in M dwarfs \citep[e.g.,][]{EmilieH2014,EmilieH2016},
which have shown that the determination of the stellar magnetic topology enable us 
to filter the RV activity jitter in stars with moderate rotation.

We modeled the combined HARPS and SOPHIE RVs 
running two sequential  quasi-periodic Gaussian processes (QP GP).
We first ran a GP on the SPIRou $B_\ell$ time series. 
Then, using the decay time, smoothing, and cycle length (i.e., rotation period) 
posterior distributions of the $B_\ell$ GP as priors, 
we ran a second GP on the optical RV series. 
For the $B_\ell$ GP, 
we used uniform distributions for the priors of the amplitude, decay time, smoothing, cycle length,
and uncorrelated noise (see further details in Table~\ref{GP_table}). 
This modeling strategy has the underlying assumption that the RV effects and the magnetic field have the same temporal behavior.
Our principal aim is to check how well the optical RVs
could be described by a QP GP model in which  the stable component is 
associated with rotation and the variable component is associated with changes in magnetic activity,
and to determine the typical timescale of that variable component.

For our analysis we used the GP package \texttt{george}~\citep[][]{2015ascl.soft11015F} which uses a quasi-periodic kernel of the form
\begin{equation}
    k\left(\Delta t\right) = A\exp{\left(-\frac{\Delta t^2}{2 l^2}\right)}
    \exp{\left(-\Gamma \sin^2{\left(\frac{\pi}{P}\Delta t\right)}\right)} 
    + \sigma^2 \delta\left(\Delta t\right),
\end{equation}
where $A$ is the amplitude, 
$\Gamma$ is the smoothing parameter, 
$l$ is the decay time in days, 
$P$ is the period (i.e., the cycle length) in days,
and $\sigma$ is the uncorrelated noise (s in the corner plots).
We sampled the distribution of the hyperparameters using a  Markov chain Monte Carlo (MCMC) 
framework using the \texttt{EMCEE} tool \citep[][]{2013ascl.soft03002F} with 100000 samples and 10000 burn-in samples. 
Figure~\ref{fig:Bl_GP} and Table~\ref{GP_table} summarize the results of the GP modeling of the $B_\ell$ time series. 
The corner plots are in Figure \ref{fig:corner_bell} in the Appendix. 
The main result of the $B_\ell$ GP is an improved determination of 
the stellar rotational period to $2.2305^{+0.0016}_{-0.0013}$ days.
It is a similar value to that retrieved using the $B_\ell$ periodogram analysis,
however, it has a smaller uncertainty. 
The decay time obtained in the $B_\ell$ time series is $216^{+82}_{-54}$ days.

\begin{figure}
    \centering
    \includegraphics[width=0.49\textwidth]{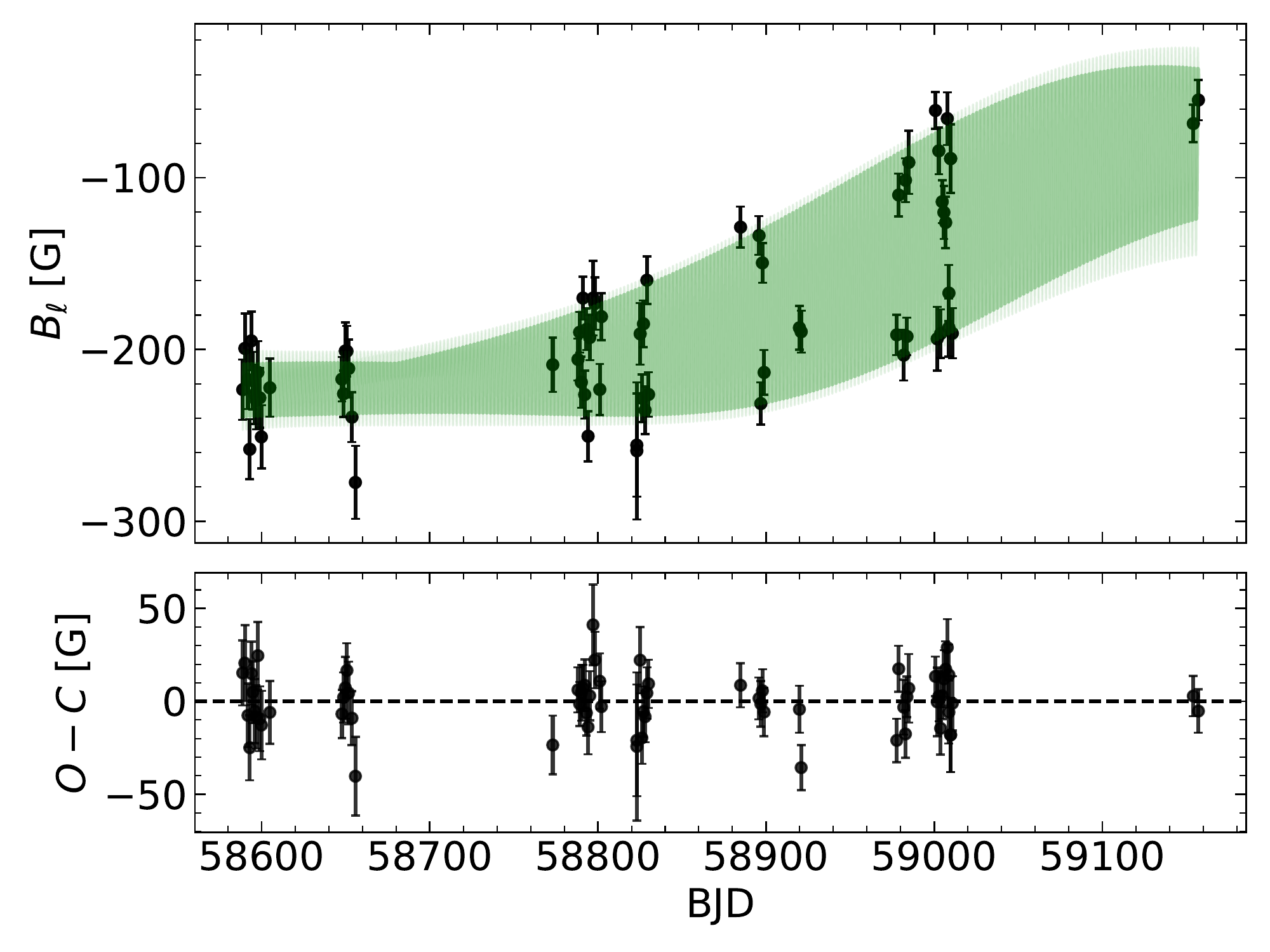}
    \caption{
    Longitudinal magnetic field ($B_\ell$) time series (black dots) and its QP GP model fit (green-line). 
    Details of the GP are given in Table~\ref{GP_table}. 
    The corner plot of the fit is given in Figure~\ref{fig:corner_bell} in the Appendix.
    }
   \label{fig:Bl_GP}
\end{figure}

\begin{figure*}
   \begin{center}
   \includegraphics[width=0.98\textwidth]{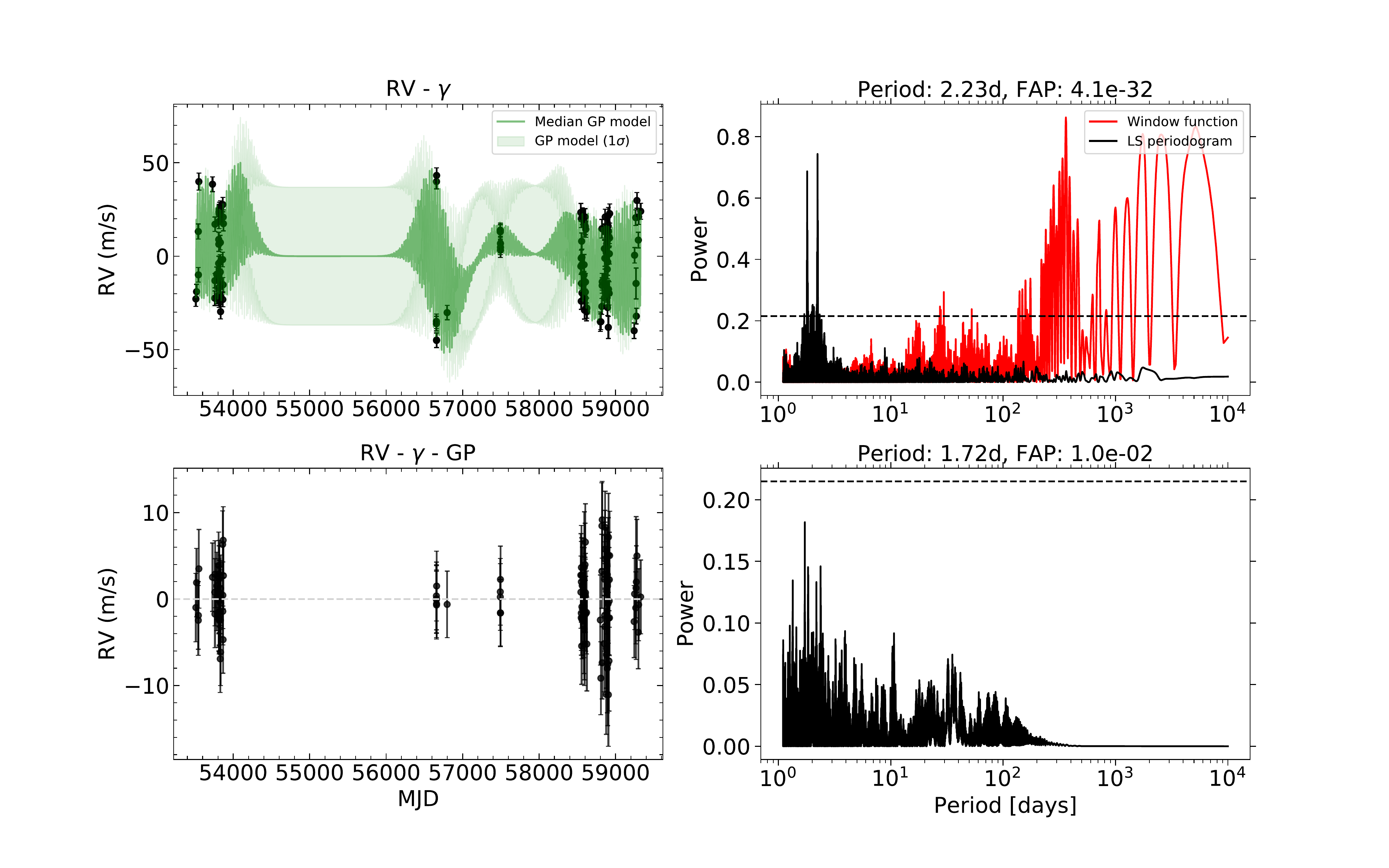}
    \caption{Summary of the GP analysis of the optical RV time series using SPIRou's $B_\ell$ as an activity proxy. 
    {\it Upper left:} QP GP  fit (green line) to the HARPS and SOPHIE optical RV time series (black points). 
    The median value of the raw RV is denoted by $\gamma$. 
    {\it Upper right:} Periodogram of the RV data and window function before the QP GP fit. 
    The horizontal line is the FAP=0.1\%. 
    {\it Lower left:} Residuals in the optical RVs after the subtraction of the QP GP fit.
     {\it Lower right:} Periodogram of the RV residuals after subtraction of the QP GP fit. 
     In the top of each periodogram, the period, and FAP of the strongest peak are given.
     Details of the GP model are summarized in Table~\ref{GP_table}. 
     The corner plot of the fit is given in Figure~\ref{fig:corner_rv} in the Appendix.
    }
     \label{fig:QPGP}
    \end{center}
\end{figure*} 

The results of the GP model  of the optical RVs are given in Figure~\ref{fig:QPGP}
and in Table~\ref{GP_table}. 
The GP corner plots are provided in Figure~\ref{fig:corner_rv} in the Appendix.
We obtained in the GP of the RVs a period of  $2.2288^{+0.0008}_{-0.0009}$ days,
which is a similar value within the error bars to that obtained in the $B_\ell$ GP.
The decay time obtained in the RV GP is $205^{+36}_{-32}$ days, 
a value that is also consistent within the error bars with the decay time obtained in the $B_\ell$ GP.
The decay time of $\sim$200 days of the RV  is shorter than the interval of time that separates the HARPS1, HARPS2, and SOPHIE observing campaigns.
Therefore, each campaign can be considered as independent with respect to the variable component of the activity 
in the RV jitter.
The similarity of the period and the decay time of the $B_\ell$ GP and RV GP 
suggests that the RV jitter is related to changes in the magnetic activity,
but it can also be due to the very fact
the $B_\ell$ GP has been used to set the priors of the RV GP model.
The bottom panels of Figure~\ref{fig:QPGP} display the residuals of the optical RV time series 
and their periodogram  after subtraction of the RV GP model.
After GP correction, the rms of the time series is 3.7 m s$^{-1}$ and there is no evidence for periodic signals. 

The main conclusion from our GP modeling is that the RV model anchored in the $B_\ell$ 
as an activity proxy provides a good description of the observed optical RV time series variability.
This suggests that the rotationally modulated activity jitter observed in the optical is indeed 
probably linked to magnetic activity.

\begin{figure}
   \begin{center}
   \includegraphics[width=0.5\textwidth]{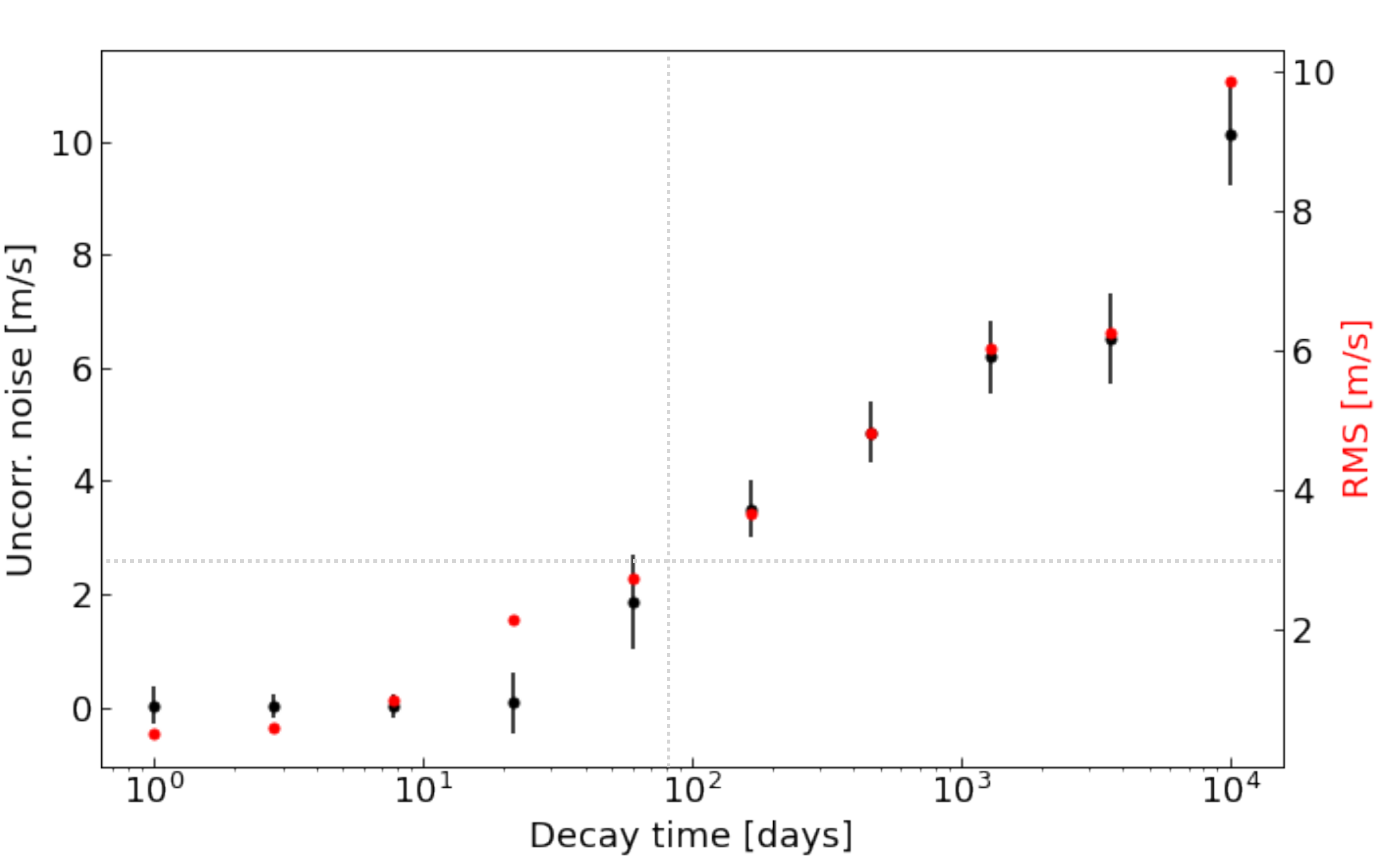}
    \caption{Uncorrelated noise and RV rms as a function of the decay time in the QP GP model of the optical HARPS and SOPHIE RV time series. 
    The horizontal line in gray dots shows the typical accuracy of SOPHIE RVs (3 m s$^{-1}$). 
    The vertical line in gray dots indicates the decay time of 80 days. 
     }
         \label{fig:rms_vs_decay}
    \end{center}
\end{figure} 

\subsubsection{Gaussian process modeling with a fixed decay time.}
In the previous analysis, we have enabled the decay time to be variable in the GP.
The consequence of this assumption is that for short decay times,
the GP by construction does not keep the phase constant between the various sets of optical RV data. 
This analysis leaves the question unanswered of whether is it still possible to fit all the 
optical RV data with a slowly evolving pattern that does not lose phase from run to run 
(i.e., whether the RV time series can be described with a GP with a long decay time),
in addition to
the question of how long the decay time can be before the uncorrelated noise and the RV rms start to increase. 
To explore the influence of the decay time in the uncorrelated noise and the RV rms,
and to test whether a very long decay time ($>$1000 days) could  describe the data,
we ran a series of QP GPs on the optical RV data sets, 
by fixing in each the decay time to different values from 1 to 10,000 days.
Our results are summarized in Figure~\ref{fig:rms_vs_decay}
where we show the evolution of the uncorrelated noise 
and the RV rms as a function of the decay time.
We  observe that as soon the decay time becomes larger than 10 days, 
the uncorrelated noise and the RV rms increase.
We recall, however, that the mean RV error in the SOPHIE RVs is $\sim$2.6~m~s$^{-1}$;
therefore, models displaying a lower RV rms  are, in fact, fitting the noise.
Figure~\ref{fig:rms_vs_decay} indicates that the RV rms becomes higher than 3~m~s$^{-1}$ rms for decay times larger than 80 days.
The decay time on the order of hundreds of days suggested both by the variable and fixed decay time GP modeling 
indicates that the RV jitter is evolving with time on a timescale of less than a year, conversely to what \citet[][]{Tuomi2018} claim.

\begin{figure}
   \centering         
   		\includegraphics[width=0.3\textwidth]{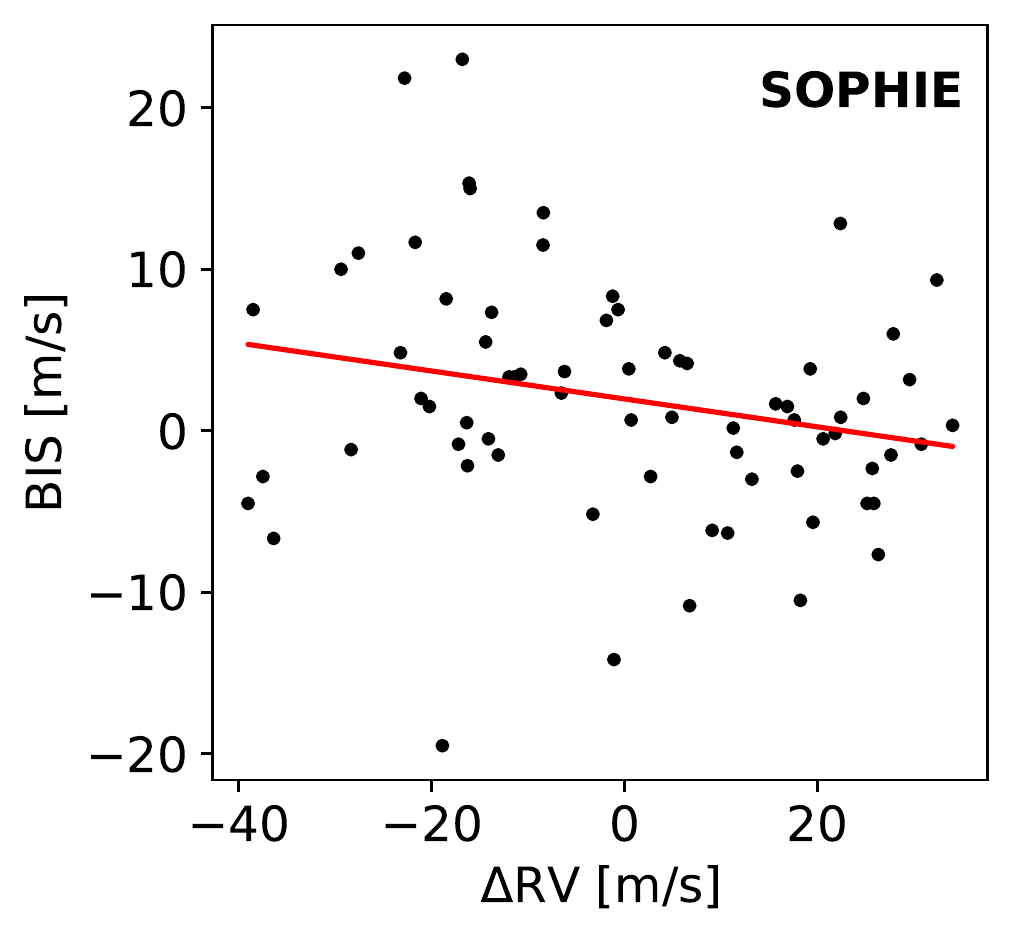}
		\caption{BIS versus RV for the SOPHIE data.
		 The red line indicates the linear fit between these variables. 
		The Pearson's coefficient is $-$0.23, which indicates a weak anticorrelation.}
                 \label{SOPHIE:RV:BIS}
 \end{figure} 
 \begin{figure}
   \centering     	
		\includegraphics[width=0.49\textwidth]{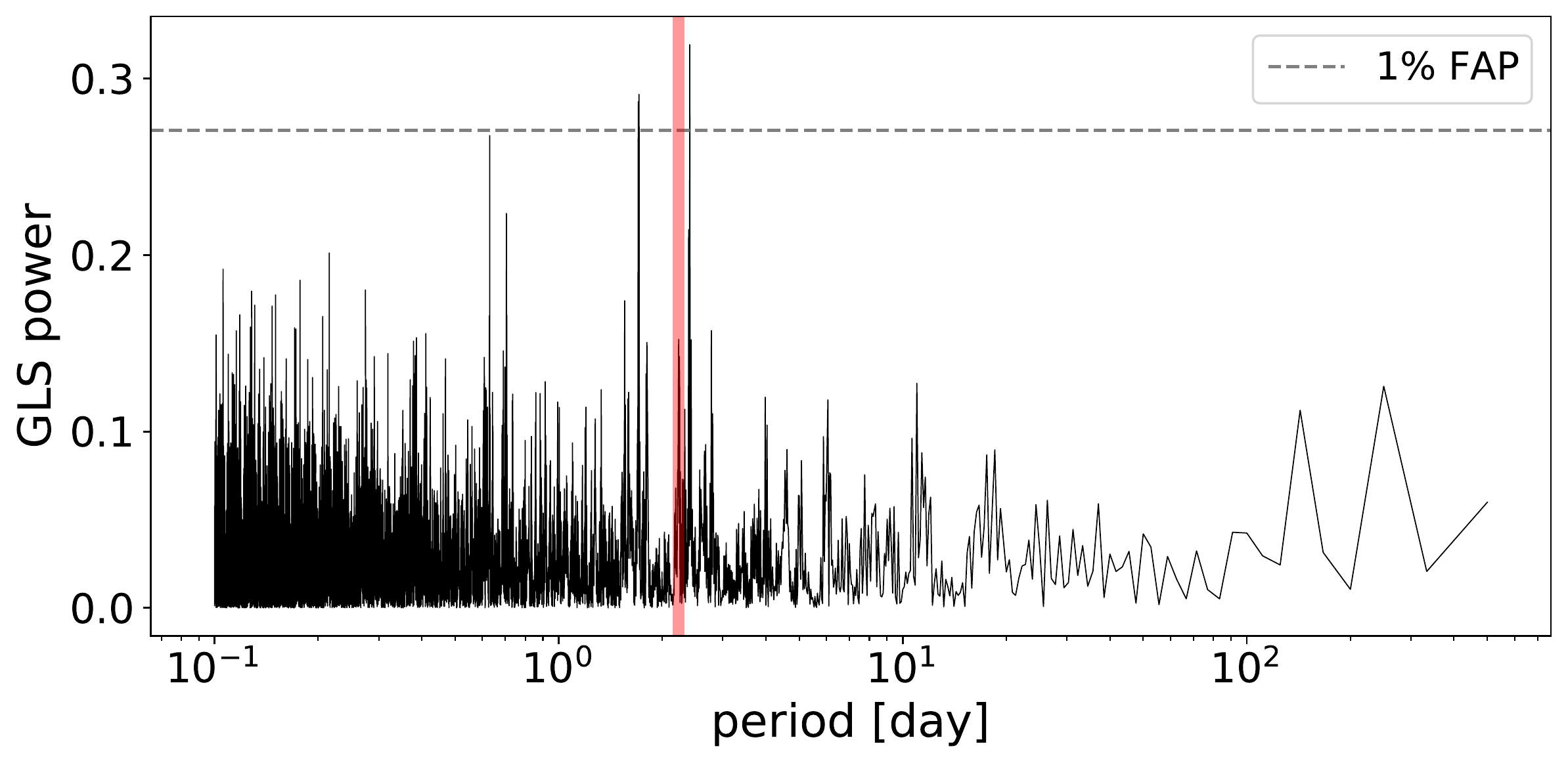}
	         \caption{Generalized Lomb-Scargle periodogram of the SOPHIE bisector time series. 
	         The gray dashed line marks the 1\% FAP. 
	         The highest peak in the periodogram corresponds to a 2.4~day period. 
	         The red vertical line indicates the location of the 2.23~day period.}
	          \label{SOPHIE:BIS:Periodogram}
 \end{figure}

\subsection{Line shape analysis}
Stellar activity changes the spectral line profiles, which in turn induces RV variations. 
The bisector inverse slope (BIS) is a line shape metric defined as the difference between 
the velocity at the top and the bottom of the cross-correlation function \citep[][]{Queloz2001}. 
Figure~\ref{SOPHIE:RV:BIS} plots the  BIS of the cross-correlation function against the RV 
for the SOPHIE data set.
The two quantities are slightly anticorrelated  with a Pearson coefficient $R$ of $-$0.23 (p-value = 0.05).
 This anticorrelation is weaker than the $R$ of $-$0.81 obtained by \citet[][]{Bonfils2013} with HARPS. 
This difference may be due to the higher spectral resolution of HARPS (R=120\,000) in comparison with SOPHIE (R=70\,000) 
since the BIS  strongly depends on this spectrograph feature \citep{Desort2007}.
Figure~\ref{SOPHIE:BIS:Periodogram} shows the GLS periodogram of the SOPHIE bisector time series, with a highest peak at 2.4 days.

Within the LBL framework which we used for the SPIRou spectra, 
we adopted the second and third derivatives of the LBL line profiles as line-shape diagnostics \citep[][Cortez-Zuleta, in prep.]{Artigau2022}. 
The second derivative can be interpreted as a proxy for the full-width half-maximum (FWHM) of the CCF, and the third derivative as a proxy for its BIS. 
The GLS periodograms of the time series of these second and third derivatives show 
no peak with a FAP below 1\% (see Figure~\ref{Activity:Periodograms}).

\subsection{Activity indicators}
The S index and the $ \rm H\alpha $ index are widely used tracers of stellar chromospheric activity in the optical domain. 
They were measured from the SOPHIE spectra following the methods of  \citet[][]{Boisse2009} and \citet[][]{Boisse2010}, respectively. 
We computed GLS periodograms for both indices (Figure~\ref{Activity:Periodograms}) and searched for periodic signals. 
None of the peaks in either periodogram rise to the 1\% FAP line. 
We note that one of the highest peaks of the $ \rm H\alpha $ periodogram is located at 4.2 days, 
which is close to the $\rm 2\times P$ subharmonic of the 2.23~day rotation period, but this might be coincidental.

Activity indicators are not as well established in the near-IR domain.
We analyzed  the temporal evolution of time series of the 
pseudo-equivalent width (pEW) of four atomic lines that were previously tested in the optical domain and/or for Sun-like stars: 
Aluminum I ($\lambda$~13154~\AA) \citep{Spina2020}, 
Potassium I ($\lambda$~12435~\AA) \citep{Barrado2001,Robertson2016,Terrien2022}, 
Iron I ($\lambda$~11693~\AA) \citep{Yana2019,Cretignier2020}, and 
Titanium I ($\lambda$~10499~\AA) \citep{Spina2020}. 
The details of the method will be described 
in Cortes-Zuleta et al. (in prep).

Figure~\ref{Activity:Periodograms} displays the GLS periodogram of the pEW of these four near-IR lines. 
There is no high significant periodicity for the periodograms of Iron, Potassium, and Titanium that are not related
to the spectral window function.
The Aluminum periodogram is the only one where peaks rise above the 1\% FAP line. 
These peaks are concentrated around the ~1 day peak due to the sampling and produce the observed the long-period (>10 days) aliases.

Due to the low inclination of Gl~388, one pole is within $\sim$20~degrees of the observer's line of sight. 
This strongly reduces the modulation of activity indicators by stellar rotation. 
Activity features, such as spots, close to the observed pole will be in view for their full lifetime; 
thus, their pEW and projected velocity change little with rotation.} 
Most of the stellar activity indicators tested in this work indeed do not vary on the 2.23~day stellar rotation period, 
which is unambiguously determined by the analysis of the longitudinal magnetic field (Section~\ref{sec:magfield}). 
The lack of variability on the stellar rotation period is therefore expected, 
particularly for the near-IR SPIRou data which have much lower RV 
jitter from activity than SOPHIE or HARPS in the optical \citep[e.g., this work and][]{Reiners2010}. 
Besides SPIRou $B_\ell$ which is, in fact, an activity indicator modulated with the stellar rotation, 
the  SOPHIE BIS is the only activity indicator in the spectra to be found to vary on a period (2.4~days) close to the stellar rotation period.
The SOPHIE BIS exhibits a periodic variation with the stellar rotation because the BIS 
is sensitive to small changes in the line profile symmetry.

\begin{figure*}
   \centering         
   		\includegraphics[width=0.96\textwidth]{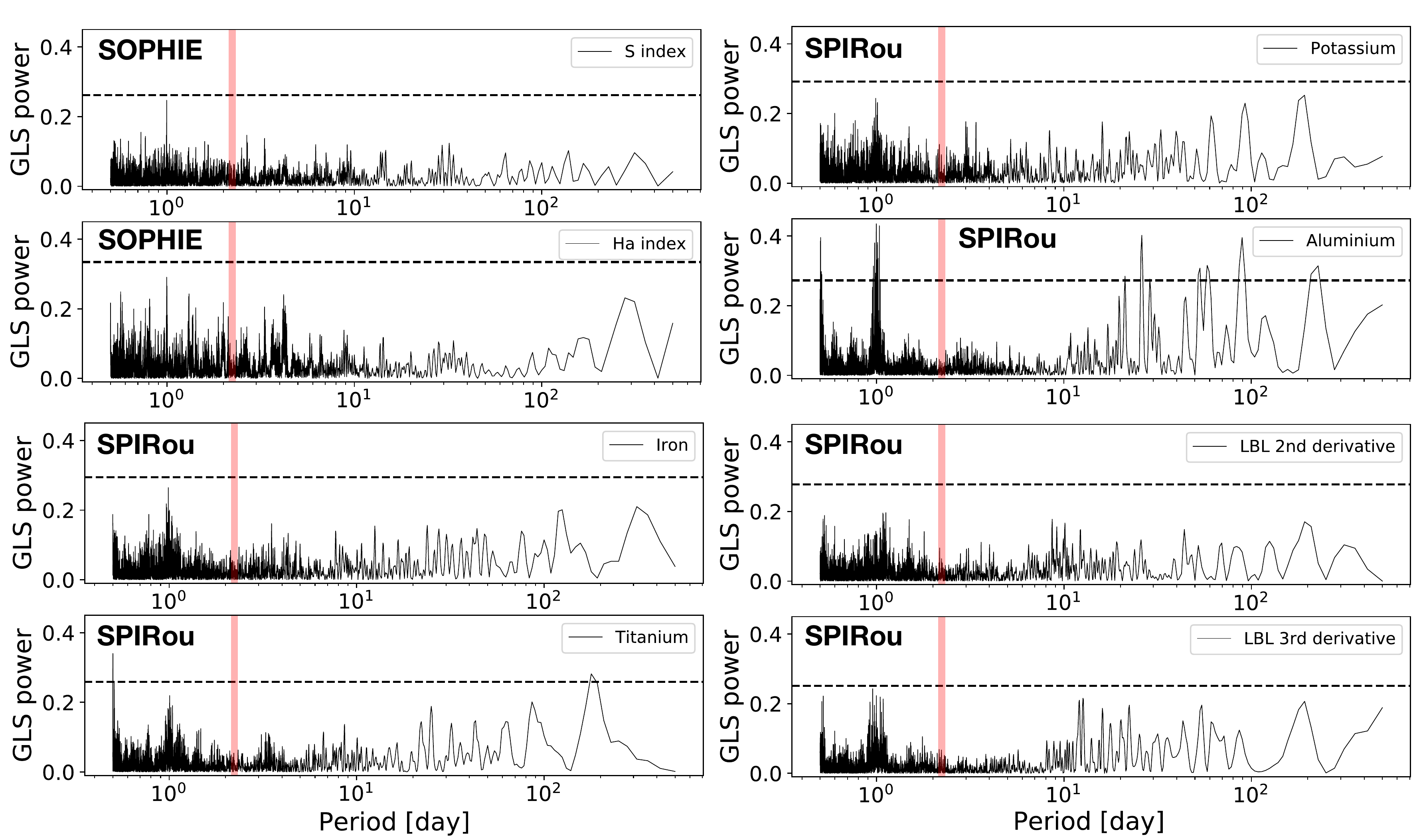}
		\caption{Generalized Lomb-Scargle periodograms of the activity  indicators measured in the SOPHIE and SPIRou spectra. 
		The dashed line marks the 1\% FAP. The vertical red line indicates the 2.23 day period.}
                 \label{Activity:Periodograms}
 \end{figure*} 		
 
\section{Conclusions}
   
Radial velocity searches for planets around nearby M dwarfs are our best chance to identify rocky planets in the habitable zone that can
be characterized by future instruments \citep[e.g.,][]{doyon2014, beichman2014, snellen2015, lovis2017}.
M dwarfs have great potential as targets because the expected RV 
signature of an Earth-like planet in the habitable zone is within the sensitivity limits of current ground-based high-resolution velocimeters. 
However, many nearby M dwarfs are magnetically active. 
The spots and plages in their atmospheres present a challenge to RV planet searches
since they can induce RV signatures that mimic those of a planet. 

In this paper, 
we have presented contemporaneous  new RV and spectropolarimetry observations 
of the magnetically active M dwarf Gl~388 obtained between 2019 and 2021,
with the optical high-resolution spectrometer SOPHIE at OHP 
and the near-IR high-resolution spectropolarimeter  SPIRou at CHFT.
Our observations were obtained ten to fifteen years after 
previous RV measurements with HARPS at la Silla (2005 and 2006)
and  optical spectropolarimetry  observations with ESPaDOnS at CFHT  (2007 and 2008).

The new  SOPHIE RVs vary on a  2.23$\pm$0.01~day period with 23.6$\pm$0.5~m~s$^{-1}$ amplitude, 
which is consistent with the earlier HARPS observations.
The two optical data sets can be jointly fit with a single Keplerian solution, 
which however has larger residuals than separate fits to the individual data sets. 
This can be explained by a subtle amplitude or phase change between the HARPS and SOPHIE epochs 
due to interactions between stellar differential rotation and changes in the average latitude of the spots. 
The HARPS data alone have long provided
strong evidence that the optical RV signal is due to stellar activity \citep[][]{Bonfils2013, Reiners2013}. 
The new SPIRou data display no periodic signal and their 5~m~s$^{-1}$ dispersion
is well below the 23.6$\pm$0.5 amplitude of the optical RV signal, 
removing any possible doubt that the latter is indeed due to activity, 
and it cannot be from the Doppler effect.
The HARPS and SOPHIE data demonstrate that activity can induce a RV signal 
that is largely stable well over a decade, even in a star as active as Gl~388.
The long-term stability of the Gl~388 RV signal shows how easily an activity signature in optical 
RV measurements can be misinterpreted  as evidence for a planet \citep[e.g.,][]{Tuomi2018}.

The SPIRou campaign on Gl 388  spectacularly demonstrates the lower sensitivity of the near-IR RVs 
(at least those obtained by an LBL method)  to stellar magnetic activity. 
This opens the way to perform exoplanet searches around stars which so far have been hardly studied 
or excluded from studies in the optical because of their activity. 
This also indicates that in some cases, when it is difficult to separate an RV activity effect from a Keplerian one, 
the use of near-IR RVs in addition to optical RVs is a crucial tool to remove spurious signals. 
For Gl~205, a star quieter than AD Leo, \citet[][]{Cortes-Zuleta2023} demonstrate that the RV jitter can 
also be present in the IR but at a different phase or with a different shape than in the optical, possibly allowing filtering.
The use of the near-IR RVs is of high relevance for the search of Earth-like planets in the habitable zone around M dwarfs 
\citep[e.g.,][]{Artigau2022}.
With lower planetary masses, the combination of data from several instruments, 
and the need for activity filtering,
a growing number of spurious signals is possible. 
To use near-IR RVs in addition to optical RVs it is important to consolidate optical RV surveys with long-term follow-up campaigns.

Our SPIRou observations additionally reiterate
how effective spectropolarimetry is at determining the stellar rotation period from the variations of the longitudinal magnetic field. 
We measured a rotation period of 2.2305$\pm$0.0016 days from the SPIRou $B_\ell$ time series.
ESPaDOnS and SPIRou, 
using the same technique at different wavelengths, 
produce essentially identical  measurements of the stellar rotation period. 
Spectropolarimetry can determine the rotation period independently of photometry, 
which is particularly relevant for stars with a complex photometric time series behavior, 
and it can independently measure the rotation period and the inclination.

\begin{acknowledgements}
Based on observations obtained at the Canada-France-Hawaii Telescope (CFHT) which is 
operated from the summit of Maunakea by the National Research Council of Canada, the Institut National des Sciences de l'Univers of the Centre 
National de la Recherche Scientifique of France, and the University of Hawaii. 
Based on observations obtained with SPIRou, an international project led by Institut de Recherche en Astrophysique et Plan\'etologie, Toulouse, France.
Based on observations obtained with the spectrograph SOPHIE 
at the Observatoire de Haute-Provence (OHP) in France,
operated by Institut National des Sciences de l'Univers of the Centre 
National de la Recherche Scientifique of France.
We acknowledge funding from the French ANR under contract number ANR\-18\-CE31\-0019 (SPlaSH).
This work is supported by the French National Research Agency in the framework of the Investissements d'Avenir program (ANR-15-IDEX-02), 
through the funding of the ``Origin of Life" project of the Grenoble-Alpes University.
JFD acknowledges funding from the European Research Council (ERC) under the H2020 research \& innovation program (grant agreement \#740651 NewWorlds).
This publication makes use of the Data \& Analysis Center for Exo-planets (DACE), which is a facility based at the University of Geneva (CH) dedicated to extrasolar planets data visualization, 
exchange and analysis. DACE is a platform of the Swiss National Centre of Competence in Research (NCCR) PlanetS, 
federating the Swiss expertise in Exoplanet research. The DACE platform is available at https://dace.unige.ch.
The project leading to this publication has received funding from the french government under the "France 2030" investment plan 
managed by the French National Research Agency (reference : ANR-16-CONV-000X / ANR-17-EURE-00XX) and 
from Excellence Initiative of Aix-Marseille University - A*MIDEX (reference AMX-21-IET-018). This work was supported by the "Programme National de Plan\'etologie" (PNP) of CNRS/INSU.
E.M. acknowledges funding from FAPEMIG under project number APQ-02493-22 and research productivity grant number 309829/2022-4 awarded by the CNPq, Brazil.
This work has made use of data from the European Space Agency (ESA) mission
{\it Gaia} (\url{https://www.cosmos.esa.int/gaia}), processed by the {\it Gaia}
Data Processing and Analysis Consortium (DPAC,
\url{https://www.cosmos.esa.int/web/gaia/dpac/consortium}). Funding for the DPAC
has been provided by national institutions, in particular the institutions
participating in the {\it Gaia} Multilateral Agreement.
The observations at the Canada-France-Hawaii Telescope were 
performed with care and respect from the summit of Maunakea which is a significant cultural and historic site.
 \end{acknowledgements}

%
\bibliographystyle{aa} 
\bibliography{Gl388.bib} 
%

\begin{appendix}

\section{SOPHIE and SPIRou chromaticity}

\begin{figure}[h]
   \centering         
   		\includegraphics[width=0.49\textwidth]{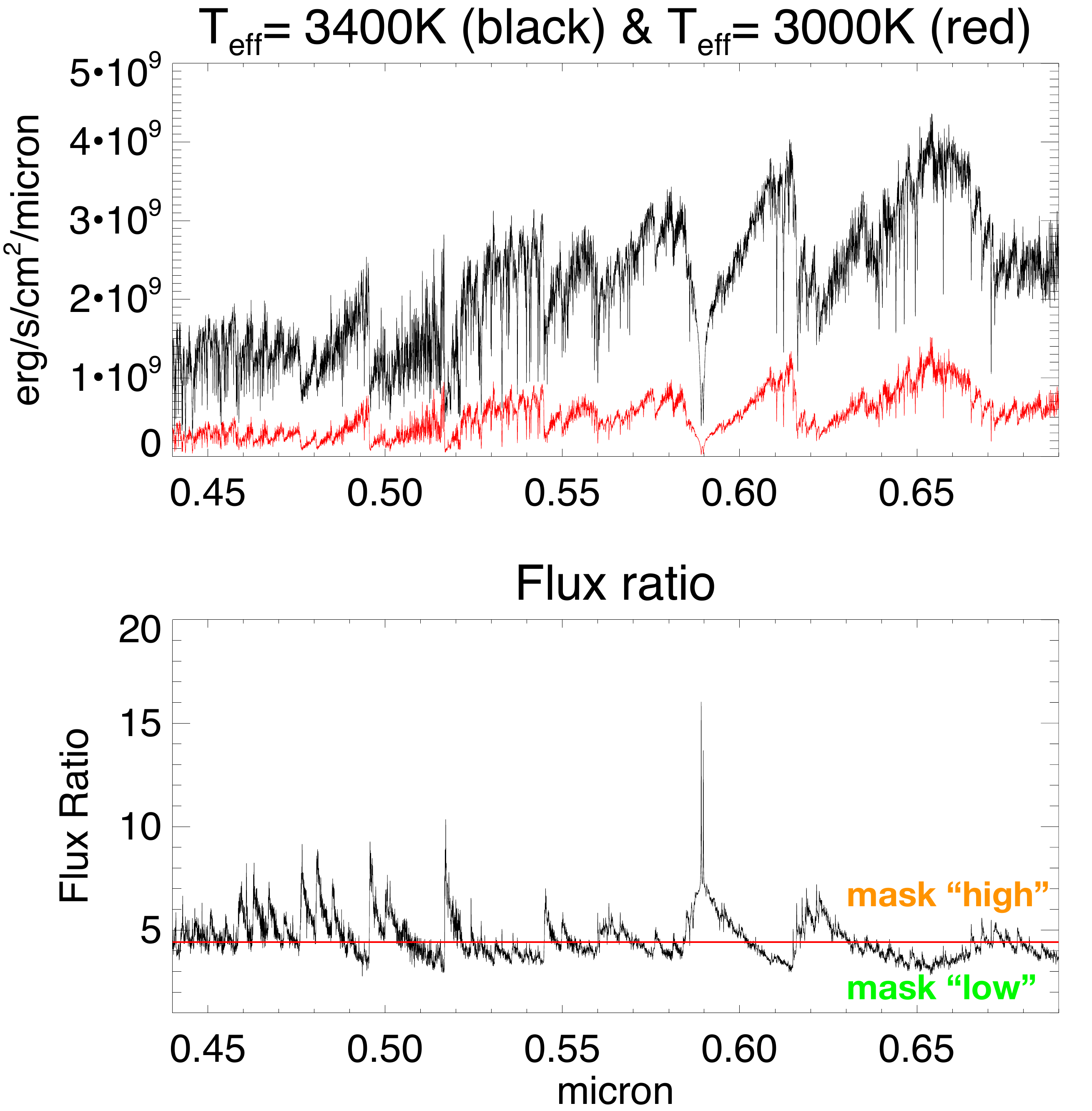}
		\caption{
		Regions in the optical spectra used to define the CCF masks to study the chromaticity of the SOPHIE RVs.
		{\it Upper panel:} 
		Theoretical PHOENIX medium resolution models of stellar atmospheres with $T_{\rm eff}=3400$ and $3000$ K in the 
                 SOPHIE wavelength range. 
                 {\it Lower panel:}  
                 Flux ratio of the atmosphere models of the upper panel. 
                 The horizontal red line is the median flux ratio.
                 Lines of the SOPHIE M-star mask in the regions over the median flux ratio constitute the high-contrast mask, 
                 and the lines located below the median constitute the low-contrast mask.
                 }
                 \label{SOPHIE_high_low_regions}
\end{figure} 
\begin{figure}[h]
                 \includegraphics[width=0.5\textwidth]{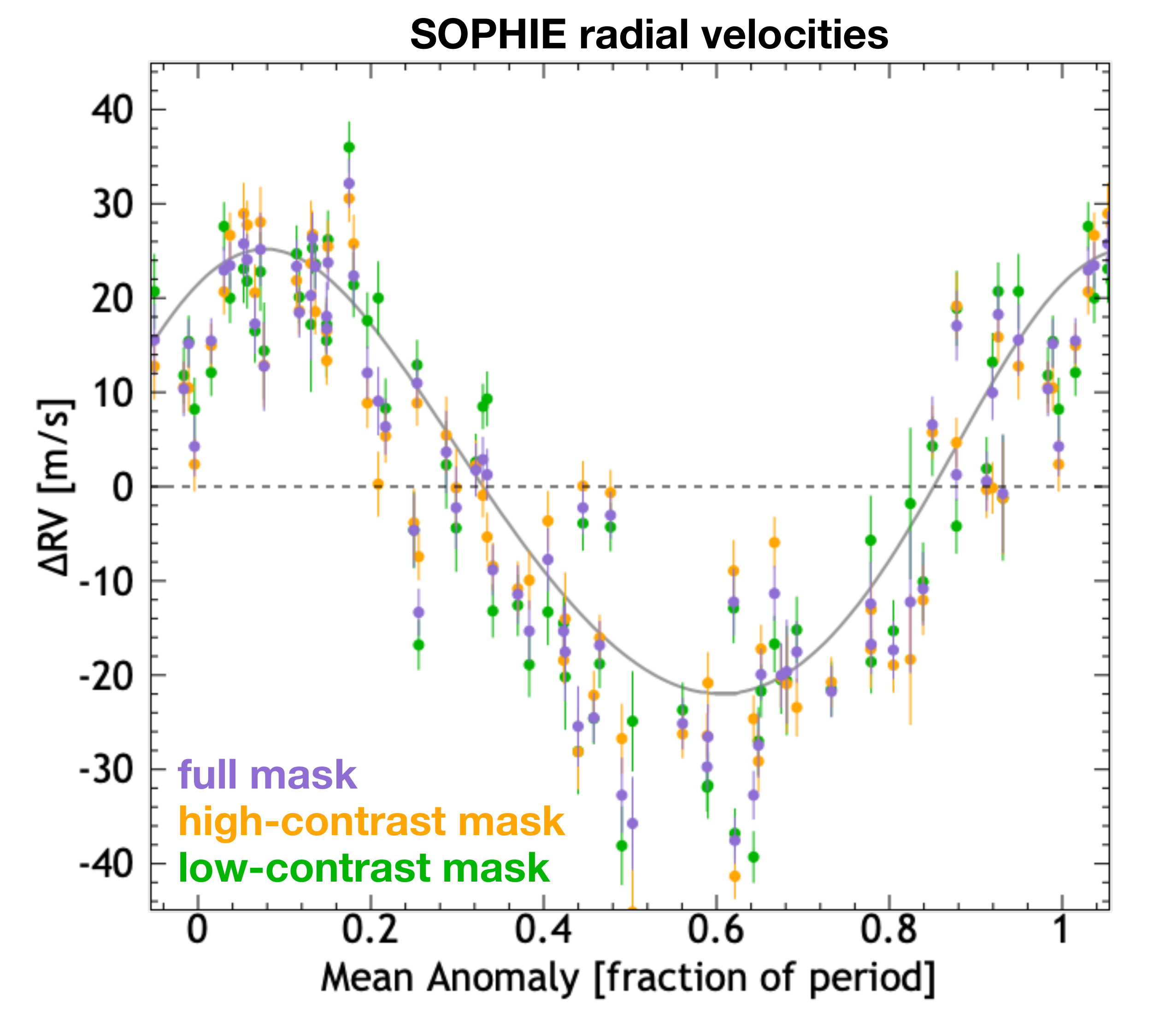}
                 \caption{Phase-folded RVs measured with SOPHIE using the full, high-contrast, and  low-contrast  SOPHIE M-type star CCF masks.
                 The model is the Keplerian model derived from SOPHIE's full-mask RVs.}
                 \label{SOPHIE_high_low_RV}
   \end{figure}   

   \begin{figure}[h]
   \centering         
                 \includegraphics[width=0.5\textwidth]{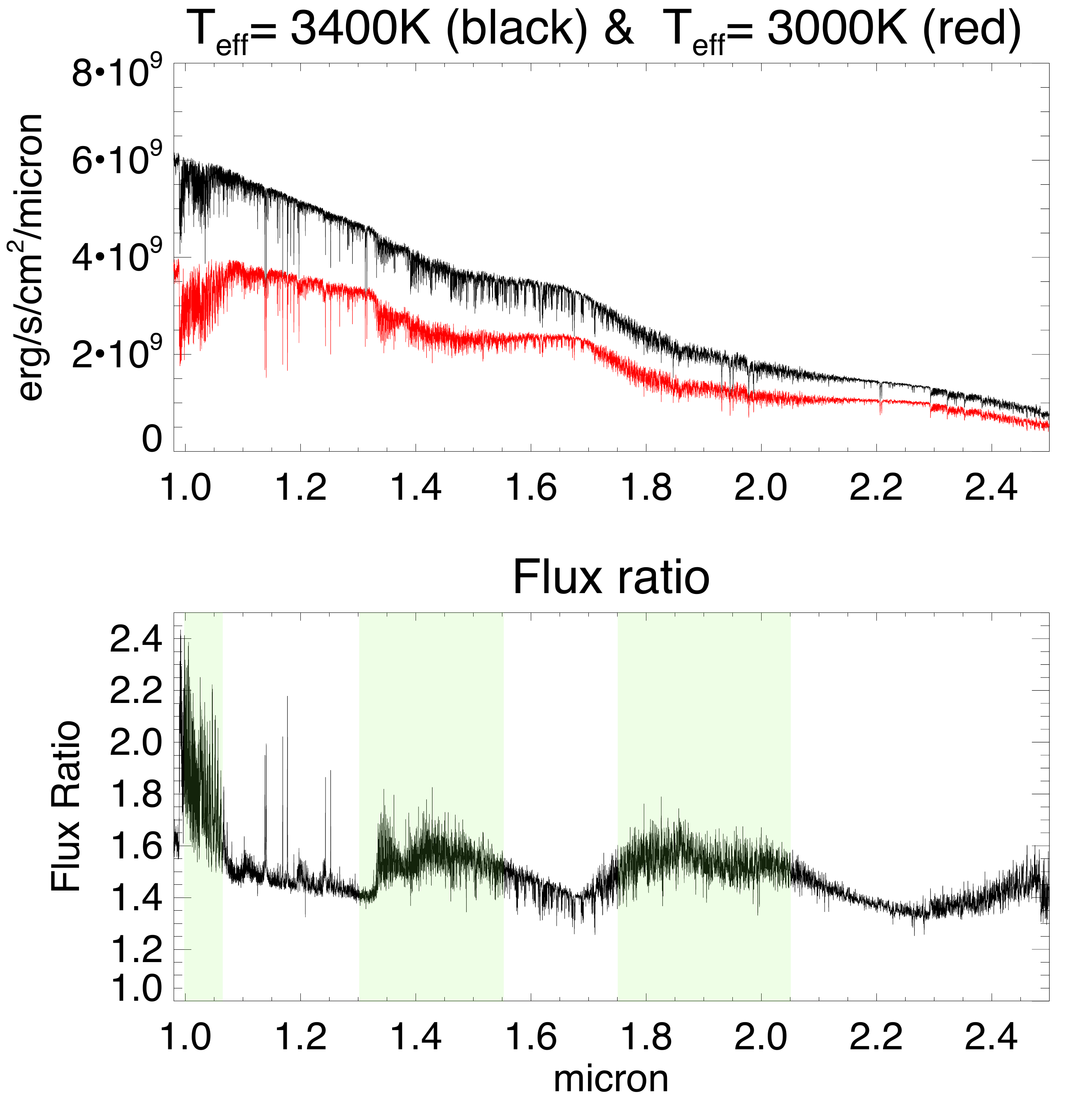}
                 \caption{
                 Regions in the near-IR spectra used to select the LBL lines to study the chromaticity of the SPIRou RVs. 
                 {\it Upper panel:} 
                 Theoretical PHOENIX medium resolution models of stellar atmospheres with $T_{\rm eff}=3400$ and $3000$ K in the 
                 SPIRou wavelength range. 
                 {\it Lower panel:}  
                 Flux ratio of the atmosphere models of the upper panel. 
                 The regions highlighted in light green are the high-contrast flux ratio regions used for filtering the LBL RV measurements. 
                 }
                   \label{SPIROU_spot_3400_3000}
\end{figure}    
\begin{figure}[h]
   \centering         
                 \includegraphics[width=0.5\textwidth]{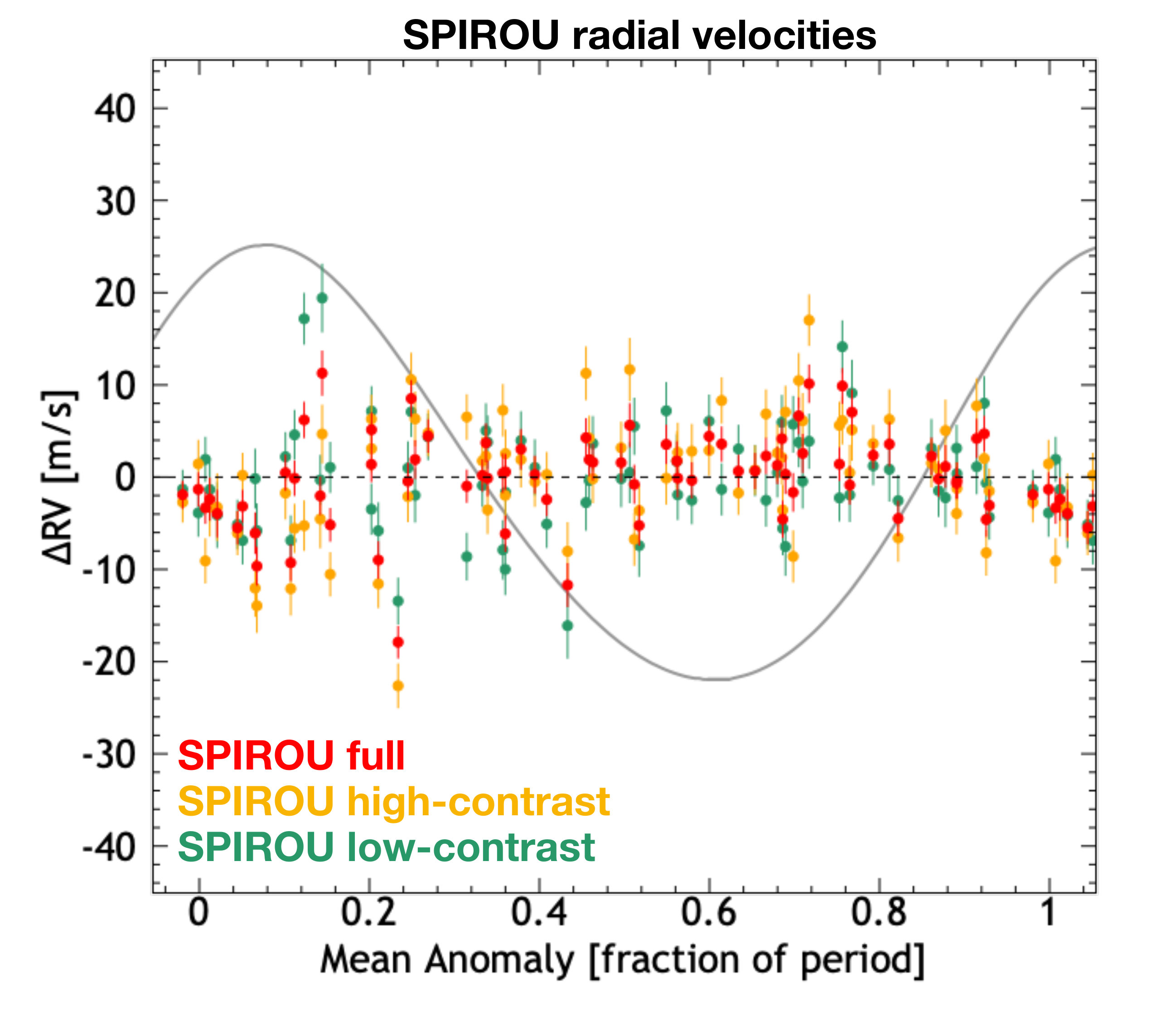}
                 \caption{
                 Phase-folded SPIRou RVs for the LBL calculation including the full, high-contrast, and low-contrast lines.
                 The model is the Keplerian model for the SOPHIE full-mask data.}
                   \label{SPIROU_high_low_RV}
\end{figure}

\begin{figure*}[]
   \centering         
                 \includegraphics[width=0.87\textwidth]{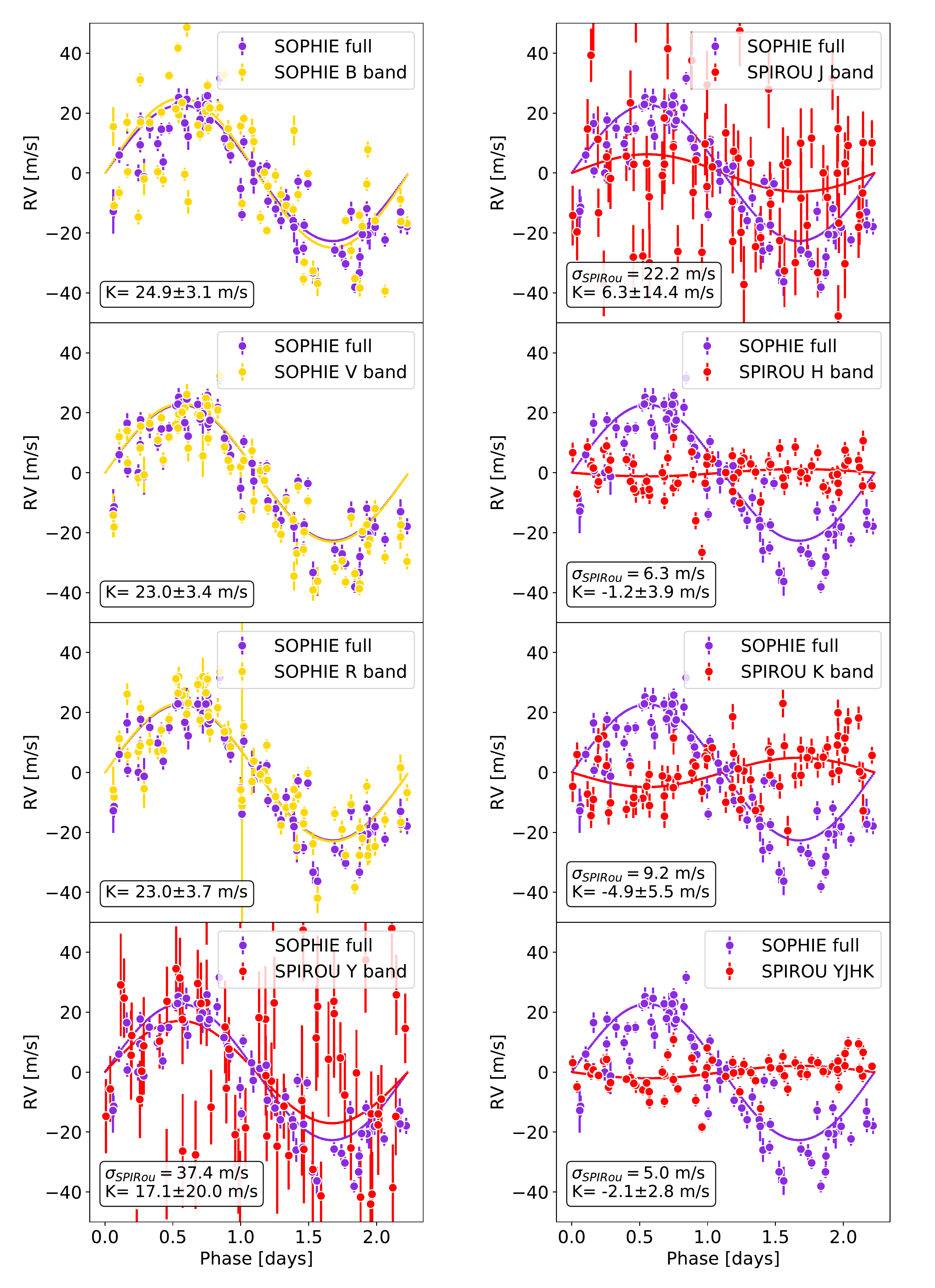}
                 \caption{
                 Phase-folded RVs in the B, V, and R bands for SOPHIE (yellow points) and in the Y, J, H, and K bands for SPIRou (red points).
                 The best sinusoidal fit per band is given in a continuous line following the same color convention.
                 For comparison, in each panel, we include (blue color) the phase-folded RVs and sinusoidal fit obtained in the full SOPHIE range. 
                 For the sinusoidal fit and RV phase-folding, 
                 the phase (i.e., T$_0$ ) and period (2.23 days) are the same for all the time series.   
                 It is important to note that although a sinusoidal fit is possible in the SPIRou data, 
                 the periodograms show that there is no signal at the 2.23 day period 
                 (see Figure~\ref{SPIROU_periodogram_bands}). 
                 The SPIRou data are compatible with a straight line with $K=0$~m~s$^{-1}$. 
                 The sinusoidal fit to the SPIRou data enabled us to set an upper limit on the RV dispersion that is smaller than the rms value.}
                   \label{RV_chromaticity_sinfit}
\end{figure*} 

\begin{figure*}[]   
\centering    
                 \includegraphics[width=0.77\textwidth]{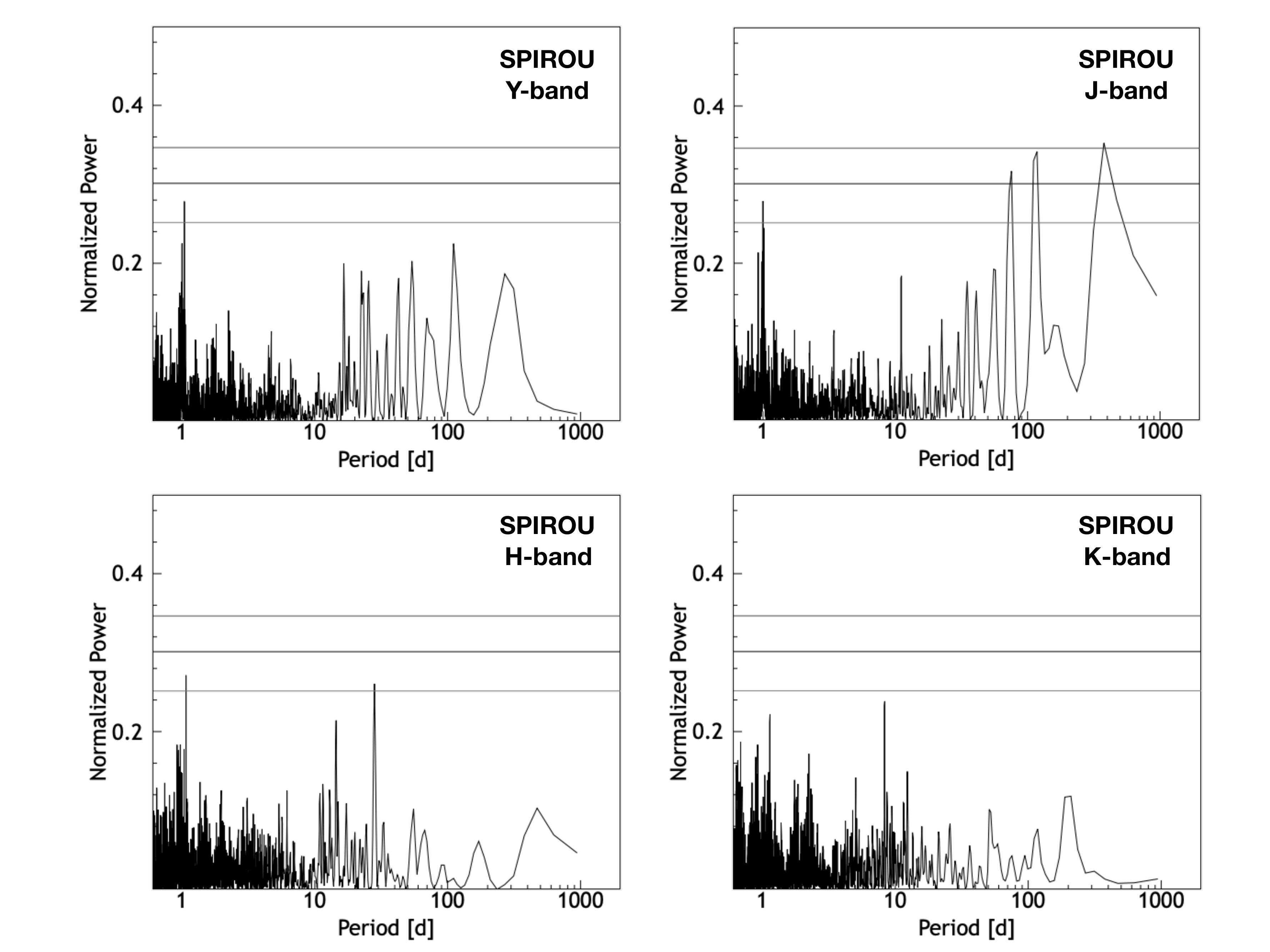}
                 \caption{Periodograms of the SPIRou RVs in the Y, J, H, and K bands. The horizontal lines mark the FAP levels of 0.1, 1, and 10\%. No periodic signal is observed in any of the bands at 2.23 days. }
                   \label{SPIROU_periodogram_bands}
\end{figure*}  

\clearpage

\onecolumn
\section{Corner plots of the quasi-periodic Gaussian process analysis.}
\begin{figure}[h]
    \centering
    \includegraphics[width=\textwidth]{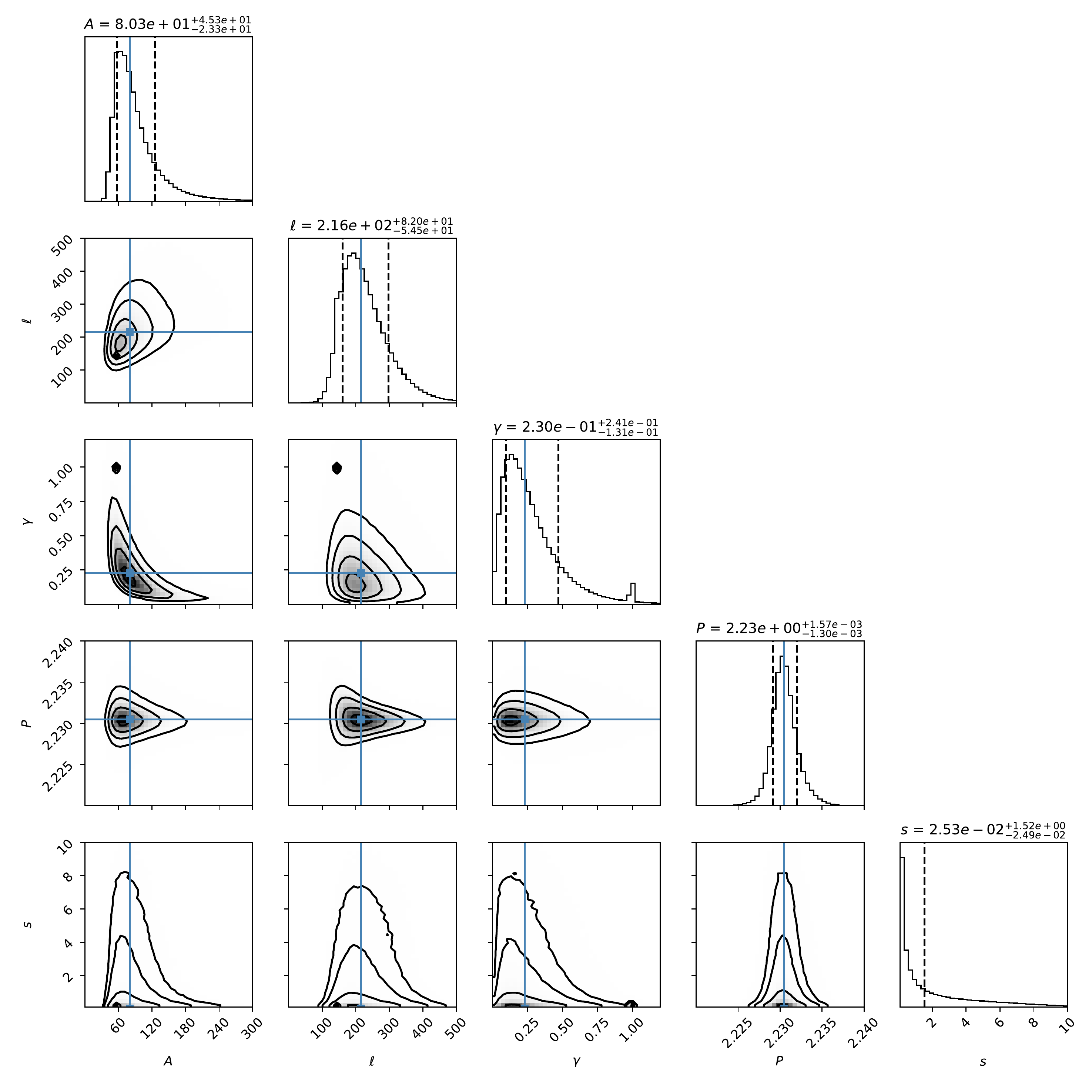}
    \caption{Corner plot of the quasi-periodic GP fit of the SPIRou longitudinal magnetic field (B$_\ell$) time series. 
    The description of  the meaning of the variables is given in Table~\ref{GP_table}.}
    \label{fig:corner_bell}
\end{figure}

\begin{figure}
    \centering
    \includegraphics[width=\textwidth]{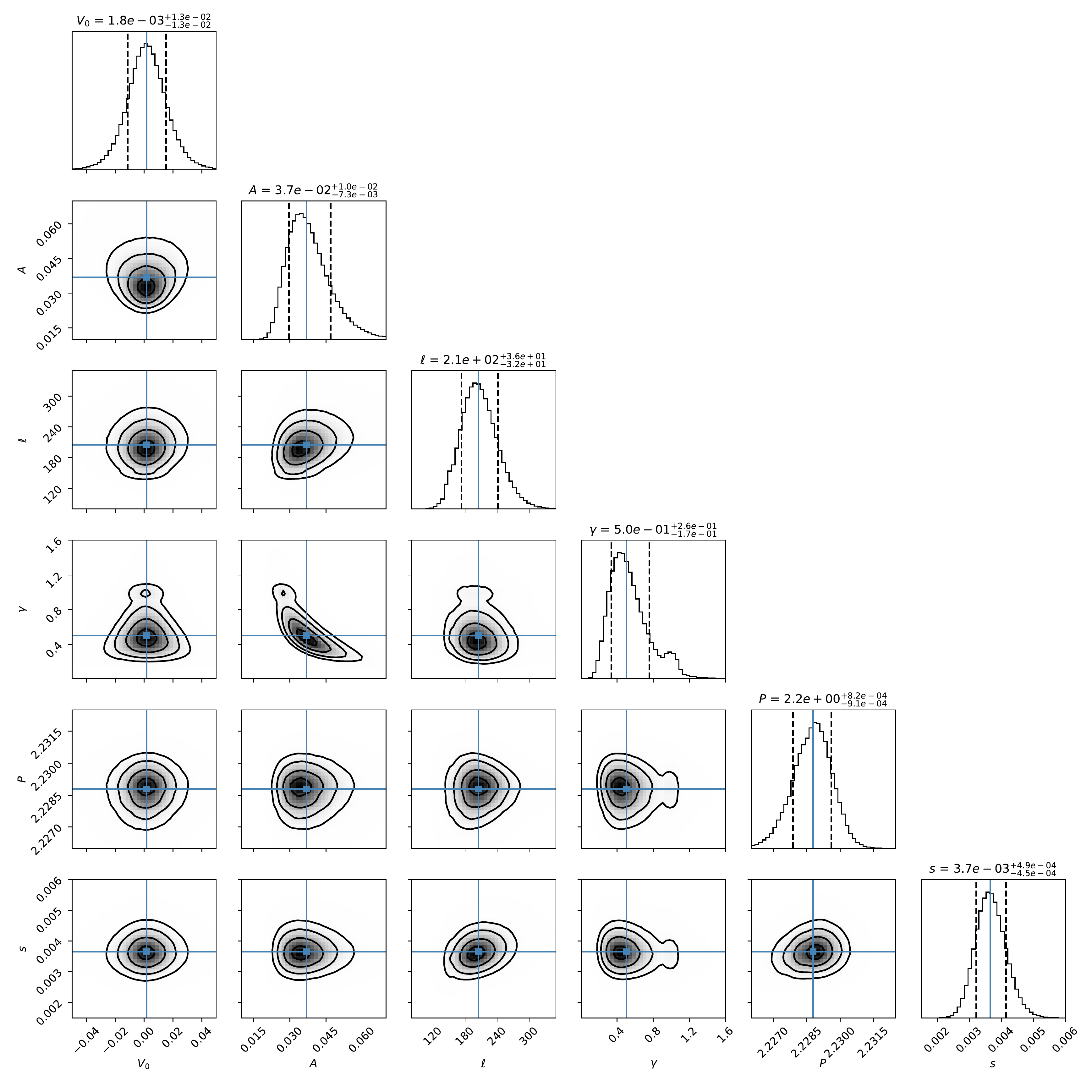}
    \caption{Corner plot of the quasi-periodic GP fit of HARPS and SOPHIE RV time series. 
    The description of  the meaning of the variables is given in Table~\ref{GP_table}.}
    \label{fig:corner_rv}
\end{figure}

\clearpage

\section{Radial velocity measurements}

\begin{longtable}{c c c c c}
\caption{\label{GL388_HARPS_RVtable} HARPS RV measurements of Gl 388}\\
\hline\hline 
DATE & BJD  $-$ 2400000 & $\delta$ RV  &  $\sigma_{\delta{\rm RV}}$  \\[1pt]
[yyyy-mm-dd] & [days] & [m/s]  & [m/s] \\[1pt]
\endfirsthead
\caption{continued.}\\
\hline \hline 
DATE & BJD  $-$ 2400000  & $\delta$ RV & $\sigma_{\delta{\rm RV}}$  \\[1pt]
[yyyy-mm-dd] & [days] & [m/s]  & [m/s] \\[1pt]
\hline
\endhead
\hline
\endfoot
\hline
2005-05-21 &  53511.54684 &      -21.1  &  1.5 \\                                                                
2005-05-30 &  53520.52047 &      -17.2  &  1.5 \\                                                                
2005-06-21 &  53543.48083 &       15.0  &  1.7 \\                                                                
2005-06-22 &  53544.45178 &       -8.2  &  1.4 \\                                                                
2005-06-28 &  53550.45964 &       41.7  &  2.7 \\                                                                
2005-12-24 &  53728.86477 &       40.3  &  1.5 \\                                                                
2006-01-23 &  53758.75397 &      -20.7  &  1.3 \\                                                                
2006-01-25 &  53760.75468 &      -11.3  &  1.3 \\                                                                
2006-01-26 &  53761.78002 &       18.9  &  1.3 \\                                                                
2006-02-17 &  53783.72555 &       -7.9  &  1.3 \\                                                                
2006-02-19 &  53785.72637 &      -19.8  &  1.3 \\                                                                
2006-03-15 &  53809.65993 &       -2.1  &  1.2 \\                                                                
2006-03-16 &  53810.67672 &       10.7  &  1.2 \\                                                                
2006-03-17 &  53811.67496 &        8.2  &  1.2 \\                                                                
2006-03-18 &  53812.66370 &       -8.8  &  1.5 \\                                                                
2006-03-19 &  53813.65843 &       20.4  &  1.2 \\                                                                
2006-03-20 &  53814.65398 &      -20.7  &  1.4 \\                                                                
2006-03-21 &  53815.57020 &       25.5  &  1.3 \\                                                                
2006-03-21 &  53815.62108 &       25.6  &  1.2 \\                                                                
2006-03-21 &  53815.73541 &       24.9  &  1.2 \\                                                                
2006-03-22 &  53816.54328 &      -14.8  &  1.3 \\                                                                
2006-03-22 &  53816.65524 &      -19.2  &  1.2 \\                                                                
2006-03-22 &  53816.72113 &      -21.8  &  1.3 \\                                                                
2006-03-23 &  53817.54998 &       24.3  &  1.2 \\                                                                
2006-03-23 &  53817.67492 &       26.8  &  1.3 \\                                                                
2006-04-04 &  53829.61461 &       -1.2  &  1.2 \\                                                                
2006-04-05 &  53830.53967 &       -6.7  &  1.3 \\                                                                
2006-04-06 &  53831.66939 &        9.1  &  1.2 \\                                                                
2006-04-07 &  53832.64967 &      -11.6  &  1.2 \\                                                                
2006-04-08 &  53833.62688 &       23.8  &  1.2 \\                                                                
2006-04-09 &  53834.61035 &      -27.9  &  1.2 \\                                                                
2006-04-10 &  53835.65510 &       27.5  &  1.2 \\                                                                
2006-04-11 &  53836.61650 &      -19.7  &  1.2 \\                                                                
2006-05-06 &  53861.59344 &        0.0  &  1.3 \\                                                                
2006-05-08 &  53863.56874 &      -21.4  &  1.3 \\                                                                
2006-05-09 &  53864.53404 &       29.4  &  1.2 \\                                                                
2006-05-12 &  53867.54169 &      -13.5  &  1.3 \\                                                                
2006-05-13 &  53868.51770 &       22.7  &  1.3 \\                                                                
2006-05-16 &  53871.56217 &       19.2  &  1.2 \\                                                                
2013-12-30 &  56656.84935 &      -34.3  &  1.4 \\                                                                
2013-12-30 &  56656.86018 &      -33.1  &  1.4 \\                                                                
2013-12-31 &  56657.85223 &       41.7  &  1.4 \\                                                                
2014-01-01 &  56658.86437 &      -43.3  &  1.4 \\                                                                
2014-01-01 &  56658.87555 &      -43.1  &  1.4 \\                                                                
2014-01-02 &  56659.85808 &       45.0  &  1.7 \\                                                                
2014-05-20 &  56797.51177 &      -28.4  &  1.2 \\                                                                
2016-04-14 &  57492.59638 &       14.8  &  1.2 \\                                                                
2016-04-14 &  57492.60715 &       15.7  &  1.2 \\                                                                
2016-04-16 &  57494.53836 &        8.6  &  1.2 \\                                                                
2016-04-16 &  57494.54915 &        5.0  &  1.2 \\                                                                
2016-04-16 &  57494.60770 &        6.5  &  1.3 \\
\hline 
\end{longtable}

\clearpage

\begin{longtable}{c c c c c c}
\caption{\label{GL388_SOPHIE_RVtable} SOPHIE RV measurements of Gl 388}\\
\hline\hline 
DATE             & BJD  $-$ 2400000   & RV     & $\sigma_{\rm RV}$ & $\delta$ RV & $\sigma_{\delta{\rm RV}}$   \\[1pt]
[yyyy-mm-dd] & [days] & [km/s] & [km/s]        & [m/s]            & [m/s]         \\[1pt]
\endfirsthead
\caption{continued.}\\
\hline \hline 
DATE             & BJD  $-$ 2400000    & RV     & $\sigma_{\rm RV}$ & $\delta$ RV & $\sigma_{\delta{\rm RV}}$   \\[1pt]
[yyyy-mm-dd] & [days] & [km/s] & [km/s]        & [m/s]             & [m/s]         \\[1pt]

\hline
\endhead
\hline
\endfoot
\hline
2019-03-02 & 58545.47800 &    12.5304 &   0.0031 &     25.2 &  3.1 \\
2019-03-03 & 58546.42477 &    12.5016 &   0.0020 &     -3.6 &  2.0 \\
2019-03-08 & 58550.56659 &    12.5059 &   0.0023 &      0.7 &  2.3 \\
2019-03-08 & 58551.45481 &    12.4829 &   0.0023 &    -22.3 &  2.3 \\
2019-03-09 & 58552.45192 &    12.5270 &   0.0029 &     21.8 &  2.9 \\
2019-03-10 & 58553.43267 &    12.4924 &   0.0032 &    -12.8 &  3.2 \\
2019-03-13 & 58556.47387 &    12.5150 &   0.0025 &      9.8 &  2.5 \\
2019-03-15 & 58557.50396 &    12.5024 &   0.0023 &     -2.8 &  2.3 \\
2019-03-16 & 58559.45769 &    12.5064 &   0.0024 &      1.2 &  2.4 \\
2019-03-18 & 58560.53512 &    12.4873 &   0.0027 &    -17.9 &  2.7 \\
2019-03-21 & 58563.53618 &    12.5284 &   0.0025 &     23.2 &  2.5 \\
2019-04-12 & 58586.40481 &    12.4969 &   0.0030 &     -8.3 &  3.0 \\
2019-04-13 & 58587.37284 &    12.4938 &   0.0036 &    -11.4 &  3.6 \\
2019-04-14 & 58588.39794 &    12.5024 &   0.0041 &     -2.8 &  4.1 \\
2019-04-17 & 58591.48120 &    12.4850 &   0.0053 &    -20.2 &  5.3 \\
2019-04-18 & 58592.48354 &    12.5249 &   0.0067 &     19.7 &  6.7 \\
2019-04-20 & 58593.50800 &    12.4781 &   0.0031 &    -27.1 &  3.1 \\
2019-04-26 & 58600.38700 &    12.4846 &   0.0031 &    -20.6 &  3.1 \\
2019-04-27 & 58601.36690 &    12.5280 &   0.0024 &     22.8 &  2.4 \\
2019-04-30 & 58604.37889 &    12.4878 &   0.0021 &    -17.4 &  2.1 \\
2019-05-01 & 58605.40743 &    12.5229 &   0.0025 &     17.7 &  2.5 \\
2019-05-02 & 58606.39785 &    12.4932 &   0.0027 &    -12.0 &  2.7 \\
2019-05-06 & 58610.36469 &    12.5214 &   0.0023 &     16.2 &  2.3 \\
2019-05-16 & 58620.39921 &    12.4772 &   0.0031 &    -28.0 &  3.1 \\
2019-05-20 & 58624.39406 &    12.4792 &   0.0040 &    -26.0 &  4.0 \\
2019-11-13 & 58800.68402 &    12.4719 &   0.0037 &    -33.3 &  3.7 \\
2019-11-20 & 58807.71413 &    12.4719 &   0.0021 &    -33.3 &  2.1 \\
2019-12-03 & 58820.68227 &    12.4801 &   0.0022 &    -25.1 &  2.2 \\
2019-12-04 & 58821.61889 &    12.5217 &   0.0034 &     16.5 &  3.4 \\
2019-12-07 & 58824.69070 &    12.4913 &   0.0020 &    -13.9 &  2.0 \\
2019-12-08 & 58825.60877 &    12.4933 &   0.0025 &    -11.9 &  2.5 \\
2020-01-04 & 58852.61815 &    12.4922 &   0.0042 &    -13.0 &  4.2 \\
2020-01-05 & 58853.59626 &    12.5110 &   0.0026 &      5.8 &  2.6 \\
2020-01-06 & 58854.56547 &    12.4847 &   0.0023 &    -20.5 &  2.3 \\
2020-01-09 & 58857.60659 &    12.5201 &   0.0019 &     14.9 &  1.9 \\
2020-01-11 & 58859.52973 &    12.5059 &   0.0023 &      0.7 &  2.3 \\
2020-01-12 & 58860.56334 &    12.4958 &   0.0022 &     -9.4 &  2.2 \\
2020-01-13 & 58861.69770 &    12.5112 &   0.0026 &      6.0 &  2.6 \\
2020-01-16 & 58864.56517 &    12.5280 &   0.0020 &     22.8 &  2.0 \\
2020-01-29 & 58877.63433 &    12.5089 &   0.0028 &      3.7 &  2.8 \\
2020-01-31 & 58879.67830 &    12.5052 &   0.0028 &      0.0 &  2.8 \\
2020-02-02 & 58881.61068 &    12.4879 &   0.0028 &    -17.3 &  2.8 \\
2020-02-03 & 58882.56870 &    12.5137 &   0.0033 &      8.5 &  3.3 \\
2020-02-07 & 58886.53954 &    12.5198 &   0.0021 &     14.6 &  2.1 \\
2020-02-07 & 58886.71058 &    12.5219 &   0.0028 &     16.7 &  2.8 \\
2020-02-08 & 58887.50692 &    12.4893 &   0.0021 &    -15.9 &  2.1 \\
2020-02-09 & 58888.61378 &    12.5146 &   0.0025 &      9.4 &  2.5 \\
2020-02-12 & 58891.58107 &    12.5000 &   0.0036 &     -5.2 &  3.6 \\
2020-02-13 & 58892.56896 &    12.4871 &   0.0029 &    -18.1 &  2.9 \\
2020-02-14 & 58893.58562 &    12.5227 &   0.0033 &     17.5 &  3.3 \\
2020-02-15 & 58894.50449 &    12.4795 &   0.0023 &    -25.7 &  2.3 \\
2020-02-18 & 58897.56111 &    12.5039 &   0.0061 &     -1.3 &  6.1 \\
2020-02-22 & 58902.43552 &    12.5231 &   0.0022 &     17.9 &  2.2 \\
2020-02-25 & 58904.53132 &    12.5287 &   0.0023 &     23.5 &  2.3 \\
2020-02-26 & 58905.52570 &    12.4689 &   0.0047 &    -36.3 &  4.7 \\
2020-02-27 & 58906.52174 &    12.5202 &   0.0035 &     15.0 &  3.5 \\
2020-02-27 & 58907.48930 &    12.4893 &   0.0029 &    -15.9 &  2.9 \\
2020-03-04 & 58913.49506 &    12.5174 &   0.0045 &     12.2 &  4.5 \\
2020-03-08 & 58916.50231 &    12.4871 &   0.0050 &    -18.1 &  5.0 \\
2020-03-09 & 58918.42554 &    12.5083 &   0.0041 &      3.1 &  4.1 \\
2020-03-11 & 58920.45204 &    12.5167 &   0.0024 &     11.5 &  2.4 \\
2020-03-13 & 58922.40620 &    12.5298 &   0.0036 &     24.6 &  3.6 \\
2021-01-28 & 59242.53312 &    12.4671 &   0.0020 &    -38.1 &  2.0 \\
2021-02-03 & 59248.57311 &    12.5075 &   0.0019 &      2.3 &  1.9 \\
2021-02-18 & 59263.51565 &    12.5276 &   0.0020 &     22.4 &  2.0 \\
2021-02-22 & 59267.51704 &    12.4924 &   0.0074 &    -12.8 &  7.4 \\
2021-02-25 & 59271.45256 &    12.4749 &   0.0020 &    -30.3 &  2.0 \\
2021-03-05 & 59279.44906 &    12.5368 &   0.0021 &     31.6 &  2.1 \\
2021-03-23 & 59297.46535 &    12.5156 &   0.0021 &     10.4 &  2.1 \\
2021-03-27 & 59301.44277 &    12.5281 &   0.0020 &     22.9 &  2.0 \\
2021-04-23 & 59328.41663 &    12.5310 &   0.0022 &     25.8 &  2.2 \\
\hline 
\end{longtable}

\clearpage

\begin{longtable}{c c c c c c c c}
\caption{\label{SPIRouRVBlTable} SPIRou RV and B$_\ell$ measurements of Gl 388}\\
\hline\hline 
DATE           & BJD  $-$ 2400000 & RV     & $\sigma_{\rm RV}$ & $\delta$ RV & $\sigma_{\delta{\rm RV}}$ & B$_\ell$   & $\sigma_{{\rm B}_\ell}$\\[1pt]
[yy-mm--dd] & [days]                    & [m/s]   & [m/s]       &  [m/s]           & [m/s]        & [Gauss] & [Gauss] \\[1pt]
\hline
\endfirsthead
\caption{continued.}\\
\hline\hline  
DATE           & BJD  $-$ 2400000 & RV     & $\sigma_{\rm RV}$ & $\delta$ RV & $\sigma_{\delta{\rm RV}}$ & B$_\ell$   & $\sigma_{{\rm B}_\ell}$\\[1pt]
[yy-mm--dd] & [days]                    & [m/s]   & [m/s]       &  [m/s]           & [m/s]        & [Gauss] & [Gauss] \\[1pt]
\hline
\endhead
\hline
\endfoot
\hline
2019-02-14  &  58529.07706 & 12.6943  &  0.0021 & -2.1  &  2.1  &  -  &  -\\
2019-02-16  &  58530.97429 & 12.6995  &  0.0021 & 3.1  &  2.1  &  -  &  -\\
2019-02-17  &  58532.13061 & 12.6863  &  0.0021 & -10.1  &  2.1  &  -  &  -\\
2019-02-18  &  58532.82316 & 12.6990  &  0.0022 & 2.6  &  2.2  &  -  &  -\\
2019-02-25  &  58539.81077 & 12.6952  &  0.0021 & -1.2  &  2.1  &  -  &  -\\
2019-02-26  &  58540.99986 & 12.6904  &  0.0018 & -5.9  &  1.8  &  -  &  -\\
2019-04-15  &  58588.76421 & 12.6976  &  0.0020 & 1.2  &  2.0  &  -223.4  &  17.5  \\
2019-04-16  &  58590.00871 & 12.6920  &  0.0026 & -4.4  &  2.6  &  -199.4  &  20.4  \\
2019-04-18  &  58591.88226 & 12.6982  &  0.0020 & 1.8  &  2.0  &  -217.5  &  17.1  \\
2019-04-19  &  58592.94805 & 12.6958  &  0.0020 & -0.5  &  2.0  &  -258.1  &  17.5  \\
2019-04-20  &  58593.87858 & 12.7058  &  0.0019 & 9.5  &  1.9  &  -195.0 &  17.1  \\
2019-04-21  &  58594.74432 & 12.7072  &  0.0025 & 10.8  &  2.5  &  -218.2  &  16.7  \\
2019-04-22  &  58595.87970 & 12.6966  &  0.0020 & 0.2  &  2.0  &  -226.1  &  17.3  \\
2019-04-23  &  58596.79859 & 12.6899  &  0.0023 & -6.5  &  2.3  &  -229.5  &  17.0  \\
2019-04-24  &  58597.75866 & 12.6975  &  0.0021 & 1.1  &  2.1  &  -213.3  &  18.2  \\
2019-04-25  &  58598.90970 & 12.6935  &  0.0023 & -2.8  &  2.3  &  -228.2  &  17.4  \\
2019-04-26  &  58599.84821 & 12.6842  &  0.0024 & -12.1  &  2.4  &  -250.9  &  18.4  \\
2019-05-01  &  58604.87986 & 12.6962  &  0.0021 & -0.1  &  2.1  &  -222.3  &  16.9  \\
2019-06-13  &  58647.77284 & 12.7006  &  0.0019 & 4.3  &  1.9  &  -217.3  &  13.0  \\
2019-06-14  &  58648.74045 & 12.6964  &  0.0021 & 0.0  &  2.1  &  -225.8  &  13.5  \\
2019-06-15  &  58649.75406 & 12.6995  &  0.0024 & 3.1  &  2.4  &  -200.9  &  16.6  \\
2019-06-16  &  58650.74030 & 12.6978  &  0.0021 & 1.5  &  2.1  &  -201.0  &  14.7 \\
2019-06-17  &  58651.75684 & 12.6985  &  0.0021 & 2.1  &  2.1  &  -211.1  &  16.9  \\
2019-06-19  &  58653.74160 & 12.7004  &  0.0020 & 4.0  &  2.0  &  -239.4  &  14.5  \\
2019-06-21  &  58655.76329 & 12.7016  &  0.0024 & 5.2  &  2.4  &  -277.4  &  21.2  \\
2019-10-16  &  58773.14644 & 12.6939  &  0.0021 & -2.5  &  2.1  &  -208.9  &  15.8  \\
2019-10-31  &  58788.14872 & 12.6957  &  0.0019 & -0.6  &  1.9  &  -205.8  &  12.2  \\
2019-11-01  &  58789.14130 & 12.6950  &  0.0018 & -1.4  &  1.8  &  -190.0 &  11.9  \\
2019-11-02  &  58790.15166 & 12.7030  &  0.0025 & 6.6  &  2.5  &  -219.1  &  14.9  \\
2019-11-03  &  58791.12146 & 12.7011  &  0.0019 & 4.7  &  1.9  &  -170.1  &  12.5  \\
2019-11-04  &  58792.15630 & 12.6982  &  0.0020 & 1.9  &  2.0  &  -226.3  &  13.9  \\
2019-11-05  &  58793.14964 & 12.6958  &  0.0019 & -0.6  &  1.9  &  -188.3  &  12.1  \\
2019-11-06  &  58794.15242 & 12.6977  &  0.0022 & 1.3  &  2.2  &  -250.5  &  14.6  \\
2019-11-07  &  58795.12735 & 12.6946  &  0.0019 & -1.8  &  1.9  &  -193.1  &  13.2  \\
2019-11-09  &  58797.08716 & 12.6971  &  0.0023 & 0.7  &  2.3  &  -170.1  &  21.7  \\
2019-11-10  &  58798.16258 & 12.6898  &  0.0020 & -6.6  &  2.0  &  -173.3  &  15.2  \\
2019-11-13  &  58801.11817 & 12.7001  &  0.0021 & 3.7  &  2.1  &  -223.3  &  14.9  \\
2019-11-14  &  58802.09522 & 12.7021  &  0.0020 & 5.8  &  2.0  &  -180.9  &  13.7  \\
2019-12-05  &  58823.14593 & 12.6958  &  0.0021 & -0.5  &  2.1  &  -255.71; -259.0  &  30.0; 39.9  \\
2019-12-07  &  58825.13544 & 12.7002  &  0.0021 & 3.8  &  2.1  &  -190.9  &  17.9  \\
2019-12-08  &  58826.10827 & 12.6956  &  0.0020 & -0.8  &  2.0  &  -228.4  &  13.9  \\
2019-12-09  &  58827.10307 & 12.6997  &  0.0021 & 3.3  &  2.1  &  -185.1  &  13.6  \\
2019-12-10  &  58828.05731 & 12.6951  &  0.0021 & -1.3  &  2.1  &  -235.4  &  13.8  \\
2019-12-11  &  58829.12854 & 12.6955  &  0.0020 & -0.9  &  2.0  &  -159.7  &  13.8  \\
2019-12-12  &  58830.11152 & 12.6914  &  0.0020 & -5.0  &  2.0  &  -226.2  &  13.0  \\
2020-01-26  &  58875.01623 & 12.6915  &  0.0019 & -4.9  &  1.9  &  - &  -\\
2020-01-27  &  58876.01327 & 12.7004  &  0.0018 & 4.0  &  1.8  &  - &  -\\
2020-01-28  &  58877.01259 & 12.7061  &  0.0021 & 9.7  &  2.1  &  - &  -\\
2020-02-05  &  58884.85491 & 12.6780  &  0.0018 & -18.3  &  1.8  &  -128.8  &  11.9  \\
2020-02-16  &  58895.82736 & 12.6908  &  0.0018 & -5.6  &  1.8  &  -133.7  &  11.3  \\
2020-02-17  &  58896.89804 & 12.6966  &  0.0018 & 0.2  &  1.8  &  -231.5  &  12.3  \\
2020-02-18  &  58897.82751 & 12.6928  &  0.0018 & -3.6  &  1.8  &  -149.6  &  11.6  \\
2020-02-19  &  58898.86775 & 12.6907  &  0.0022 & -5.7  &  2.2  &  -213.4  &  13.0  \\
2020-03-11  &  58919.84910 & 12.6913  &  0.0019 & -5.0  &  1.9  &   -187.4  &   12.7 \\
2020-03-12  &  58920.89500 & 12.6962  &  0.0021 & -0.1  &  2.1  &  -189.7  &  12.2  \\
2020-05-08  &  58977.75348 & 12.6953  &  0.0017 & -1.1  &  1.7  &  -191.5  &  11.7  \\
2020-05-09  &  58978.90858 & 12.6935  &  0.0018 & -2.9  &  1.8  &  -110.1  &  12.4  \\
2020-05-12  &  58981.90631 & 12.6973  &  0.0018 & 1.0  &  1.8  &  -203.3 &  14.8  \\
2020-05-13  &  58982.90871 & 12.6973  &  0.0019 & 1.0  &  1.9  &  -101.5  &  12.8  \\
2020-05-14  &  58983.82758 & 12.6995  &  0.0019 & 3.2  &  1.9  &  -192.4  &  10.9  \\
2020-05-15  &  58984.91350 & 12.6964  &  0.0020 & 0.0  &  2.0  &  -91.1  &  18.4  \\
2020-05-31  &  59000.76772 & 12.6870  &  0.0020 & -9.4  &  2.0  &  -60.9  &  10.7  \\
2020-06-01  &  59001.81430 & 12.6972  &  0.0020 & 0.9  &  2.0  &  -193.8  &  18.5  \\
2020-06-02  &  59002.76806 & 12.6867  &  0.0020 & -9.7  &  2.0  &  -84.4  &  13.5  \\
2020-06-03  &  59003.82018 & 12.6956  &  0.0020 & -0.8  &  2.0  &  -190.8  &  14.1  \\
2020-06-04  &  59004.77426 & 12.6926  &  0.0017 & -3.7  &  1.7  &  -114.0  &  12.5  \\
2020-06-05  &  59005.78111 & 12.6978  &  0.0018 & 1.5  &  1.8  &  -120.2  &  15.4  \\
2020-06-06  &  59006.83047 & 12.6929  &  0.0017 & -3.5  &  1.7  &  -126.1  &  15.0  \\
2020-06-07  &  59007.79158 & 12.6965  &  0.0020 & 0.1  &  2.0  &  -65.6  &  15.3  \\
2020-06-08  &  59008.75599 & 12.6983  &  0.0015 & 1.9  &  1.5  &  -167.2; -187.3  &  16.4; 16.9 \\
2020-06-09  &  59009.77433 & 12.7045  &  0.0020 & 8.1  &  2.0  &  -88.9  &  20.0  \\
2020-06-10  &  59010.79073 & 12.7026  &  0.0020 & 6.2  &  2.0  &  -190.6  &  14.6  \\
2020-10-31  &  59154.13047 & 12.6940  &  0.0015 & -2.3  &  1.5  &  -68.5  &  10.9  \\
2020-11-03  &  59157.14680 & 12.6961  &  0.0017 & -0.2  &  1.7  &  -54.8  &  11.7  \\
\hline 
\end{longtable}

\end{appendix}

\end{document}